\documentclass{pasj00}
\draft

\begin{document}
\SetRunningHead{Imanishi et al.}{Akari spectroscopy of ULIRGs}
\Received{2008/05/06}%{yyyy/mm/dd}
\Accepted{2008/??/??}%{yyyy/mm/dd}

\title{Systematic infrared 2.5--5 $\mu$m spectroscopy of nearby
ultraluminous infrared galaxies with AKARI}

%%% begin:list of authors
% Do NOT capitalize all letters in "textsc".
\author{Masatoshi \textsc{Imanishi} %
  \thanks{Department of Astronomy, School of Science, Graduate
University for Advanced Studies, Mitaka, Tokyo 181-8588, Japan}}
\affil{National Astronomical Observatory, 2-21-1, Osawa, Mitaka, Tokyo
181-8588, Japan}
\email{masa.imanishi@nao.ac.jp}

\author{Takao \textsc{Nakagawa}, Yoichi \textsc{Ohyama} 
  \thanks{Present address: Institute of Astronomy and Astrophysics
Academia Sinica, P.O. Box 23-141, Taipei 10617, Taiwan, R.O.C.}, 
Mai {\sc Shirahata}, Takehiko {\sc Wada}} 
\affil{Institute of Space and Astronautical Science, Japan Aerospace
Exploration Agency, 3-1-1 Yoshinodai, Sagamihara, Kanagawa 229-8510,
Japan}

\author{Takashi \textsc{Onaka}} 
\affil{
Department of Astronomy, Graduate School of Science, University of
Tokyo, Bunkyo-ku, Tokyo 113-0033, Japan}

\and

\author{Nagisa \textsc{Oi}
\thanks{National Astronomical Observatory, 2-21-1, Osawa, Mitaka, Tokyo
181-8588, Japan}
} 
\affil{
Department of Astronomy, School of Science, Graduate
University for Advanced Studies, Mitaka, Tokyo 181-8588, Japan}

%%% end:list of authors

%%% Please use the following style in case that sorting by 
%%% affilation is impossible. 
%
% \author{%
%   D-Firstname \textsc{D-Familyname}\altaffilmark{1}
%   E-Firstname \textsc{E-Familyname}\altaffilmark{1,2}
%   and
%   F-Firstname \textsc{F-Familyname}\altaffilmark{2}}
% \altaffiltext{1}{Address of Institute}
% \email{ddddd@xxx.xxx.xx.xx}
% \email{eeeee@xxx.xxx.xx.xx}
% \altaffiltext{2}{Address of Institute}

%% `\KeyWords{}' always has to be placed before `\maketitle'.
 %Do NOT move this preamble from here!

\maketitle

\begin{abstract}
We report on the results of systematic infrared 2.5--5 $\mu$m
spectroscopy of 45 nearby ultraluminous infrared galaxies (ULIRGs) at $z
<$ 0.3 using the IRC infrared spectrograph onboard the
AKARI satellite. This paper investigates whether the luminosities of
these ULIRGs are dominated by starburst activity, or alternatively,
whether optically elusive buried active galactic nuclei (AGNs) are
energetically important.  Our criteria include the strengths of the 3.3
$\mu$m polycyclic aromatic hydrocarbon (PAH) emission features and the
optical depths of absorption features at 3.1 $\mu$m due to ice-covered
dust grains and at 3.4 $\mu$m from bare carbonaceous dust
grains. Because of the AKARI IRC's spectroscopic capability in the full
2.5--5 $\mu$m wavelength range, unaffected by Earth's atmosphere, we
can apply this energy diagnostic method to ULIRGs at $z >$ 0.15. We
estimate the intrinsic luminosities of extended (several kpc), modestly
obscured (A$_{\rm V}$ $<$ 15 mag) starburst activity based on the
observed 3.3 $\mu$m PAH emission luminosities measured in AKARI IRC
slitless spectra, and confirm that such starbursts are energetically
unimportant in nearby ULIRGs. In roughly half of the observed ULIRGs
classified optically as non-Seyferts, we find signatures of luminous
energy sources that produce no PAH emission and/or are more centrally
concentrated than the surrounding dust. We interpret these energy
sources as buried AGNs. The fraction of ULIRGs with detectable buried
AGN signatures is higher in ULIRGs classified optically as LINERs than
HII-regions, and increases with increasing infrared luminosity. 
Our overall results support the scenario that luminous buried AGNs are
important in many ULIRGs at $z <$ 0.3 classified optically as
non-Seyferts, and that the optical undetectability of such buried AGNs
occurs merely because of a large amount of nuclear dust, which can make
the sightline of even the lowest dust column density opaque to the
ionizing radiation of the AGNs. 
\end{abstract}

\section{Introduction}

Ultraluminous infrared galaxies (ULIRGs) have large luminosities ($L>
10^{12}L_{\odot}$) that are radiated mostly as infrared dust emission
\citep{san88a,sam96}. This means that very luminous energy sources ($L>
10^{12}L_{\odot}$) are present hidden behind dust, so that most of the
energetic photons from these energy sources are once absorbed by the
surrounding dust and the heated dust then emits strongly in the
infrared.
The ULIRG population dominates the bright end of the luminosity function
in the local universe ($z <$ 0.3) \citep{soi87}. The contribution from
ULIRGs to the total infrared radiation density increases rapidly with
increasing redshift \citep{lef05} and becomes important at $z >$ 1.3
\citep{per05}. Distinguishing whether the dust-obscured energy sources
of ULIRGs are dominated by starburst activity (nuclear fusion inside
stars), or whether active galactic nuclei (AGNs; active mass accretion
onto compact supermassive black holes with M $>$ 10$^7$M$_{\odot}$) are
also energetically important, is closely related to unveiling the
history of dust-obscured star formation and supermassive black hole
growth in the universe.    

Identifying the dust-obscured energy sources in ULIRGs is difficult
because a large amount of molecular gas and dust is
concentrated in the nuclear regions of ULIRGs
\citep{dow98,ink06,ima07b}, and can easily {\it bury} (= obscure in all
directions) the putative AGNs because the emitting regions of AGNs are
spatially very compact. Unlike AGNs obscured by torus-shaped dust, which
are classified optically as Seyfert 
2s, such {\it buried} AGNs are elusive to conventional optical
spectroscopy \citep{mai03}, and thus very difficult to detect. However,
quantitative determination of the energetic importance of buried AGNs is
indispensable in understanding the true nature of the ULIRG population. 

To study such optically elusive buried AGNs in ULIRGs, observing them at
wavelengths of low dust extinction is important. Infrared 3--4 $\mu$m
(rest-frame) spectroscopy is one such wavelength. Dust extinction at
3--4 $\mu$m is as low as that at 5--13 $\mu$m \citep{lut96}. More
importantly, starburst and AGN emission are distinguishable from
spectral shapes based on the following two arguments. 
(1) A normal starburst galaxy should always show large equivalent width
polycyclic aromatic hydrocarbons (PAH) emission at rest-frame 3.3
$\mu$m, while an AGN shows a PAH-free continuum originating in
AGN-heated hot dust emission \citep{moo86,imd00,idm06}. Hence, if the
equivalent width of the 3.3 $\mu$m PAH emission is substantially smaller
than that of starburst-dominated galaxies, then a significant
contribution from an AGN to the observed 3--4 $\mu$m flux is the most
natural explanation \citep{imd00,idm01,im03,idm06,ima06}.  
(2) In a normal starburst, the stellar energy sources and dust are
spatially well mixed (Figure 1a), while in a buried AGN, the energy
source (= a compact mass accreting supermassive black hole) is more
centrally concentrated than the surrounding dust (Figure 1b). In a
normal starburst with mixed dust/source geometry, the optical depths of
dust absorption features at 3.1 $\mu$m by ice-covered dust and at 3.4
$\mu$m by bare carbonaceous dust cannot exceed certain thresholds, but
can be arbitrarily large in a buried AGN \citep{im03,idm06}. Therefore,
detection of strong dust absorption features whose optical depths are
substantially larger than the upper limits set by mixed dust/source
geometry suggests the buried-AGN-like centrally concentrated energy
source geometry \citep{im03,idm06}.  

Infrared 3--4 $\mu$m slit spectroscopy of a large sample of nearby
ULIRGs using infrared spectrographs attached to ground-based large
telescopes was performed, and signatures of intrinsically luminous
buried AGNs were found in a significant fraction of optically
non-Seyfert ULIRGs
\citep{imd00,idm01,im03,idm06,ris06,ima06,ima07b,san08}. However, the
observed sample was restricted to $z <$ 0.15 because above this
redshift, a part of the above-mentioned spectral features pass beyond
the Earth's atmospheric window ($L$-band; 2.8--4.1 $\mu$m), making this
3--4 $\mu$m energy diagnostic method impossible. Because of the 2.5--5
$\mu$m spectroscopic capability of the IRC 
\citep{ona07,ohy07}, which was mounted onboard the AKARI satellite
\citep{mur07} and thus unaffected by Earth's atmosphere, we can now
extend this successful approach to more distant ULIRGs at $z> $ 0.15. As
AKARI IRC spectroscopy from space is quite sensitive, given the lack of 
large background emission from Earth's atmosphere, this extension to higher
redshift is feasible in terms of sensitivity. Additionally, the AKARI
IRC employs slitless spectroscopy, so that the bulk of the extended
starburst emission in host galaxies ($>$ several kpc scale) is covered,
unless it is extended more than 1 $\times$ 1 arcmin$^{2}$ (Onaka et
al. 2007). The observed 3.3 $\mu$m PAH emission luminosities in the
AKARI IRC spectra can thus be used to roughly estimate the intrinsic
luminosities of modestly obscured (A$_{\rm V}$ $<$ 15 mag) extended
starbursts in the ULIRG host galaxies, which is impossible with
ground-based spectroscopy using a narrow ($<$ a few arcsec) slit. Based
on observations of several sources, nearby ULIRGs are argued to 
be energetically dominated by highly obscured {\it compact} 
($<$1 kpc) nuclear cores, with small contributions from {\it extended},
modestly obscured starbursts in host galaxies \citep{soi00,fis00}. The
AKARI IRC spectra can directly test whether this argument holds for the
majority of nearby ULIRGs.   

In this paper, we present the results of systematic 2.5--5 $\mu$m
slitless spectroscopy of a large number of nearby ULIRGs at $z <$ 0.3
using the AKARI IRC. Throughout this paper, $H_{0}$ $=$ 75 km s$^{-1}$
Mpc$^{-1}$, $\Omega_{\rm M}$ = 0.3, and $\Omega_{\rm \Lambda}$ = 0.7 are
adopted to be consistent with our previous papers
\citep{idm06,ima06}. The physical scale of 1$''$ is 0.74 kpc at $z =$
0.040 (the nearest source), 2.44 kpc at $z =$ 0.15, and 3.84 kpc at $z
=$ 0.268 (the farthest source).  

\section{Targets}

Our targets are basically selected from the list of ULIRGs in the {\it
IRAS} 1 Jy sample \citep{kim98}. Since AKARI employs the Sun-synchronous
polar orbit, flying along the day--night border \citep{mur07}, assigning
pointed observations of objects with high ecliptic latitudes is easy due
to high visibility, but is difficult for low ecliptic latitude
objects. Hence, our sample is biased toward ULIRGs at high ecliptic
latitudes, but there is no obvious bias with respect to the physical
nature of the sources. 
ULIRGs at $z >$ 0.15 are higher priority than those at $z <$
0.15 because rest-frame 3--4 $\mu$m spectroscopy of the $z >$ 0.15
sources is possible only with the AKARI IRC. However, ULIRGs at $z <
0.15$ are also included as second priority sources for the following
reasons: first, using AKARI IRC slitless spectra, we can measure the
total 3.3 $\mu$m PAH emission luminosity from entire galactic regions,
and then test the previous argument that extended, modestly obscured
starbursts in host galaxies are indeed energetically insignificant in
ULIRGs (see $\S$1). Second, the AKARI IRC has higher sensitivity at
$>$2.5 $\mu$m and wider wavelength coverage than ground-based $L$-band
spectra. Thus, even for ULIRGs at $z <$ 0.15, with available
ground-based $L$-band spectra, we can better determine the continuum
level for the broad 3.1 $\mu$m absorption features in higher-quality
AKARI IRC spectra. Additionally, absorption features at $\lambda_{\rm
rest}$ $>$ 4 $\mu$m (rest-frame), such as CO (4.67 $\mu$m) and CO$_{2}$
(4.26 $\mu$m), can be newly investigated, unless they are redshifted to
$\lambda_{\rm obs}$ $>$ 5 $\mu$m (observed frame). 
 
Since our primary scientific aim was to find optically elusive buried
AGNs ($\S$1), ULIRGs classified optically as non-Seyferts (i.e., LINER,
HII-region, and unclassified) were our main targets. Based on the
optical classification by \citet{vei99}, 15 LINER, 12 HII-region, and 6
unclassified ULIRGs at $z >$0.15 exist in the IRAS 1 Jy sample. Among
these, 11 LINER, 10 HII-region, and 5 unclassified ULIRGs were observed.
In total, AKARI IRC spectra of 26 (of 33) non-Seyfert ULIRGs at $z >$
0.15 in the IRAS 1 Jy sample were taken. At $z <$ 0.15, 28 LINER, 20
HII-region, and 4 unclassified ULIRGs are listed, of which 8 LINER, 5
HII-region, and 0 unclassified ULIRGs were observed. 
For ULIRGs optically classified as Seyfert 2s, the presence of
energetically significant AGNs surrounded by torus-shaped dust is almost
certain, and so these Seyfert 2 ULIRGs were set as a low priority in our
project.  However, due to good target visibility, four Seyfert 2 ULIRGs
were observed. No ULIRGs classified optically as Seyfert 1s (i.e.,
unobscured broad-line AGN) were observed.  

In addition to these ULIRGs in the IRAS 1 Jy sample, the two well-known
ULIRGs, UGC 5101 and IRAS 19245$-$7245 (Super-Antennae), are also
included in the sample. The optical classification of UGC 5101 and IRAS
19254$-$7245 is LINER \citep{vei95} and Seyfert 2 \citep{mir91},
respectively. Both of these sources show clear 3.4 $\mu$m absorption
features in ground-based $L$-band spectra
\citep{idm01,im03,ris03}. Thus, 
further investigations of absorption features using high-quality AKARI
IRC spectra with wider wavelength coverage are warranted. 

In total, 43 ULIRGs in the IRAS 1 Jy sample (39 non-Seyferts and 4
Seyferts) and 2 additional ULIRGs of interest were observed. Table 1
summarizes the basic information and {\it IRAS}-based infrared
properties of the observed ULIRGs.  

\section{Observations and Data Analysis}

Spectroscopic observations were made using the IRC infrared spectrograph
\citep{ona07} onboard the AKARI infrared satellite \citep{mur07} between
October 2006 and August 2007. All data were taken in the mission program
named ``AGNUL''. Table 2 summarizes the observation log. We used the NG
grism mode, which can cover 2.5--5.0 $\mu$m 
simultaneously with an effective spectral resolution of R $\sim$ 120 at
3.6 $\mu$m for point sources \citep{ona07}. The 1 $\times$ 1
arcmin$^{2}$ NIR slit was used to minimize the contamination of the
spectra by source overlap \citep{ona07,ohy07}. The pixel scale of AKARI
IRC is 1$\farcs$46 $\times$ 1$\farcs$46. Only one pointing was assigned
for each ULIRG, but since we employed the observing mode of IRC04, one
pointing consisted of 8 or 9 independent frames \citep{ona07} that could
be used to remove the effects of cosmic ray hits and obtain spectra of
the objects. The total net on-source exposure time was $\sim$6 min for
each ULIRG. 

The spectral analysis was performed in a standard manner using the IDL
package prepared for the reduction of AKARI IRC spectra 
%-----
\footnote{
The actual software package used for our data reduction is ``IRC
Spectroscopy Toolkit Version 20070913'', which can be found at
http://www.ir.isas.jaxa.jp/ASTRO-F/Observation/DataReduction/IRC/}. 
%-----
Detailed descriptions are contained in \citet{ohy07}. Each frame was
dark-subtracted, linearity-corrected, and flat-field corrected. In AKARI
IRC slitless spectra, emission more extended than 3--5 pixels (= 3--5
$\times$ 1$\farcs$46) was not clearly recognizable, confirming that
very extended starburst emission is insignificant. We thus
integrated signals within 3--5 pixels for spectral extraction, depending
on the actual signal profile of each ULIRG. The background signal level
is estimated from the data points at the both sides of the object
position and is subtracted. Wavelength and flux calibrations were made
within this package. The wavelength calibration accuracy is taken to be
$\sim$1 pixel or $\sim$0.01 $\mu$m \citep{ohy07}. The absolute flux
calibration accuracy is $\sim$10\% at the central wavelength of the
spectra, and can be as large as $\sim$20\% at the edge of the NG spectra
(close to 2.5 $\mu$m and 5.0 $\mu$m) \citep{ohy07}. To reduce the
scatter of data points, appropriate binning of spectral elements was
performed, particularly for faint ULIRGs. The resulting spectral
resolution can be as low as R $\sim$ 50 at some wavelength range for
such faint ULIRGs.  

\section{Results}

Figure 2 presents final infrared 2.5--5.0 $\mu$m spectra of ULIRGs
obtained with the AKARI IRC. For IRAS 17028+5817, the western (W) and
eastern (E) nuclei were resolved with the AKARI IRC, and spectra of both
nuclei were of sufficient quality. Spectra of the W (LINER) and E
(HII-region) nuclei are thus presented separately, totaling 46 spectra
in Figure 2. 

Figure 3 compares the spectrum of IRAS 08572+3915 taken with the AKARI
IRC and the ground-based IRTF 3m telescope \citep{idm06}. Since the
infrared dust emission of this ULIRG is dominated by a compact ($<$
300pc) nuclear core \citep{soi00}, with a very small contribution from
extended 
($>$several kpc) starburst activity in the host galaxy, this is an
appropriate source to compare AKARI IRC slitless spectrum with the
ground-based narrow-slit (1$\farcs$6-wide) spectrum
\citep{idm06}. Overall, the spectra agree very well in the overlapped
region ($\lambda_{\rm obs}$ = 2.9--4.1 $\mu$m), demonstrating the good
calibration accuracy of AKARI IRC spectroscopy. However, when we compare
the spectra in more detail, slight discrepancies become apparent at
shorter wavelengths; in particular, the flux of the ground-based slit 
spectrum is smaller than the AKARI IRC slitless spectrum. This
discrepancy is larger than the statistical errors of both spectra. 
In IRAS 08572+3915, although infrared dust emission is dominated by the
compact nuclear energy source, {\it stellar} emission in the extended
host galaxy, which becomes more important at shorter wavelengths, may be
non-negligible. The slight discrepancy at shorter wavelengths could thus
be explained by the scenario that extended stellar emission is covered
only by the AKARI IRC slitless spectrum, but not by the ground-based
slit spectrum, because of aperture losses.  

In Figure 2, although the spectral shapes are rich in variety, most of
the observed ULIRGs show emission features at $\lambda_{\rm obs}$ $\sim$
(1 + $z$) $\times$ 3.29 $\mu$m. We ascribe these features to 3.3 $\mu$m
PAH emission because contribution from the Pf$\delta$ ($\lambda_{\rm
rest}$ $\sim$ 3.3 $\mu$m; $\sim$10\% strength of Br$\alpha$;
Wynn-Williams 1984) is usually negligible in ULIRGs not classified
optically as Seyfert 1s (i.e., unobscured broad-line AGNs)
\citep{idm06}. To estimate the 3.3 $\mu$m PAH emission strength, we make
the reasonable assumption that the profiles of the 3.3 $\mu$m PAH
emission in these ULIRGs are similar to those of Galactic star-forming
regions and nearby starburst galaxies (type-1 sources in Tokunaga et
al. 1991), following \citet{idm06}. The adopted profile reproduces the
observed 3.3 $\mu$m PAH emission features of the ULIRGs reasonably well
given the adopted continuum levels (shown as solid lines in Figure 2 for
PAH-detected ULIRGs) and spectral resolution of the AKARI IRC. Table 3
summarizes the fluxes, luminosities, and rest-frame equivalent widths
(EW$_{\rm 3.3PAH}$) of the 3.3 $\mu$m PAH emission feature. The
uncertainties of the 3.3 $\mu$m PAH fluxes coming from fitting errors
and continuum choice are unlikely to exceed $\sim$30\% as long as we
adopt reasonable continuum levels.   

In addition to the 3.3 $\mu$m PAH emission features, broad
absorption features from ice-covered dust, centered at $\lambda_{\rm
rest}$ = 3.05--3.1 $\mu$m and extending from $\lambda_{\rm rest}$ $\sim$
2.75 $\mu$m to $\sim$3.55 $\mu$m, are clearly detected in many
ULIRGs. Since spectra as short as $\lambda_{\rm obs}$ = 2.5 $\mu$m are
covered (as compared to $\lambda_{\rm obs}$ = 2.8 $\mu$m in ground-based
$L$-band spectra; Imanishi et al. 2006a), the continuum level at the
shorter wavelength side of the 3.1 $\mu$m ice absorption feature is well
determined. To estimate the optical depths of the 3.1 $\mu$m absorption
features ($\tau_{3.1}$), we adopt linear continua (shown as dashed lines
in Figure 2 for 3.1$\mu$m-absorption-detected ULIRGs) following
\citet{idm06}. We consider the 3.1 $\mu$m absorption feature to be
clearly detected only in ULIRGs that display both a significant flux
depression below the adopted continuum level at $\lambda_{\rm rest}$ =
3.05--3.1 $\mu$m, and a spectral gap at the 3.3um PAH emission feature
in which the flux at the shorter wavelength side of this 3.3 $\mu$m PAH
feature is smaller than that at the longer wavelength side
\citep{idm06}. 
The derived $\tau_{3.1}$ values are summarized in column 2 of Table
4. Due to the spectrally broad nature of the 3.1 $\mu$m ice absorption
feature, the continuum determination has some ambiguity. However, we
estimate the typical uncertainty of $\tau_{3.1}$, originating in continuum
determination ambiguity and spectral element noise close to the
absorption peak, as $\sim$0.1. Since this 3.1 $\mu$m ice absorption
feature is spectrally intrinsically broad, not consisting of spectrally
unresolved narrow lines \citep{smi89}, the measured $\tau_{3.1}$ values
in AKARI IRC's low-resolution spectra are taken as true ones. 
For ULIRGs observed with both the AKARI IRC and ground-based telescopes,
the presence of the 3.1 $\mu$m ice absorption features is generally
clearer in AKARI IRC spectra than in ground-based $L$-band spectra
because of the AKARI IRC's higher spectral quality and wider wavelength
coverage (e.g., the LINER ULIRG IRAS 09539+0857; Figure 2 of this paper
and Imanishi et al. 2006a). IRAS 10091+4704 (LINER), 21477+0502 (LINER),
and 04313$-$1649 (unclassified) may display the 3.1 $\mu$m absorption
feature, but we do not classify these sources as
3.1$\mu$m-absorption-detected ULIRGs because the continuum ambiguity is
large.  

In addition to the 3.1 $\mu$m ice absorption, absorption features at
$\lambda_{\rm rest}$ = 3.4 $\mu$m by bare carbonaceous dust
\citep{pen94} are found in four ULIRGs: IRAS 08572+3915, 17044+6720, UGC
5101, and IRAS 19254$-$7245. To estimate the optical depths of the 3.4
$\mu$m absorption features ($\tau_{3.4}$) for these sources, we adopt
linear continua, which are shown as dashed lines in Figure 2. The
estimated $\tau_{3.4}$ values for the four ULIRGs are summarized in
column 3 of Table 4.  

Br$\alpha$ ($\lambda_{\rm rest}$ = 4.05 $\mu$m) emission is detected in
many ULIRGs. However, the Br$\alpha$ emission in most cases is
spectrally unresolved at AKARI IRC's resolution (R $\sim$ 120),
resulting in large uncertainty for the derived flux and
luminosity. Br$\alpha$ emission will not be used for our detailed
quantitative discussions.  

For ULIRGs at $z <$ 0.1, AKARI IRC spectral coverage extends to the
redshifted CO absorption features centered at $\lambda_{\rm rest}$
$\sim$ 4.67 $\mu$m, extending between $\lambda_{\rm rest}$ = 4.4--5.0
$\mu$m \citep{spo04,san08}.  These features are detected in a few ULIRG
spectra (e.g., IRAS 08572+3915, UGC 5101, IRAS 19254$-$7245). This CO
absorption feature consists of many narrow features \citep{geb06} that
are spectrally unresolved with the AKARI IRC (R $\sim$120). Hence, the
apparent optical depths are only lower limits. In IRAS 08572+3915,
CO$_{2}$ absorption features at $\lambda_{\rm rest}$ $\sim$ 4.26 $\mu$m
are also discernible. Detailed discussions of these CO and CO$_{2}$
absorption features will be made by Shirahata et al. (2008).  

\section{Discussion}

We investigate energy sources of observed ULIRGs based on the 3.3
$\mu$m PAH emission and 3.1 $\mu$m and 3.4 $\mu$m absorption
features. 

\subsection{Modestly obscured starbursts}
In most of the observed ULIRGs, the presence of starbursts is evident
from detection of the 3.3 $\mu$m PAH emission. Since dust extinction at
3.3 $\mu$m is only $\sim$1/15 of that in the optical $V$-band ($\lambda$
= 0.6 $\mu$m; Rieke \& Lebofsky 1985; Lutz et al. 1996), the flux
attenuation of 3.3 $\mu$m PAH emission with dust extinction of A$_{\rm
V}$ $<$ 15 mag is less than 1 mag. Thus, the observed 3.3 $\mu$m PAH
emission luminosities can be used to quantitatively derive the intrinsic
luminosities of modestly obscured (A$_{\rm V}$ $<$ 15 mag) starburst
activity. The observed 3.3 $\mu$m PAH to infrared luminosity ratios
(L$_{\rm 3.3PAH}$/L$_{\rm IR}$) are summarized in column 4 of Table 3.  

Figures 4a and 4b respectively compare the observed 3.3 $\mu$m PAH
luminosity (L$_{\rm 3.3PAH}$) and its rest-frame equivalent width
(EW$_{\rm 3.3PAH}$) measured in ground-based slit spectra (abscissa) and
in AKARI IRC slitless spectra (ordinate). The abscissa and ordinate of
Figure 4a trace nuclear ($<$kpc) and total starburst luminosities,
respectively. If {\it extended} (several kpc) starbursts in the host
galaxies of ULIRGs are energetically much more important than modestly
obscured {\it nuclear} starbursts, then both the L$_{\rm 3.3PAH}$ and
EW$_{\rm 3.3PAH}$ values in the ordinate are expected to be
substantially larger than the abscissa. However, the difference in both
L$_{\rm 3.3PAH}$ and EW$_{\rm 3.3PAH}$ is relatively small, a factor of
a few at most.  
We see no evidence that the spatially {\it extended}, modestly obscured
starbursts in host galaxies are substantially (more than an order of
magnitude) more luminous than the modestly obscured {\it nuclear}
starbursts in ULIRGs. 
Even including extended starbursts in host galaxies, the L$_{\rm 3.3
PAH}$/L$_{\rm IR}$ ratios in ULIRGs are factors of 2.5 to more than 10
times smaller than those found in less infrared-luminous starbursts
($\sim$10$^{-3}$; Mouri et al. 1990; Imanishi 2002). Taken at face
value, the detected modestly obscured starbursts can account for only
$<$10\% to at most 40\% of the infrared luminosities of the observed
ULIRGs. Our AKARI IRC spectra reinforce the previous arguments, based on
a small sample \citep{soi00}, that nearby ULIRGs at $z <$ 0.3 are not
energetically dominated by extended modestly obscured starburst activity
in host galaxies.   

\subsection{Buried AGNs with weak starbursts}
The deficit in observed 3.3 $\mu$m PAH luminosity relative to the
infrared luminosity requires energy sources in addition to the modestly
obscured (A$_{\rm V}$ $<$ 15 mag) PAH-emitting starbursts. The first
possibility is very highly obscured (A$_{\rm V}$ $>>$ 15 mag)
PAH-emitting starbursts, in which the 3.3 $\mu$m PAH flux is severely
attenuated, while the flux of longer wavelength infrared emission
(8--1000 $\mu$m) may not be highly attenuated. The second possibility is
that an AGN exists that can produce large infrared dust emission
luminosities with no PAH emission (see $\S$1), decreasing the observed
L$_{\rm 3.3 PAH}$/L$_{\rm IR}$ ratios. 
These two scenarios are difficult to differentiate based on the absolute
PAH luminosities but can be distinguished by the {\it equivalent width}
of emission or absorption features. 
  
In a normal starburst galaxy, where HII regions, molecular gas,
and photodissociation regions are spatially well mixed (Figure 1a), the
equivalent width of the 3.3 $\mu$m PAH-emission feature is insensitive
to dust extinction \citep{idm06}. If PAH-free AGN emission contributes
significantly to the observed 2.5--5 $\mu$m flux, then the EW$_{\rm
3.3PAH}$ value should decrease. 
The EW$_{\rm 3.3PAH}$ values in starbursts have an average value of
EW$_{\rm 3.3PAH}$ $\sim$ 100 nm, with some scatter, but never become
lower than 40 nm \citep{moo86}. Thus, we adopt EW$_{\rm 3.3PAH}$
$\lesssim$ 40 nm as a strong signature of significant AGN contribution
to an observed flux. 

Among ULIRGs classified optically as non-Seyferts in the IRAS 1 Jy
sample, three LINER ULIRGs, IRAS 23129+2548, 08572+3915, and 17044+6720,
and one HII-region ULIRG, IRAS 22088$-$1831, have EW$_{\rm 3.3PAH}$ $<$
20 nm (Table 3, column 5), more than a factor of 5 less than typical
starburst galaxies. These ULIRGs are strong buried AGN candidates. 
The following non-Seyfert ULIRGs are also taken to contain
luminous buried AGNs because the EW$_{\rm 3.3PAH}$ values are $<$40 nm:
the three LINER ULIRGs (IRAS 04074$-$2801, 11180+1623, and 21477+0502),
two HII-region ULIRGs (IRAS 14202+2615 and 17028+5817E), and one
optically unclassified ULIRG (IRAS 08591+5248).   

\subsection{Buried AGNs with coexisting strong starbursts}

Based on the EW$_{\rm 3.3 PAH}$ values, we can easily detect buried AGNs
with very weak starbursts. Even if strong starburst activity is present,
{\it weakly obscured} AGNs are detectable because weakly attenuated
PAH-free continua from the AGNs can dilute the 3.3 $\mu$m PAH emission
considerably.  However, detecting {\it deeply buried} AGNs with
coexisting surrounding strong starbursts is very difficult (Figure
1c). Even if the intrinsic luminosities of a buried AGN and surrounding
less-obscured starbursts are similar, the AGN flux will be more highly
attenuated by dust extinction than the starburst emission. When a buried
AGN is obscured by {\it ice-covered} dust grains, the AGN flux at
$\lambda_{\rm rest}$ = 3.3 $\mu$m is attenuated even more severely by
the strong, broad 3.1 $\mu$m absorption feature, making the EW$_{\rm
3.3PAH}$ values in observed spectra apparently large. 

To determine
whether a deeply buried AGN is present in addition to strong starbursts,
we use the optical depths of dust absorption features found in the
2.5--5 $\mu$m spectra. As described in $\S$1 and in \citet{idm06} in
more detail, these values can be used to distinguish whether the energy
sources are spatially well mixed with dust (a normal starburst), or are
more centrally concentrated than the dust (a buried AGN).  For a normal
starburst with the mixed dust/source geometry in a ULIRG's core,
$\tau_{3.1}$ cannot exceed 0.3, while a buried AGN can produce
$\tau_{3.1} >$ 0.3 \citep{im03,idm06}. Therefore, detection of
$\tau_{3.1} >$ 0.3 can be used to argue for the presence of a buried AGN
with a centrally concentrated energy source geometry. Considering the
uncertainty of $\tau_{3.1}$ with $\sim$0.1 ($\S$4), we classify ULIRGs
with $\tau_{3.1}$ $>$ 0.4 as buried AGN candidates.    

Aside from the above ULIRGs with low EW$_{\rm 3.3PAH}$, the following
non-Seyfert ULIRGs in the IRAS 1 Jy sample are newly classified as
buried AGNs: ten LINER ULIRGs (IRAS 05020$-$2941, 09463+8141,
11028+3130, 14121$-$0126, 16333+4630, 00482$-$2721, 09539+0857,
10494+4424, 16468+5200, 17028+5817W), four HII-region ULIRGs (IRAS
01199$-$2307, 01355$-$1814, 17068+4027, 11387+4116), and one optically
unclassified ULIRG (IRAS 01494$-$1845). This large $\tau_{3.1}$ method
is sensitive to deeply buried AGNs but obviously misses weakly obscured
AGNs, which are more easily detected with the above low EW$_{\rm
3.3PAH}$ method. Hence, the large $\tau_{3.1}$ and low EW$_{\rm 3.3PAH}$
methods are used in complementary fashion to detect AGN signatures.  

We can also investigate the dust/source geometry from the $\tau_{3.4}$ value.  
ULIRGs with $\tau_{3.4} >$ 0.2 can be used to argue for the presence of
buried AGNs \citep{im03,idm06}. Among the optically non-Seyfert ULIRGs
in the IRAS 1 Jy sample, only the LINER ULIRG IRAS 08572+3915 displays
$\tau_{3.4}$ $>$ 0.2. The LINER ULIRG of interest, UGC 5101, and the
Seyfert 2 ULIRG, IRAS 19254$-$7245, also show $\tau_{3.4}$ $>$
0.2. However, all of these three ULIRGs have already been classified as
having luminous AGNs based on their low EW$_{\rm 3.3PAH}$ values.  

Imanishi et al. (2006a, 2007a) commented that exceptionally centrally
concentrated starbursts (Figure 1d) and normal starbursts with mixed
dust/source geometry obscured by foreground dust in edge-on host
galaxies (Figure 1e) can also produce large $\tau_{3.1}$ and
$\tau_{3.4}$ values, but argued that it is very unlikely for the bulk of
ULIRGs with $\tau_{3.1} >$ 0.3 and/or $\tau_{3.4} >$ 0.2 to correspond 
to these non-AGN cases.   

For the remaining non-Seyfert ULIRGs with EW$_{\rm 3.3PAH}$ $>$ 40 nm,
$\tau_{3.1}$ $\lesssim$ 0.4, and $\tau_{3.4}$ $\lesssim$ 0.2, no obvious
buried AGN signatures were observed in the 2.5--5 $\mu$m spectra. Their
spectra can be explained by either of the following scenarios: normal
starbursts with mixed dust/source geometry are energetically
dominant, or AGNs are present, but the AGN emission is so highly
attenuated that its contribution to the observed 2.5--5 $\mu$m flux is
not significant.  We have no way of distinguishing between these two
scenarios. However, some examples are known (e.g., NGC 4418) in which
buried AGN signatures are found only at wavelengths longer than 5 $\mu$m
\citep{ima04,dud97,spo01}. Thus, the detected buried AGN fraction in the
2.5--5 $\mu$m AKARI IRC spectra is only a lower limit. 

\subsection{Dust extinction and intrinsic AGN luminosities} 
In a buried AGN with centrally concentrated energy source geometry, dust
at a temperature of 1000 K, which is close to the innermost dust
sublimation radius, produces continuum emission with a peak at $\lambda
\sim$ 3 $\mu$m, assuming approximately blackbody emission. Since a
foreground screen dust distribution model is applicable to a buried AGN
(Imanishi et al. 2006a, 2007a), the $\tau_{3.1}$ (ice-covered dust) and
$\tau_{3.4}$ (bare dust) values reflect the dust column density toward
the 3 $\mu$m continuum-emitting region, which is almost equal to the
column density toward the buried AGN itself (Imanishi et al. 2006a,
2007a).  

The 3.1 $\mu$m absorption feature is detectable if ice-covered dust
grains are present in front of the 3--4 $\mu$m continuum emitting energy
source. Such ice-covered dust grains are usually found deep inside
molecular gas, where ambient UV radiation is sufficiently shielded
\citep{whi88,tan90,smi93,mur00}. 
The 3.4 $\mu$m absorption feature should be detected if the 3--4 $\mu$m
continuum emitting energy source is obscured by bare carbonaceous dust
grains \citep{pen94,ima96,raw03}, but is undetected if the absorbing
dust is ice-covered \citep{men01}. Since absorbing dust consists of both
bare and ice-covered dust grains, ULIRGs with obscured energy sources
should show both features, and the true dust column density is derivable
from a proper combination of $\tau_{3.1}$ and $\tau_{3.4}$.

However, despite the detection of the 3.1 $\mu$m ice absorption features
in many ULIRGs, only two ULIRGs in the IRAS 1 Jy sample, IRAS 08572+3915
and 17044+6720, show clear 3.4 $\mu$m carbonaceous dust absorption
features. Even including the two sources of interest, UGC 5101 and IRAS
19254$-$7245, only four ULIRGs display clearly detectable 3.4 $\mu$m
absorption features. The difference in the detection rate largely comes
from the intrinsically smaller oscillator strength of the 3.4 $\mu$m
carbonaceous dust absorption feature ($\tau_{3.4}$/A$_{\rm V}$ =
0.004--0.007; Pendleton et al. 1994) compared to the 3.1 $\mu$m ice
absorption feature ($\tau_{3.1}$/A$_{\rm V}$ = 0.06; Tanaka et al. 1990;
Smith et al. 1993; Murakawa et al. 2000). Even if a modestly large
amount of bare carbonaceous dust grains is present in front of the
continuum-emitting energy source, the $\tau_{3.4}$ value is small,
making the detection of the 3.4 $\mu$m dust absorption feature
difficult.  
Additionally, the 3.3 $\mu$m PAH emission feature is often accompanied
by a sub-peak at 3.4 $\mu$m \citep{tok91,imd00}, and this sub-peak may
dilute the 3.4 $\mu$m dust absorption feature at the same wavelength. In
fact, all the four ULIRGs with detectable 3.4 $\mu$m absorption
features are limited to relatively weak PAH emitters (EW$_{\rm 3.3PAH}$
$<$ 35 nm). Finally, when the spectrally broad 3.1 $\mu$m absorption
feature is strong, the absorption feature extends to the longer
wavelength side of
the 3.3 $\mu$m PAH emission feature, making it difficult to distinguish
the origin of the apparent flux depression at $\lambda_{\rm rest}$
$\sim$ 3.4 $\mu$m. Among the four sources with detectable 3.4 $\mu$m
absorption features, IRAS 08572+3915, 17044+6720, and 19254$-$7245
indeed display only weak or undetectable 3.1 $\mu$m absorption
features. The remaining source, UGC 5101, shows large $\tau_{3.1}$, but
we can recognize the 3.4 $\mu$m absorption feature, primarily because
UGC 5101 is one of the brightest sources and the signal to noise ratios
are among the highest. 
Detection of the 3.4 $\mu$m absorption features in the remaining ULIRGs
with large EW$_{\rm 3.3PAH}$, large $\tau_{3.1}$, and limited
signal-to-noise ratios in the continuum is basically
difficult. Therefore, while the total dust column density can be
estimated in a reasonably reliable way for ULIRGs with both detectable
3.1 $\mu$m and 3.4 $\mu$m absorption features, the estimated dust column
densities are only lower limits, and should be much smaller than the
actual values for ULIRGs with only detectable 3.1 $\mu$m absorption
features. The estimated dust column densities are summarized in column 4
of Table 4.  

In a buried AGN, the surrounding dust has a strong temperature gradient
in that inner dust, close to the central energy source, has higher
temperature than outer dust.
Luminosity is transferred at each temperature, and the intrinsic
luminosity of inner hot dust emission at 3--4 $\mu$m ($\nu$F$_\nu$)  
should be comparable to that of outer cool dust emission at 
60 $\mu$m, the wavelength which dominates the observed infrared emission
of ULIRGs \citep{san88a}.
Thus, if the intrinsic AGN's 3--4 $\mu$m luminosity ($\nu$F$_\nu$) 
is comparable to the observed infrared luminosities of ULIRGs, then we
can argue that the buried AGN is energetically important. 
For ULIRGs with low EW$_{\rm 3.3PAH}$ ($<$40 nm), we can roughly extract
the AGN's PAH-free continuum at 3--4 $\mu$m, based on the assumption
that starburst activity intrinsically shows EW$_{\rm 3.3PAH}$ $\sim$ 100
nm. That is, for ULIRGs with EW$_{\rm 3.3PAH}$ = 30 nm (30\% of the
typical starburst value), we consider that 70\% of the 3--4 $\mu$m
continuum comes from AGN's PAH-free continuum emission. Thus, we can
estimate the {\it observed }  
3--4 $\mu$m flux of AGN emission. Next, for ULIRGs with 
both detectable 3.1 $\mu$m and 3.4 $\mu$m absorption features (IRAS
08572+3915, UGC 5101, and IRAS 19254$-$7245), and for IRAS 17044+6720,
which displays only the 3.4 $\mu$m absorption feature, we can estimate
dust column density, or {\it dust extinction}, toward the 3--4 $\mu$m
continuum emitting regions based on $\tau_{3.1}$ and $\tau_{3.4}$. When
we combine the {\it observed} AGN flux at 3--4 $\mu$m and {\it dust
extinction} toward the 3--4 $\mu$m continuum emitting regions, we can
derive the dust-extinction-corrected {\it intrinsic} AGN flux, and thus
the intrinsic AGN luminosity. If we adopt A$_{\rm 3-4 \mu m}$/A$_{\rm
V}$ $\sim$ 0.058 \citep{rie85}, then we find $\tau_{3.1}$/A$_{\rm 3-4
\mu m}$ $\sim$ 1 and 
$\tau_{3.4}$/A$_{\rm 3-4 \mu m}$ $\sim$ 0.069--0.12 for the Galactic
interstellar medium. Assuming this relationship, we estimate the
intrinsic AGN luminosities at 3--4 $\mu$m ($\nu$F$_{\nu}$) to be L
$\gtrsim$ 10$^{12}$L$_{\odot}$ in all the 3.4$\mu$m-absorption-detected
ULIRGs. The putative AGN activity is therefore energetically sufficient
to quantitatively account for the bulk of the infrared luminosities of
these ULIRGs (L$_{\rm IR}$ $\sim$ 10$^{12}$L$_{\odot}$).    

\subsection{Dependence of the buried AGN fraction on optical spectral type:
LINER vs. HII-region} 
In total, AKARI IRC 2.5--5 $\mu$m spectra of 19 LINER, 16 HII-region,
and 5 optically unclassified ULIRG's nuclei in the {\it IRAS} 1 Jy
sample were obtained. The low EW$_{\rm 3.3PAH}$ method suggests six
LINER and three HII-region ULIRG's nuclei contain luminous buried AGNs
($\S$5.2). In addition to these ULIRGs, the large $\tau_{3.1}$ method
classifies ten LINER and four HII-regions ULIRG's nuclei as sources with
deeply obscured buried AGNs ($\S$5.3).   
In total, the detected buried AGN fraction is 16/19 (84\%) for LINER
ULIRGs, and 7/16 (44\%) for HII-region ULIRGs (Table 4, columns 5 and 6). 
Since the selection of the observed ULIRGs is based solely on the
target's visibility from the AKARI satellite and should be unbiased in
terms of their energy sources ($\S$2), we argue that the detected buried
AGN fraction is higher in LINER ULIRGs than in HII-region ULIRGs. The same
result was found from ground-based $L$-band spectroscopy and Spitzer
5--35 $\mu$m spectroscopy of ULIRGs at $z <$ 0.15 (Imanishi et
al. 2006a, 2007a), and also from a VLA radio observational
search for compact radio core emission (another good AGN indicator)
\citep{nag03}. Therefore, we confirm that a larger fraction of LINER
ULIRGs possess luminous buried AGNs than HII-region ULIRGs with a
probability of $\sim$99\%.   

The higher buried AGN fraction in optically LINER ULIRGs can be
explained qualitatively by a dustier starburst scenario
\citep{ima07a}. For starburst/buried AGN composite ULIRGs (Figure 1c),
the optical LINER or HII-region classifications are likely largely
affected by the properties of the modestly obscured starbursts at the
exteriors of the buried AGN, rather than by buried AGN-related emission,
as optical observations can probe only the surfaces of dusty objects. In
a dusty starburst, shock-related emission can be relatively important in
the optical compared to the emission from the HII-regions themselves,
resulting in optical LINER classification.  

When a luminous AGN is placed at the center of a {\it less dusty}
starburst classified optically as an HII-region, the AGN emission is
more easily detectable in the optical, making such an object an optical
Seyfert. In contrast, when a luminous AGN is placed at the center of a
{\it dusty} starburst classified optically as a LINER, the AGN emission
is more elusive in the optical, so that such an object is classified as
an optical non-Seyfert. Hence, this scenario can explain the observed
higher fraction of optically elusive {\it buried} AGNs in optically
LINER ULIRGs compared to HII-region ULIRGs. In fact, the emission probed
in the optical was found to be dustier in LINER ULIRGs than in
HII-region ULIRGs (Veilleux et al. 1995, 1999). 

\subsection{The buried AGN fraction as a function of infrared luminosity}

Due to the inclusion of ULIRGs at $z >$ 0.15, we have now a large number
of ULIRGs with L$_{\rm IR}$ $\gtrsim$
10$^{12.3}$L$_{\odot}$. Specifically, only 1 of 13 observed non-Seyfert
ULIRGs at $z <$ 0.15 has L$_{\rm IR}$ $\gtrsim$ 10$^{12.3}$L$_{\odot}$,
while 16 of 26 observed non-Seyfert ULIRGs at $z >$ 0.15 have L$_{\rm
IR}$ $\gtrsim$ 10$^{12.3}$L$_{\odot}$ (see Table 1). We can thus
investigate the buried AGN fraction, separating ULIRGs into two
categories: those with L$_{\rm IR}$ $<$ 10$^{12.3}$L$_{\odot}$ and
those with L$_{\rm IR}$ $\gtrsim$ 10$^{12.3}$L$_{\odot}$. 

Among the observed non-Seyfert ULIRGs in the IRAS 1 Jy sample, 22
sources have L$_{\rm IR}$ $<$ 10$^{12.3}$L$_{\odot}$ and the remaining
17 sources have L$_{\rm IR}$ $\gtrsim$ 10$^{12.3}$L$_{\odot}$. Based on
the buried AGN signatures in Table 4, the fraction of ULIRGs with
detectable buried AGN signatures is 12/22 (= 55\%) for non-Seyfert
ULIRGs with L$_{\rm IR}$ $<$ 10$^{12.3}$L$_{\odot}$ and 12/17 (= 71\%)
for non-Seyfert ULIRGs with L$_{\rm IR}$ $\gtrsim$
10$^{12.3}$L$_{\odot}$. Although the total sample size is not large, we
find a higher buried AGN fraction with increasing ULIRG infrared
luminosity.   
\citet{idm06} investigated, based on ground-based $L$-band (2.8--4.1
$\mu$m) spectra, the buried AGN fraction in a larger number of ULIRGs at
$z <$ 0.15 in the IRAS 1 Jy sample than in this paper. 
Several ULIRGs at $z <$ 0.15 studied  in this paper are also included in
this ground-based study. When we divide 
the observed ULIRGs by \citet{idm06} into those with L$_{\rm IR}$ $<$
10$^{12.3}$L$_{\odot}$ and $\gtrsim$ 10$^{12.3}$L$_{\odot}$, the
detected buried AGN fraction is 16/29 (= 55\%) in non-Seyfert ULIRGs
with L$_{\rm IR}$ $<$ 10$^{12.3}$L$_{\odot}$, and 6/9 (= 67\%) in 
non-Seyfert ULIRGs with L$_{\rm IR}$ $\gtrsim$
10$^{12.3}$L$_{\odot}$. The detected buried AGN fractions 
in ground-based and AKARI IRC spectra are comparable for non-Seyfert
ULIRGs with both L$_{\rm IR}$ $<$ 10$^{12.3}$L$_{\odot}$ ($\sim$55\%)
and $\gtrsim$ 10$^{12.3}$L$_{\odot}$ ($\sim$70\%).  We therefore argue
that the detected buried AGN fraction increases with ULIRG infrared
luminosity. This trend parallels the higher detection rate of optical
Seyfert signatures in ULIRGs with higher infrared luminosities
\citep{vei99}.  

The fraction of both optical Seyferts and luminous buried AGNs is
significantly smaller in galaxies with L$_{\rm IR}$ $<$
10$^{12}$L$_{\odot}$ than ULIRGs \citep{vei99,soi01}. Hence, we conclude
that AGN activity becomes more important as the infrared luminosities of
galaxies increase. 
Recently, the so-called galaxy downsizing phenomena have been found,
where galaxies with currently larger stellar masses have finished their
major star-formation in earlier cosmic age \citep{cow96,bun05}.  
AGN feedbacks are proposed to be responsible for the galaxy downsizing
phenomena \citep{gra04,bow06,cro06}. 
Namely, in galaxies with currently larger stellar masses, AGN feedbacks
have been stronger in the past, and gas has been expelled in a shorter
time scale. 
Buried AGNs can have particularly strong feedbacks, because the AGNs are
surrounded by a large amount of nuclear gas and dust.
If we reasonably assume that galaxies with currently larger stellar
masses have previously been more infrared luminous, then the detected
higher buried AGN fraction in more infrared luminous galaxies may
support the AGN feedback scenario as the origin of the galaxy downsizing
phenomena  
\footnote{
To form more stars, more star-formation should have been occurred in
the past, producing stronger star-formation related infrared emission
in the past.  
If AGN's energetic contribution is negligible in LIRGs with
L$_{\rm IR}$ $<$ 10$^{12}$L$_{\odot}$, but important, say $\sim$50\%, in
ULIRGs with L$_{\rm IR}$ $>$ 10$^{12}$L$_{\odot}$, then AGN feedbacks can
be stronger in ULIRGs, and yet higher infrared luminosity ($\sim$5
$\times$ 10$^{11}$L$_{\odot}$) can come from star-forming activity in
ULIRGs, producing more stellar masses in ULIRGs than in LIRGs.  
}.

\subsection{Comparison with Seyfert 2 ULIRGs} 
We compare the 2.5--5 $\mu$m spectral properties of non-Seyfert ULIRGs
showing buried AGN signatures with those of Seyfert 2 ULIRGs (i.e.,
known obscured-AGN-possessing ULIRGs).  The apparent main difference
between them is that the ionizing radiation from the putative buried
AGNs in non-Seyfert ULIRGs is obscured by the surrounding dust along
virtually all lines-of-sight; in contrast, dust around the AGNs in
Seyfert 2 ULIRGs is distributed in a ``torus'', and ionizing radiation
from the AGNs can escape along the torus axis, allowing for optical
Seyfert signature detection (Figure 5).   

The four ULIRGs classified optically as Seyfert 2s in the IRAS 1 Jy
sample show no clear absorption features at 2.5--5 $\mu$m. \citet{idm06}
also found in ground-based $L$-band spectra that the fraction of
optically classified Seyfert 2 ULIRGs showing strong dust absorption
features is substantially smaller than buried AGNs in optically
non-Seyfert ULIRGs. We thus argue that the line-of-sight dust column
density toward the AGNs is lower in Seyfert 2 ULIRGs than buried AGNs in
non-Seyfert ULIRGs. Thus, buried AGNs and Seyfert 2 AGNs (obscured by
torus-shaped dust) differ not only in dust geometry, but also in dust
column density along our line-of-sight; specifically, the dust columns
toward the buried AGNs are much higher \citep{idm06}. Since the dust
covering factor around buried AGNs (almost all directions) is also
larger than Seyfert 2 AGNs (torus-shaped), the total amount of nuclear
dust must be larger in the former (Figure 5).  

Since the gas and dust in an AGN have angular momentum with respect to
the central supermassive black hole, an axisymmetric spatial
distribution is more natural than a spherical geometry. In this case,
the column densities can be high in certain directions but low in
others (Figure 5). For a fixed angular momentum, the dust column density
ratios between the highest and lowest column directions are similar
among different galaxies.  
If the total amount of nuclear dust is modest, the direction of the
lowest dust column density can be transparent to the AGN's ionizing
radiation, making Seyfert signatures detectable in the optical
spectra. As the total amount of nuclear dust increases, even the
direction of the lowest dust column density can be opaque to the AGN's
ionizing radiation, making such galaxies optically elusive buried
AGNs. Thus, all of the observed spectral properties of buried AGNs and
Seyfert-type AGNs are explicable given a larger amount of nuclear dust
in the former. Since ULIRGs contain a large amount of nuclear gas and
dust \citep{sam96}, the high buried AGN fraction is
inevitable. Understanding optically elusive buried AGNs is therefore
essential if we are to unveil the true nature of the ULIRG population. 

\subsection{Dependence on far-infrared colors}
Based on the {\it IRAS} 25 $\mu$m to 60 $\mu$m flux ratio ($f_{\rm
25}$/$f_{\rm 60}$), ULIRGs are divided into cool ($<$ 0.2) and warm ($>$
0.2) sources \citep{san88b}. Many of the non-Seyfert ULIRGs with
detectable buried AGN signatures show cool far-infrared colors (Table
1). Although AGNs classified optically as Seyferts usually show warm
far-infrared colors \citep{deg87,kee05},  cool far-infrared colors of
buried AGNs are the natural consequence of a large amount of nuclear
dust, where contributions from the outer, cooler dust components to the
infrared radiation become more important than for optical Seyferts.  

Figure 6(a) compares {\it IRAS} 25 $\mu$m to 60 $\mu$m flux ratios
(i.e., far-infrared color) with the observed 3.3 $\mu$m PAH to infrared
luminosity ratios (L$_{\rm 3.3PAH}$/L$_{\rm IR}$). Seyfert ULIRGs tend
to appear in the warmer far-infrared color range than non-Seyfert ULIRGs,
as expected from the decreased amount of nuclear dust in the former
($\S$5.7). No systematic difference in the L$_{\rm 3.3PAH}$/L$_{\rm IR}$
ratios between non-Seyfert and Seyfert ULIRGs exists.  

Figure 6(b) compares the far-infrared colors and EW$_{\rm 3.3PAH}$ for
non-Seyfert ULIRGs.  Buried AGNs appear in both the warm and cool
ranges. Although it is sometimes argued that ULIRGs with cool
far-infrared colors must be starburst-dominated, simply because
Seyfert-type AGNs show warm far-infrared colors (e.g., Downes \& Solomon
1998), we do not confirm this to be true for the heavily buried AGNs.

\section{Summary}

We present the results of infrared 2.5--5 $\mu$m AKARI IRC slitless
spectroscopy of a large sample of nearby ULIRGs at $z <$ 0.3 from the
IRAS 1 Jy sample. We mainly observed ULIRGs with no obvious optical
Seyfert signatures (i.e., LINER, HII-region, and unclassified). One
ULIRG showed bright resolvable double nuclei, and spectra of 19 LINER,
16 HII-region, and 5 optically unclassified ULIRG nuclei were obtained
in total. In addition to these non-Seyfert ULIRGs, four Seyfert 2 ULIRGs
and two well-known ULIRGs were observed. Using the 3.3 $\mu$m PAH
emission, 3.1 $\mu$m ice absorption, and 3.4 $\mu$m bare carbonaceous
dust absorption features, we investigated whether the infrared 2.5--5
$\mu$m spectra of the non-Seyfert ULIRGs could be explained solely by
starbursts, or whether the ULIRGs displayed signatures of luminous, but
optically elusive, {\it buried} AGNs.     
The AKARI IRC's 2.5--5 $\mu$m wavelength coverage, unaffected by Earth's
atmospheric window, enabled us to unambiguously distinguish between 
absorption-dominated and emission-dominated sources, even for ULIRGs at
$z >$ 0.15, which was impossible with ground-based $L$-band (2.8--4.1
$\mu$m) spectroscopy.  
 
Our main conclusions are the following.

\begin{enumerate}
\item The 3.3 $\mu$m PAH emission, the starburst probe, was detected in
all but four non-Seyfert ULIRGs. The intrinsic luminosities of
modestly obscured (A$_{\rm V}$ $<$ 15 mag) starbursts were roughly
estimated from the observed 3.3 $\mu$m PAH emission luminosities.
We found no substantial enhancement of the 3.3 $\mu$m PAH
luminosities and equivalent widths in the AKARI IRC slitless spectra
compared to ground-based slit spectra.  
The observed 3.3 $\mu$m PAH to infrared luminosity ratios were
smaller by a factor of 2.5 to $>$10 compared to those found in less
infrared-luminous starbursts, suggesting that detected modestly
obscured starbursts, even including spatially extended (several kpc)
ones in the host galaxies probed by AKARI, are energetically
insignificant as the origin of the large ULIRG's infrared dust
emission luminosities.    
\item Among the optically non-Seyfert ULIRGs in the IRAS 1 Jy sample,
six LINER, three HII-region, and one unclassified ULIRG's
nuclei showed 3.3 $\mu$m PAH emission equivalent widths 
(EW$_{\rm 3.3PAH}$), which were much smaller than typical values found in
starburst galaxies. A strong contribution from the AGN's PAH-free
continua to the observed 2.5--5 $\mu$m fluxes was suggested. 
\item Besides these low EW$_{\rm 3.3PAH}$ ULIRGs, the optical depths of
3.1 $\mu$m ice ($\tau_{3.1}$) and 3.4 $\mu$m bare carbonaceous dust
($\tau_{3.4}$) absorption features were used to determine whether the
energy sources at the cores of ULIRGs are more centrally concentrated
than the surrounding dust (as is expected for a buried AGN), or are
spatially well mixed with dust (a normal starburst). The large
optical depths in ten LINER, four HII-region, and one unclassified
ULIRG's nuclei suggested that the energy sources are more centrally
concentrated than the surrounding dust. These ULIRGs were also
classified as displaying signatures of luminous	buried AGNs in
addition to detectable starburst activity.  
\item In total, 16/19 (84\%) LINER, 7/16 (44\%) HII-region, and 2/5
(40\%) 	optically unclassified ULIRG's nuclei showed some buried AGN
signatures, based either on the low EW$_{\rm 3.3PAH}$ method or on
the large $\tau_{3.1}$ ($\tau_{3.4}$) method. The higher detection
rate of buried AGNs in LINER ULIRGs than in HII-region ULIRGs was
similar to previous results based on ground-based $L$-band (2.8--4.1
$\mu$m) spectroscopy and Spitzer 5--35 $\mu$m spectroscopy of
ULIRGs at $z <$ 0.15. This trend can be explained qualitatively by a
dustier starburst scenario in LINER ULIRGs than in HII-region
ULIRGs. 
\item We found that the fraction of buried AGNs increases with
increasing infrared luminosity.    
This may support the AGN-feedback scenario as the origin of galaxy
downsizing phenomena. 
\item We confirmed the following main trends, as previously seen in
ground-based $L$-band spectroscopy of ULIRGs at $z <$ 0.15:   
(1) The fraction of sources with large dust absorption optical
depths is higher in buried AGNs in non-Seyfert ULIRGs than ULIRGs
classified optically as Seyfert 2s. 
(2) Luminous buried AGN signatures were detected in ULIRGs with both
warm and cool far-infrared colors. These results can reasonably be
explained by the scenario that buried AGNs in optically non-Seyfert
ULIRGs contain a systematically larger amount of nuclear dust than
optical Seyfert 2s (= AGNs obscured by torus-shaped dust), making even
the direction of the lowest dust column density opaque to the AGN's
ionizing radiation. The high buried AGN fraction in ULIRGs is
inevitable, given their high nuclear dust concentration. Therefore,
understanding optically elusive buried AGNs in the dusty ULIRG
population is very important. 
\end{enumerate}

More non-Seyfert ULIRGs in the IRAS 1 Jy sample, and lower infrared
luminosity galaxies (L$_{\rm IR}$ $=$ 10$^{11-12}$L$_{\odot}$),
are scheduled to be observed spectroscopically with the AKARI IRC during
AKARI phase 3. The above conclusions will be then tested based on a
statistically larger sample size than in the present paper.   

\vspace{1cm}

%\acknowledgments

This work is based on observations with AKARI, a JAXA project with the
participation of ESA. We thank the AKARI IRC instrument team,
particularly H. Matsuhara and D. Ishihara, for making this study
possible. 
Data presented herein were obtained at the W.M. Keck Observatory from
telescope time allocated to the National Aeronautics and Space
Administration through the agency's scientific partnership with the
California Institute of Technology and the University of California.
The Observatory was made possible by the generous financial support of
the W.M. Keck Foundation. 
We thank the anonymous referee for his/her useful comments. 
M.I. is supported by Grants-in-Aid for Scientific Research
(19740109). Part of the data analysis was performed using a computer
system operated by the Astronomical Data Analysis Center (ADAC) and the
Subaru Telescope of the National Astronomical Observatory, Japan.
This research made use of the SIMBAD database, operated at CDS,
Strasbourg, France, and of the NASA/IPAC Extragalactic Database (NED)
operated by the Jet Propulsion Laboratory, California Institute of
Technology, under contract with the National Aeronautics and Space
Administration.  

\clearpage

%\appendix

\section{Appendix A}

\citet{idm06} presented ground-based $L$-band (2.8--4.1 $\mu$m) spectra
of a large number of non-Seyfert ULIRGs at $z <$ 0.15 in the IRAS 1 Jy
sample \citep{kim98}. However, the sample is not statistically complete,
and unobserved non-Seyfert ULIRGs remain in this sample. After the
publication of \citet{idm06}, ground-based $L$-band spectra of six such
unobserved non-Seyfert ULIRGs, and two ULIRGs classified optically as
Seyfert 2s, (Table 5), were obtained. 

Observations of these ULIRGs were made using the IRCS infrared
spectrograph \citep{kob00} attached to the Nasmyth focus of the Subaru
8.2-m telescope \citep{iye04}, and the NIRSPEC infrared spectrograph 
\citep{mcl98} attached to the Nasmyth focus of the Keck II 10-m
telescope. The observing log is summarized in Table 5. 
For both observing runs, the sky was clear and the seeing at $K$ (2--2.5
$\mu$m), measured in images taken before $L$-band spectroscopy, was
$\sim$0$\farcs$4--0$\farcs$9 full-width at half-maximum (FWHM). 

For Subaru IRCS observing runs, a 0$\farcs$9-wide slit and the $L$-grism
were used with a 52-mas pixel scale.  The achievable spectral resolution
was R $\sim$ 140 at $\lambda \sim$ 3.5 $\mu$m.  
The slit length was 18$''$. A standard telescope nodding
technique (ABBA pattern), with a throw of 7$''$ along the slit, was
employed to subtract background emission.  Subaru's optical guider was
used to monitor the telescope tracking.  Exposure time was 0.8--1.2 s,
and 50--60 coadds were made at each nod position. 

For Keck II NIRSPEC observing runs, low-resolution long-slit
spectroscopy was employed with a 0$\farcs$76-wide (4 pixel) slit and 
the KL (2.16--4.19 $\mu$m) filter. 
The achievable spectral resolution was R $\sim$ 1100. 
The slit length was 42$''$.
A standard telescope nodding technique (ABBA pattern), with a throw of
15$''$ along the slit, was employed to subtract background emission. The
telescope tracking was monitored based on the infrared $K$-band image on
the SCAM, the slit viewer of NIRSPEC. 
Exposure time was 0.5--0.6 s, and 40--120 coadds were made at each nod
position.  

Appropriate standard stars (Table 5) were observed, with an air mass
difference of $<$0.1 compared to individual ULIRGs to correct for the
transmission of Earth's atmosphere.  The $L$-band magnitudes of the
standard stars were estimated based on their $V$-band (0.6 $\mu$m)
magnitudes and V$-$L colors of the corresponding stellar types
\citep{tok00}.  

Standard data analysis procedures were employed using IRAF 
\footnote{IRAF is distributed by the National Optical Astronomy
Observatories, which are operated by the Association of Universities for
Research in Astronomy, Inc. (AURA), under cooperative agreement with the
National Science Foundation.}.  
%--------------
Initially, frames taken with an A (B) beam were subtracted from frames
subsequently taken with a B (A) beam, and the resulting subtracted
frames were added and divided by a spectroscopic flat image.  Then, bad
pixels and pixels hit by cosmic rays were replaced with interpolated
values of surrounding pixels.  Finally the spectra of ULIRG's nuclei and
standard stars were extracted by integrating signals over
1$\farcs$1--2$\farcs$7, depending on actual signal profiles.  Wavelength
calibration was performed, taking into account the wavelength-dependent
transmission of Earth's atmosphere.  The spectra of ULIRG's nuclei were
divided by the observed spectra of standard stars, multiplied by the
spectra of blackbodies with temperatures appropriate to individual
standard stars (Table 5).  

Flux calibration was conducted based on the signals of ULIRGs and
standard stars detected inside our slit spectra. To reduce the scatter
of data points, appropriate binning of spectral elements was performed,
particularly at $\lambda_{\rm obs}$ $<$ 3.3 $\mu$m and $>$3.9 $\mu$m,
where the scatter is large due to poor atmospheric transmission and/or
large background emission.   

Figure 7 presents the final spectra of these eight ULIRGs. Table 6
summarizes the strengths of the 3.3 $\mu$m PAH emission features.   
Among the six ULIRGs classified optically as non-Seyferts, the low
EW$_{\rm 3.3PAH}$ value of the HII-region ULIRG, IRAS 01004$-$2237
($\sim$30 nm; Table 6), suggests the significant contribution from a
buried AGN to the observed 3--4 $\mu$m flux.

\clearpage

%--- Table 1 ---%
\begin{table}[h]
\scriptsize
\caption{ULIRGs observed with the AKARI IRC and their {\it IRAS}-based
infrared emission properties} 
\begin{center}
\begin{tabular}{lcrrrrcrll}
\hline
\hline
Object & Redshift &  
f$_{\rm 12}$  & f$_{\rm 25}$  & 
f$_{\rm 60}$  & f$_{\rm 100}$  & 
log L$_{\rm IR}$ & f$_{25}$/f$_{60}$ & Optical & Remark \\
 &    & (Jy) & (Jy)  & (Jy) 
& (Jy)  & (L$_{\odot}$) &  & Class & \\
(1) & (2) & (3) & (4) & 
(5) & (6) & (7) & (8) & (9) & (10) \\ \hline
IRAS 03521+0028 & 0.152 & $<$0.11 & 0.20 & 2.52 & 3.62 & 12.5 & 0.08 (C) & LINER & $z > $ 0.15\\  
IRAS 04074$-$2801 & 0.153 & $<$0.07 & 0.07 & 1.33 & 1.72 & 12.2 & 0.05 (C) & LINER & \\  
IRAS 05020$-$2941 & 0.154 & $<$0.06 & 0.10 & 1.93 & 2.06 & 12.3 & 0.05 (C) & LINER &  \\  
IRAS 09463+8141 & 0.156 & $<$0.07 & $<$0.07 & 1.43 & 2.29 & 12.3 &
$<$0.05 (C)  & LINER &  \\  
IRAS 10091+4704 & 0.246 & $<$0.06 & $<$0.08 & 1.18 & 1.55 & 12.6 &
$<$0.07 (C) & LINER & \\  
IRAS 11028+3130 & 0.199 & $<$0.09 & 0.09 & 1.02 & 1.44 & 12.4 & 0.09 (C) & LINER & \\  
IRAS 11180+1623 & 0.166 & $<$0.08 & $<$0.19 & 1.19 & 1.60 & 12.2 &
$<$0.16 (C) & LINER & \\  
IRAS 14121$-$0126 & 0.151 & 0.06 & 0.11 & 1.39 & 2.07 & 12.3 & 0.08 (C) & LINER & \\  
IRAS 16333+4630 & 0.191 & $<$0.06 & 0.06 & 1.19 & 2.09 & 12.4 & 0.05 (C) & LINER & \\  
IRAS 21477+0502 & 0.171 & $<$0.09 & 0.16 & 1.14 & 1.46 & 12.3 & 0.14 (C) & LINER & \\  
IRAS 23129+2548 & 0.179 & $<$0.08 & 0.08 & 1.81 & 1.64 & 12.4 & 0.04 (C) & LINER & \\  \hline
IRAS 01199$-$2307 & 0.156 & $<$0.11 & $<$0.16 & 1.61 & 1.37 & 12.3 &
$<$0.1 (C)  & HII-region & $z > $ 0.15 \\   
IRAS 01355$-$1814 & 0.192 & $<$0.06 & 0.12 & 1.40 & 1.74 & 12.4 & 0.09 (C) & HII-region & \\  
IRAS 03209$-$0806 & 0.166 & $<$0.10 & $<$0.13 & 1.00 & 1.69 & 12.2 & $<$0.13 (C) & HII-region & \\  
IRAS 10594+3818 & 0.158 & $<$0.09 & $<$0.15 & 1.29 & 1.89 & 12.2 & $<$0.12 (C) & HII-region & \\  
IRAS 12447+3721 & 0.158 & $<$0.12 & 0.10 & 1.04 & 0.84 & 12.1 & 0.10 (C) & HII-region & \\  
IRAS 13469+5833 & 0.158 & $<$0.05 & 0.04 & 1.27 & 1.73 & 12.2 & 0.03 (C) & HII-region & \\  
IRAS 14202+2615 & 0.159 & 0.18 & 0.15 & 1.49 & 1.99 & 12.4 & 0.10 (C) & HII-region & \\  
IRAS 15043+5754 & 0.151 & $<$0.12 & 0.07 & 1.02 & 1.50 & 12.1 & 0.07 (C) & HII-region & \\  
IRAS 17068+4027 & 0.179 & $<$0.08 & 0.12 & 1.33 & 1.41 & 12.3 & 0.09 (C) & HII-region & \\  
IRAS 22088$-$1831 & 0.170 & $<$0.09 & 0.07 & 1.73 & 1.73 & 12.4 & 0.04 (C) & HII-region & \\  \hline
IRAS 00482$-$2721 & 0.129 & $<$0.10 & $<$0.18 & 1.13 & 1.84 & 12.0 &
$<$0.16 (C) & LINER & $z <$ 0.15 \\
IRAS 08572+3915 & 0.058 & 0.32 & 1.70 & 7.43 & 4.59 & 12.1 & 0.23 (W) & LINER &\\
IRAS 09539+0857 & 0.129 & $<$0.15 & $<$0.15 & 1.44 & 1.04 & 12.0 & $<$0.11 (C) &LINER &\\  
IRAS 10494+4424 & 0.092 & $<$0.12 & 0.16 & 3.53 & 5.41 & 12.2 & 0.05 (C) &LINER &\\  
IRAS 16468+5200   & 0.150 & $<$0.06 & 0.10 & 1.01 & 1.04 & 12.1 & 0.10 (C)& LINER &\\  
IRAS 16487+5447 & 0.104 & $<$0.07 & 0.20 & 2.88 & 3.07 & 12.1 & 0.07 (C) &LINER &\\ 
IRAS 17028+5817   & 0.106 & $<$0.06 & 0.10 & 2.43 & 3.91 & 12.1 & 0.04
(C)& LINER $^{a}$ &\\  
IRAS 17044+6720   & 0.135 & $<$0.07 & 0.36 & 1.28 & 0.98 & 12.1 & 0.28 (W)& LINER &\\  \hline  
IRAS 00456$-$2904 & 0.110 & $<$0.08 & 0.14 & 2.60 & 3.38 & 12.2 & 0.05
(C) & HII-region & $z <$0.15 \\
IRAS 01298$-$0744 & 0.136 & $<$0.12 & 0.19 & 2.47 & 2.08 & 12.3 & 0.08 (C) & HII-region &\\
IRAS 01569$-$2939 & 0.141 & $<$0.11 & 0.14 & 1.73 & 1.51 & 12.2 & 0.08 (C) & HII-region &\\
IRAS 11387+4116 & 0.149 & 0.12 & $<$0.14 & 1.02 & 1.51 & 12.2 & $<$0.14 (C) &HII-region &\\  
IRAS 13539+2920 & 0.108 & $<$0.09 & 0.12 & 1.83 & 2.73 & 12.0 & 0.07 (C)
&HII-region & \\  \hline
IRAS 01494$-$1845 & 0.158 & $<$0.08 & $<$0.15 & 1.29 & 1.85 & 12.2 &
$<$0.12 (C) & unclassified & \\  
IRAS 02480$-$3745 & 0.165 & $<$0.05 & $<$0.11 & 1.25 & 1.49 & 12.2 &
$<$0.09 (C) & unclassified & \\  
IRAS 04313$-$1649 & 0.268 & $<$0.07 & 0.07 & 1.01 & 1.10 & 12.6 & 0.07 (C) & unclassified & \\  
IRAS 08591+5248 & 0.158 & $<$0.10 & $<$0.16 & 1.01 & 1.53 & 12.2 & $<$0.16 (C) & unclassified & \\  
IRAS 10035+2740 & 0.165 & $<$0.14 & $<$0.17 & 1.14 & 1.63 & 12.3 & $<$0.15 (C) & unclassified & \\ \hline
IRAS 05189$-$2524 & 0.042 & 0.73 & 3.44 & 13.67 & 11.36 & 12.1 & 0.25
(W) & Seyfert 2 & \\  
IRAS 14394+5332 & 0.105 & 0.03 & 0.35 & 1.95 & 2.39 & 12.1 & 0.18 (C) & Seyfert 2 & \\  
IRAS 17179+5444 & 0.147 & $<$0.08 & 0.20 & 1.36 & 1.91 & 12.2 & 0.15 (C) & Seyfert 2 & \\  
IRAS 23498+2423 & 0.212 & $<$0.10 & 0.12 & 1.02 & 1.45 & 12.5 & 0.12 (C) & Seyfert 2 & \\  \hline
UGC 5101 & 0.040 & 0.25 & 1.03 & 11.54 & 20.23 & 12.0 & 0.09 (C)  & LINER & interesting \\  
IRAS 19254$-$7245 & 0.062 & 0.22 & 1.24 & 5.48 & 5.79 & 12.1 & 0.23 (W) & Seyfert 2 & \\  \hline
\end{tabular}
\end{center}
\end{table}

\clearpage

Notes.

Col.(1): Object name.

Col.(2): Redshift. 

Cols.(3)--(6): f$_{12}$, f$_{25}$, f$_{60}$, and f$_{100}$ are 
{\it IRAS} fluxes at 12$\mu$m, 25$\mu$m, 60$\mu$m, and 100$\mu$m,
respectively, taken from \citet{kim98}, except UGC 5101 and IRAS
19254$-$7245, for which {\it IRAS FSC} fluxes are shown.

Col.(7): Decimal logarithm of infrared (8$-$1000 $\mu$m) luminosity in
units of solar luminosity (L$_{\odot}$), calculated with $L_{\rm IR} =
2.1 \times 10^{39} \times$ D(Mpc)$^{2}$ $\times$ (13.48 $\times$
$f_{12}$ + 5.16 $\times$ $f_{25}$ + $2.58 \times f_{60} + f_{100}$) ergs
s$^{-1}$ \citep{sam96}. Since the calculation is based on our adopted
cosmology, the infrared luminosities differ slightly ($<$10\%) from the
values shown in Kim \& Sanders (1998, their Table 1, column 15). For
sources that have upper limits in some {\it IRAS} bands, we can derive
upper and lower limits on the infrared luminosity by assuming that the
actual flux is the {\it IRAS}-upper limit and zero value,
respectively. The difference between the upper and lower values is
usually very small, less than 0.2 dex. We assume that the infrared
luminosity is the average of these values.  

Col.(8): {\it IRAS} 25 $\mu$m to 60 $\mu$m flux ratio.
ULIRGs with f$_{25}$/f$_{60}$ $<$ 0.2 and $>$ 0.2 are
classified as cool and warm sources (denoted as ``C'' and ``W''),
respectively \citep{san88b}.

Col.(9): Optical spectral classification by \citet{vei99}, except 
UGC 5101 and IRAS 19254$-$7245.
For UGC 5101 and IRAS 19254$-$7245, optical classification is based
on \citet{vei95} and \citet{mir91}, respectively.

Col.(10): Remarks on individual objects. 

$^{a}$: IRAS 17028+5817 consists of western (W) and eastern (E) nuclei.  
IRAS 17028+5817 W and E are classified optically as a LINER and HII-region,
respectively \citep{vei99}.
The optical classification of the combined spectrum of both nuclei 
is a LINER (Veilleux et al. 1999; their Table 2).

\clearpage

%--- Table 2 ---%
\begin{table}[h]
\scriptsize
\caption{Observation log of the AKARI IRC}
\begin{center}
\begin{tabular}{lcl}
\hline
\hline
Object & Observation ID & Observation date \\
(1) & (2) & (3) \\ \hline
IRAS 03521+0028   & 1100200-001 & 2007 August 19   \\
IRAS 04074$-$2801 & 1100201-001 & 2007 August 14   \\
IRAS 05020$-$2941 & 1100003-001 & 2007 February 28 \\
IRAS 09463+8141   & 1100004-001 & 2006 October 8   \\
IRAS 10091+4704   & 1100122-001 & 2007 May 7       \\
IRAS 11028+3130   & 1100006-001 & 2006 November 26 \\
IRAS 11180+1623   & 1100202-001 & 2007 June 5      \\
IRAS 14121$-$0126 & 1100011-001 & 2007 January 22  \\
IRAS 16333+4630   & 1100013-001 & 2007 February 8  \\
IRAS 21477+0502   & 1100207-001 & 2007 May 22      \\
IRAS 23129+2548   & 1100015-001 & 2006 December 22 \\ \hline
IRAS 01199$-$2307 & 1100209-001 & 2007 July 1      \\
IRAS 01355$-$1814 & 1100018-001 & 2007 January 6   \\
IRAS 03209$-$0806 & 1100210-001 & 2007 August 8    \\
IRAS 10594+3818   & 1100021-001 & 2006 November 23 \\
IRAS 12447+3721   & 1100022-001 & 2006 December 15 \\
IRAS 13469+5833   & 1100023-001 & 2006 December 7  \\
IRAS 14202+2615   & 1100212-001 & 2007 July 15     \\
IRAS 15043+5754   & 1100213-001 & 2007 June 23     \\
IRAS 17068+4027   & 1100026-001 & 2007 February 26 \\
IRAS 22088$-$1831 & 1100214-001 & 2007 May 19      \\ \hline
IRAS 00482$-$2721 & 1100036-001 & 2006 December 21 \\
IRAS 08572+3915   & 1100049-001 & 2006 October 29  \\
IRAS 09539+0857   & 1100267-001 & 2007 May 19      \\
IRAS 10494+4424   & 1100266-001 & 2007 May 16      \\
IRAS 16468+5200   & 1100249-001 & 2007 August 10   \\
IRAS 16487+5447   & 1100247-001 & 2007 August 6    \\
IRAS 17028+5817   & 1100248-001 & 2007 August 8    \\
IRAS 17044+6720   & 1100297-001 & 2007 May 31      \\ \hline
IRAS 00456$-$2904 & 1100221-001 & 2007 June 19     \\
IRAS 01298$-$0744 & 1100226-001 & 2007 July 10     \\
IRAS 01569$-$2939 & 1100225-001 & 2007 July 7      \\
IRAS 11387+4116   & 1100269-001 & 2007 May 29      \\
IRAS 13539+2920   & 1100235-001 & 2007 July 6      \\ \hline
IRAS 01494$-$1845 & 1100215-001 & 2007 July 10     \\
IRAS 02480$-$3745 & 1100030-001 & 2007 January 14  \\
IRAS 04313$-$1649 & 1100031-001 & 2007 February 22 \\
IRAS 08591+5248   & 1100121-001 & 2007 April 21    \\
IRAS 10035+2740   & 1100216-001 & 2007 May 15      \\ \hline 
IRAS 05189$-$2524 & 1100129-001 & 2007 March 8     \\
IRAS 14394+5332   & 1100283-001 & 2007 June 25     \\ 
IRAS 17179+5444   & 1100253-001 & 2007 August 23   \\
IRAS 23498+2423   & 1100287-001 & 2007 July 1      \\ \hline 
UGC 5101          & 1100134-001 & 2007 April 22    \\
IRAS 19254$-$7245 & 1100132-001 & 2007 March 30    \\ \hline
\end{tabular}
\end{center}
\end{table}

\clearpage

Notes.

Col.(1): Object name.

Col.(2): Observation ID.

Col.(3): Observation date in UT.

\clearpage

%--- Table 3 ---%
\begin{table}[h]
\scriptsize
\caption{Properties of 3.3 $\mu$m PAH emission, derived from AKARI IRC 
spectra}
\begin{center}
\begin{tabular}{lcccc}
\hline
\hline
Object &  f$_{3.3 \rm PAH}$ & L$_{3.3 \rm PAH}$ & L$_{3.3 \rm PAH}$/L$_{\rm IR}$ & 
rest EW$_{3.3 \rm PAH}$ \\
 & ($\times$10$^{-14}$ ergs s$^{-1}$ cm$^{-2}$) & ($\times$10$^{41}$ergs s$^{-1}$) 
& ($\times$10$^{-3}$) & (nm) \\  
(1) & (2) & (3) & (4) & (5) \\ \hline
IRAS 03521+0028 & 2.2 & 12.2 & 0.1 & 108  \\
IRAS 04074$-$2801 & 0.9 & 5.0 & 0.1 & 37  \\
IRAS 05020$-$2941 & 1.0 & 5.5 & 0.07 & 53 \\
IRAS 09463+8141 & 2.2 & 12.6 & 0.2 & 149  \\
IRAS 10091+4704 & 0.7 & 11.4 & 0.08 & 75  \\
IRAS 11028+3130 & 0.6 & 6.3 & 0.07 & 100  \\
IRAS 11180+1623 & 0.5 & 3.1 & 0.05 & 36   \\
IRAS 14121$-$0126 & 2.9 & 15.4 & 0.2 & 80 \\
IRAS 16333+4630 & 2.1 & 19.0 & 0.2 & 63   \\
IRAS 21477+0502 & 0.6 & 4.0 & 0.05 & 29   \\
IRAS 23129+2548 & $<$0.4 & $<$2.5 & $<$0.03 & $<$12 \\ \hline
IRAS 01199$-$2307 & 0.8 & 4.5 & 0.07 & 76  \\
IRAS 01355$-$1814 & 0.6 & 5.2 & 0.05 & 54  \\
IRAS 03209$-$0806 & 2.2 & 14.8 & 0.25 & 68 \\
IRAS 10594+3818 & 2.6 & 15.4 & 0.25 & 79   \\
IRAS 12447+3721 & 1.3 & 7.7 & 0.15 & 187   \\
IRAS 13469+5833 & 1.4 & 8.1 & 0.15 & 40    \\
IRAS 14202+2615 & 4.2 & 25.3 & 0.25 & 32   \\
IRAS 15043+5754 & 2.0 & 10.9 & 0.2 & 95   \\
IRAS 17068+4027 & 2.2 & 17.4 & 0.2 & 144   \\
IRAS 22088$-$1831 & $<$0.2 & $<$1.3 & $<$0.02 & $<$8 \\ \hline
IRAS 00482$-$2721 & 1.3 & 4.8 & 0.1 & 99  \\
IRAS 08572+3915 & $<$4.2 & $<$2.9 & $<$0.07 & $<$5 \\
IRAS 09539+0857 & 1.2 & 4.5 & 0.1 & 73    \\
IRAS 10494+4424 & 3.6 & 6.7 & 0.1 & 55    \\
IRAS 16468+5200 & 0.6 & 3.0 & 0.07 & 78    \\
IRAS 16487+5447 & 3.3 & 8.0 & 0.15 & 90    \\
IRAS 17028+5817W & 3.3 & 8.3 & 0.15 & 82   \\
IRAS 17044+6720 & 2.3 & 9.8 & 0.2 & 14     \\ \hline
IRAS 00456$-$2904 & 4.0 & 10.9 & 0.2 & 56 \\
IRAS 01298$-$0744 & 1.7 & 7.4 & 0.1 & 133  \\
IRAS 01569$-$2939 & 1.8 & 8.4 & 0.15 & 84  \\
IRAS 11387+4116 & 1.9 & 10.2 & 0.2 & 59    \\
IRAS 13539+2920 & 4.3 & 11.2 & 0.3 & 52   \\ 
IRAS 17028+5817E & 0.4 & 1.1 & 0.02 & 34   \\ \hline
IRAS 01494$-$1845 & 2.2 & 12.8 & 0.2 & 68  \\
IRAS 02480$-$3745 & 1.5 & 9.9 & 0.15 & 265 \\
IRAS 04313$-$1649 & $<$0.4 & $<$6.3 & $<$0.05 & $<$73  \\
IRAS 08591+5248 & 1.4 & 8.4 & 0.15 & 39    \\
IRAS 10035+2740 & 1.4 & 9.1 & 0.15 & 80     \\ \hline
IRAS 05189$-$2524 & 30.0 &  10.7 & 0.2 & 10 \\
IRAS 14394+5332 & 6.2 & 15.3 & 0.4 & 71  \\
IRAS 17179+5444 & 1.2 & 5.9 & 0.1 & 16  \\
IRAS 23498+2423 & $<$1.9 & $<$21 & $<$0.25 & $<$8  \\ \hline
UGC 5101 & 22.5 & 7.3 & 0.2 & 33  \\
IRAS 19254$-$7245 & 3.8 & 3.1 & 0.07 & 5  \\ \hline
\end{tabular}
\end{center}
\end{table}

\clearpage

Col. (1): Object name. 
Col. (2): Observed flux of 3.3 $\mu$m PAH emission. 
%The second effective digit, if smaller than unity, is shown in units of 0.5. 
Col. (3): Observed luminosity of 3.3 $\mu$m PAH emission.  
Col. (4): Observed 3.3 $\mu$m PAH-to-infrared luminosity ratio in units
          of 10$^{-3}$. 
          Typical ratios for less infrared-luminous starbursts are 
          $\sim$10$^{-3}$ \citep{mou90,ima02}.   
Col. (5): Rest-frame equivalent width of 3.3 $\mu$m PAH emission.  
          Those for starbursts are typically $\sim$100 nm
          \citep{moo86,imd00}.   

\clearpage 

%--- Table 4 ---%
\begin{table}[h]
\scriptsize
\caption{Optical depths of absorption features and AGN signatures,
derived from AKARI IRC spectra}
\begin{center}
\begin{tabular}{lccc|cc}
\hline
\hline
Object & Observed $\tau_{3.1}$ & Observed $\tau_{3.4}$ & A$_{\rm V}$
(mag) & \multicolumn{2}{c}{AGN signatures} \\
 & & & & EW$_{\rm 3.3PAH}$ & $\tau_{3.1}$ or $\tau_{3.4}$ \\
(1) & (2) & (3) & (4) & (5) & (6) \\ \hline
IRAS 03521+0028   & 0.2 & --- & $>$3  & --- & --- \\
IRAS 04074$-$2801 & 0.6 & --- & $>$10 & $\bigcirc$ & $\bigcirc$\\
IRAS 05020$-$2941 & 0.5 & --- & $>$8  & --- & $\bigcirc$ \\
IRAS 09463+8141   & 0.5 & --- & $>$8  & --- & $\bigcirc$ \\
IRAS 10091+4704   & --- & --- & ---   & --- & --- \\
IRAS 11028+3130   & 0.6 & --- & $>$10 & --- & $\bigcirc$ \\
IRAS 11180+1623   & 0.7 & --- & $>$11 & $\bigcirc$ & $\bigcirc$ \\ 
IRAS 14121$-$0126 & 1.0 & --- & $>$16 & --- & $\bigcirc$ \\
IRAS 16333+4630   & 2.0 & --- & $>$33 & --- & $\bigcirc$ \\
IRAS 21477+0502   & --- & --- & ---   & $\bigcirc$ & --- \\
IRAS 23129+2548   & --- & --- & ---   & $\bigcirc$ & --- \\ \hline 
IRAS 01199$-$2307 & 0.8 & --- & $>$13 & --- & $\bigcirc$ \\
IRAS 01355$-$1814 & 1.0 & --- & $>$16 & --- & $\bigcirc$ \\
IRAS 03209$-$0806 & 0.3 & --- & $>$5  & --- & --- \\
IRAS 10594+3818   & 0.4 & --- & $>$6  & --- & --- \\
IRAS 12447+3721   & --- & --- & ---   & --- & --- \\
IRAS 13469+5833   & 0.2 & --- & $>$3  & --- & --- \\
IRAS 14202+2615   & --- & --- & ---   & $\bigcirc$ & --- \\
IRAS 15043+5754   & 0.4 & --- & $>$6  & --- & --- \\
IRAS 17068+4027   & 1.5 & --- & $>$25 & --- & $\bigcirc$ \\
IRAS 22088$-$1831 & 0.5 & --- & $>$8  & $\bigcirc$ & $\bigcirc$ \\ \hline
IRAS 00482$-$2721 & 1.0 & --- & $>$16 & --- & $\bigcirc$ \\
IRAS 08572+3915   & 0.3 & 0.8 & 119--205 & $\bigcirc$ & $\bigcirc$ \\
IRAS 09539+0857   & 1.1 & --- & $>$18 & --- & $\bigcirc$ \\
IRAS 10494+4424   & 0.8 & --- & $>$13 & --- & $\bigcirc$ \\
IRAS 16468+5200   & 1.0 & --- & $>$16 & --- & $\bigcirc$ \\
IRAS 16487+5447   & --- & --- & ---   & --- & --- \\
IRAS 17028+5817W  & 0.9 & --- & $>$15 & --- & $\bigcirc$ \\
IRAS 17044+6720   & --- & 0.15 & 21--38 & $\bigcirc$ & --- \\ \hline
IRAS 00456$-$2904 & 0.4 & --- & $>$6  & --- & --- \\
IRAS 01298$-$0744 & --- & --- & ---   & --- & --- \\
IRAS 01569$-$2939 & --- & --- & ---   & --- & --- \\
IRAS 11387+4116   & 0.5 & --- & $>$8  & --- & $\bigcirc$ \\
IRAS 13539+2920  & 0.2 & --- & $>$3  & --- & --- \\ 
IRAS 17028+5817E  & --- & --- & ---   & $\bigcirc$ & --- \\ \hline
IRAS 01494$-$1845 & 0.7 & --- & $>$11 & --- & $\bigcirc$ \\
IRAS 02480$-$3745 & --- & --- & ---   & --- & --- \\
IRAS 04313$-$1649 & --- & --- & ---   & --- & --- \\
IRAS 08591+5248   & 0.3 & --- & $>$5  & $\bigcirc$ & --- \\
IRAS 10035+2740   & 0.3 & --- & $>$5  & ---  & --- \\ \hline
IRAS 05189$-$2524 & --- & --- & & $\bigcirc$ & --- \\
IRAS 14394+5332   & --- & --- & & --- & --- \\
IRAS 17179+5444   & --- & --- & & $\bigcirc$ & --- \\
IRAS 23498+2423   & --- & --- & & $\bigcirc$ & --- \\ 
\hline
UGC 5101          & 1.0 & 0.6 & 102--166 & $\bigcirc$ & $\bigcirc$ \\
IRAS 19254$-$7245 & 0.2 & 0.5 & 74--128  & $\bigcirc$ & $\bigcirc$ \\ \hline
\end{tabular}
\end{center}
\end{table}

\clearpage

Col. (1): Object name. 
Col. (2): Observed optical depth of 3.1 $\mu$m absorption features
due to ice-covered dust grains.
Col. (3): Observed optical depth of 3.4 $\mu$m absorption features 
due to bare carbonaceous dust grains.
Col. (4): Dust extinction (A$_{\rm V}$) toward the 3--4 $\mu$m
continuum emission regions, derived from the absorption optical depths,
assuming the Galactic dust extinction curve and a foreground screen dust
distribution model. 
For those with detected 3.1 $\mu$m absorption, but no measurable
$\tau_{3.4}$ values, the derived A$_{\rm V}$ values should be taken as
lower limits (see text in $\S$5.4).  
Col. (6): Signatures of an AGN, based on the low value of the 
          rest-frame equivalent width of the 3.3 $\mu$m PAH emission.
          $\bigcirc$: present.
Col. (7): Signatures of an AGN, based on the large optical depth of the 
          3.1 $\mu$m and/or 3.4 $\mu$m absorption features.
          $\bigcirc$: present.
\clearpage 

%--- Table 5 ---%
\begin{table}[h]
\scriptsize
\caption{Observed ULIRGs with ground-based telescopes and
observation log}
\begin{center}
\begin{tabular}{llcllcrlccc}
\hline
\hline
Object & Redshift & Spectral type & Date & Telescope & Integration &
P.A. \footnote{a} & \multicolumn{4}{c}{Standard Stars} \\
(1) & (2) & (3) & (4) & (5) & (6) & (7) & (8) & (9) & (10) & (11) \\ \hline
IRAS 00482$-$2721 & 0.129 & LINER & 2007 September 25 & Keck NIRSPEC &
32 & 0 & HR 173 & 4.6 & G3V & 5800 \\
IRAS 04103$-$2838 & 0.118 & LINER & 2007 September 24 & Keck NIRSPEC &
48 & 0 & HR 1179 & 5.2 & F8V & 6000 \\
IRAS 00456$-$2904 & 0.110 & HII & 2006 July 19 & Subaru IRCS & 38 & 90 &
HR 210 & 4.0 & G3V & 5800 \\ 
IRAS 22491$-$1808 & 0.076 & HII & 2006 July 18 & Subaru IRCS & 60 & 110
& HR 8544 & 5.1 & G2V & 5830 \\
IRAS 01004$-$2237 & 0.118 & HII & 2007 September 24 & Keck NIRSPEC & 48
& 15 & HR 173 & 4.6 & G3V & 5800 \\
IRAS 02021$-$2013 & 0.116 & Unclassified & 2007 January 15 & Subaru IRCS
& 32 & 90 & HR 695 & 3.7 & G0V & 5930 \\
IRAS 23233+2817 & 0.114 & Sy2 & 2007 September 25 & Keck NIRSPEC & 40 &
$-$16 & HR 8792 & 4.9 & F7V & 6240 \\
IRAS 23389+0300 & 0.145 & Sy2 & 2007 September 25 & Keck NIRSPEC & 64 &
10 & HR 8931 & 5.1 & F8V & 6000 \\
\hline
\end{tabular}
\end{center}
\end{table}

Notes: 

Col.(1): Object name.
Col.(2): Redshift.
Col.(3): Optical spectral classification by \citet{vei99}.
Col.(4): Observation date in UT.
Col.(5): Telescope and instrument. 
Col.(6): Net on-source integration time in minutes.
Col.(7): Position angle of the slit.
Col.(8): Standard star name.
Col.(9): Adopted $L$-band magnitude.
Col.(10): Stellar spectral type.
Col.(11): Effective temperature.

$^{a}$: 0$^{\circ}$ corresponds to the north-south direction.
Position angle increases counterclockwise on the sky plane.

%--- Table 6 ---%
\begin{table}[h]
\scriptsize
\caption{Properties of 3.3 $\mu$m PAH emission for ULIRGs
observed with ground-based telescopes}
\begin{center}
\begin{tabular}{lccc}
\hline
\hline
Object &  f$_{3.3 \rm PAH}$ & L$_{3.3 \rm PAH}$ & rest EW$_{3.3 \rm PAH}$ \\
 & ($\times$10$^{-14}$ ergs s$^{-1}$ cm$^{-2}$) & ($\times$10$^{41}$ergs s$^{-1}$) 
& (nm) \\  
(1) & (2) & (3) & (4)  \\ \hline
IRAS 00482$-$2721 & 0.9 & 3.6 & 115 \\
IRAS 04103$-$2838 & 3.3 & 10.2 & 45 \\
IRAS 00456$-$2904 & 1.8 & 4.8 & 125 \\
IRAS 22491$-$1808 & 3.5 & 4.3 & 180 \\
IRAS 01004$-$2237 & 1.6 & 5.1 & 30 \\
IRAS 02021$-$2013 & 1.4 & 4.4 & 55 \\ 
IRAS 23233+2817   & 1.7 & 4.9 & 30 \\
IRAS 23389+0300   & 2.3 & 11.1 & 90 \\ \hline
\end{tabular}
\end{center}
\end{table}

Notes. 

Col.(1): Object name. 
Col.(2): Observed flux of 3.3 $\mu$m PAH emission. 
Col.(3): Observed luminosity of 3.3 $\mu$m PAH emission.  
Col.(4): Rest-frame equivalent width of 3.3 $\mu$m PAH emission.  

\clearpage 

%---  Figure 1 ---% 
\begin{figure}
\FigureFile(80mm,80mm){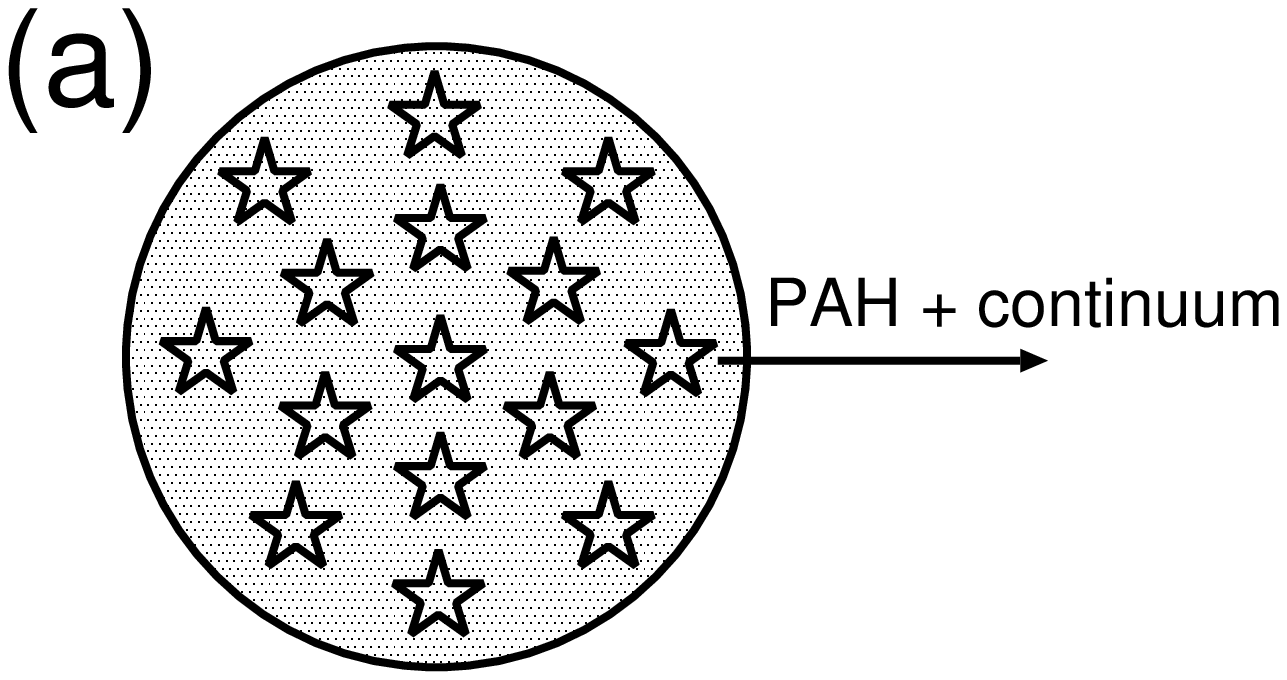}
\FigureFile(80mm,80mm){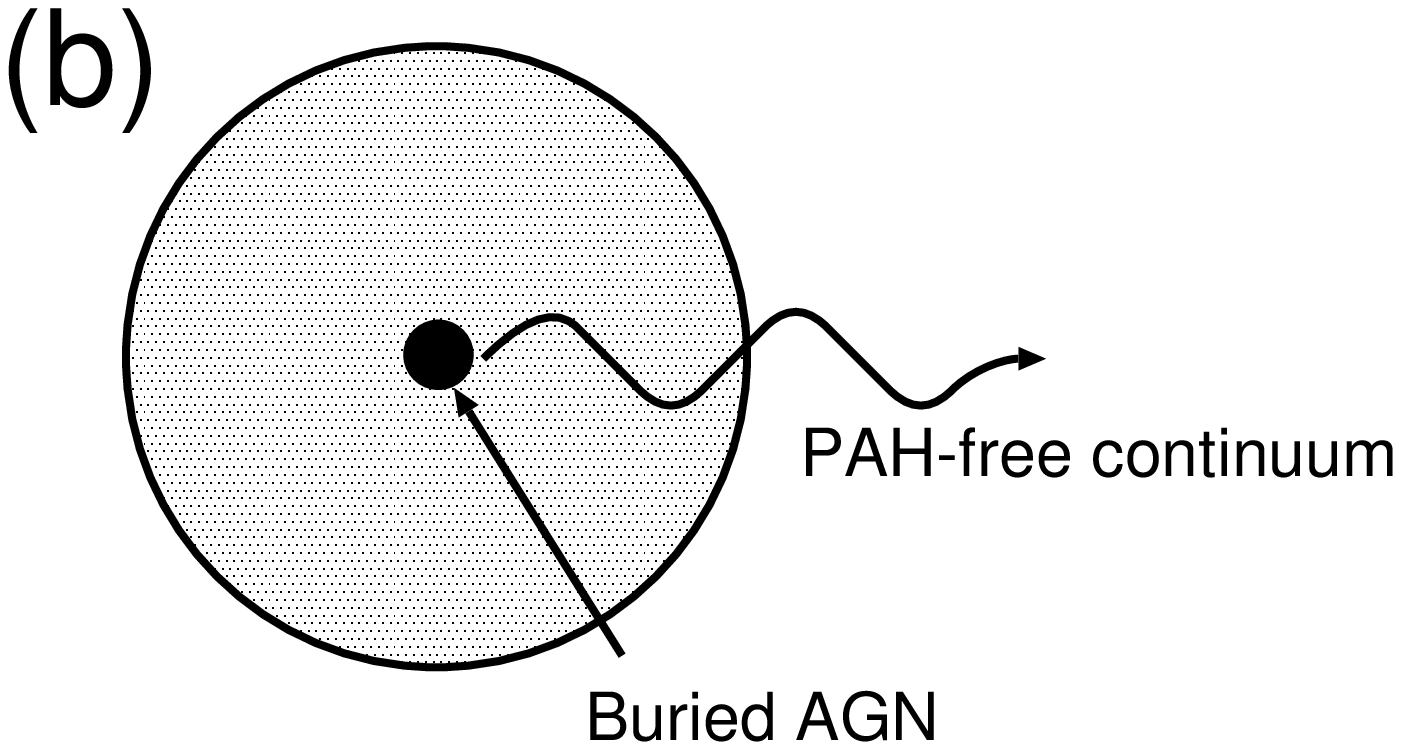} 
\FigureFile(80mm,80mm){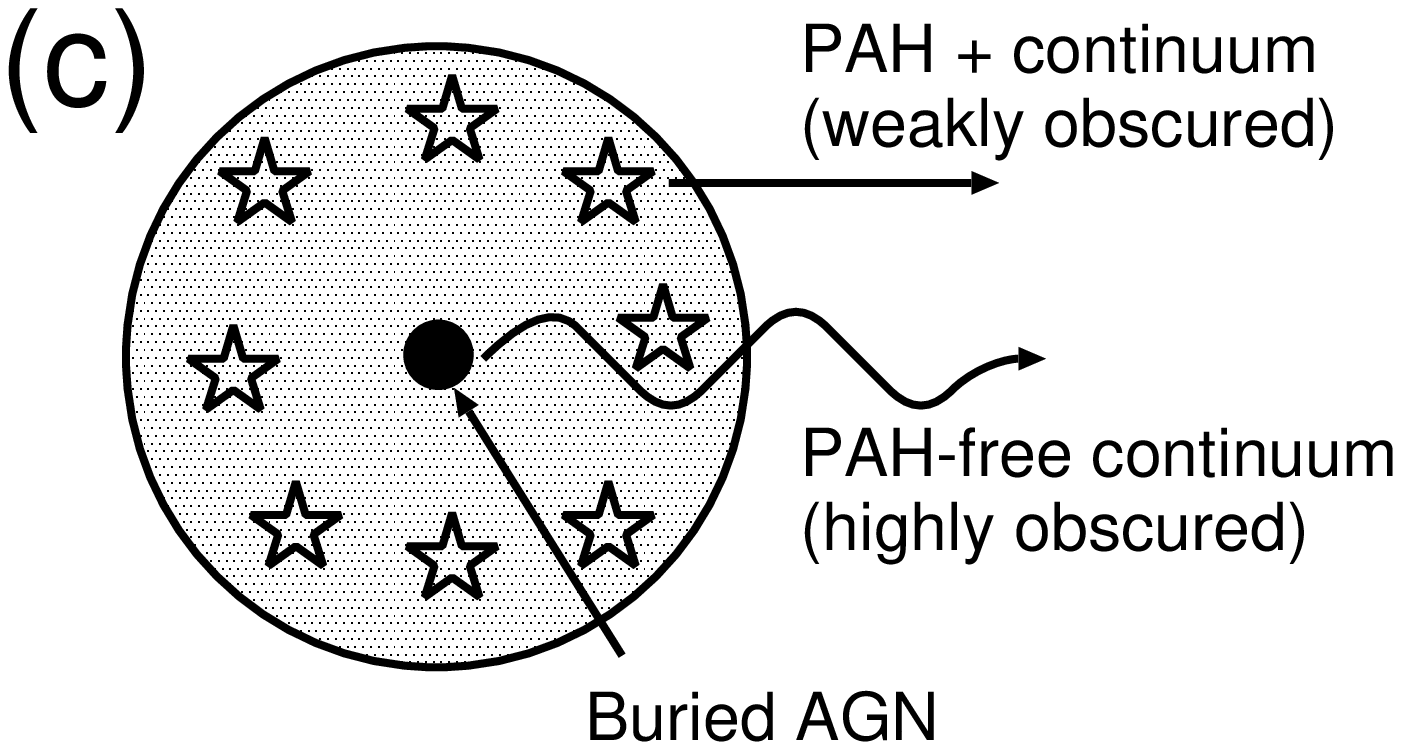}
\FigureFile(80mm,80mm){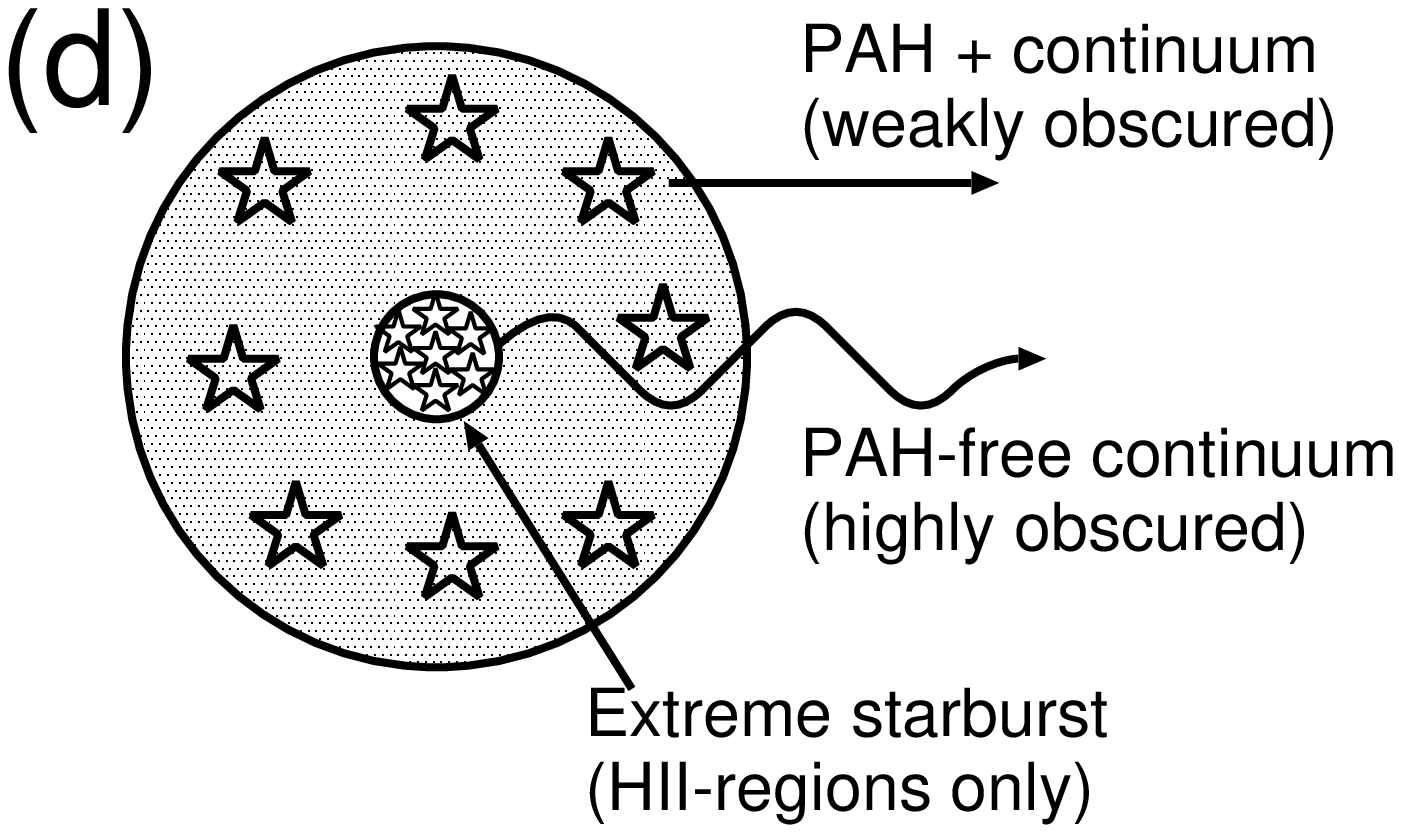} 
\FigureFile(110mm,110mm){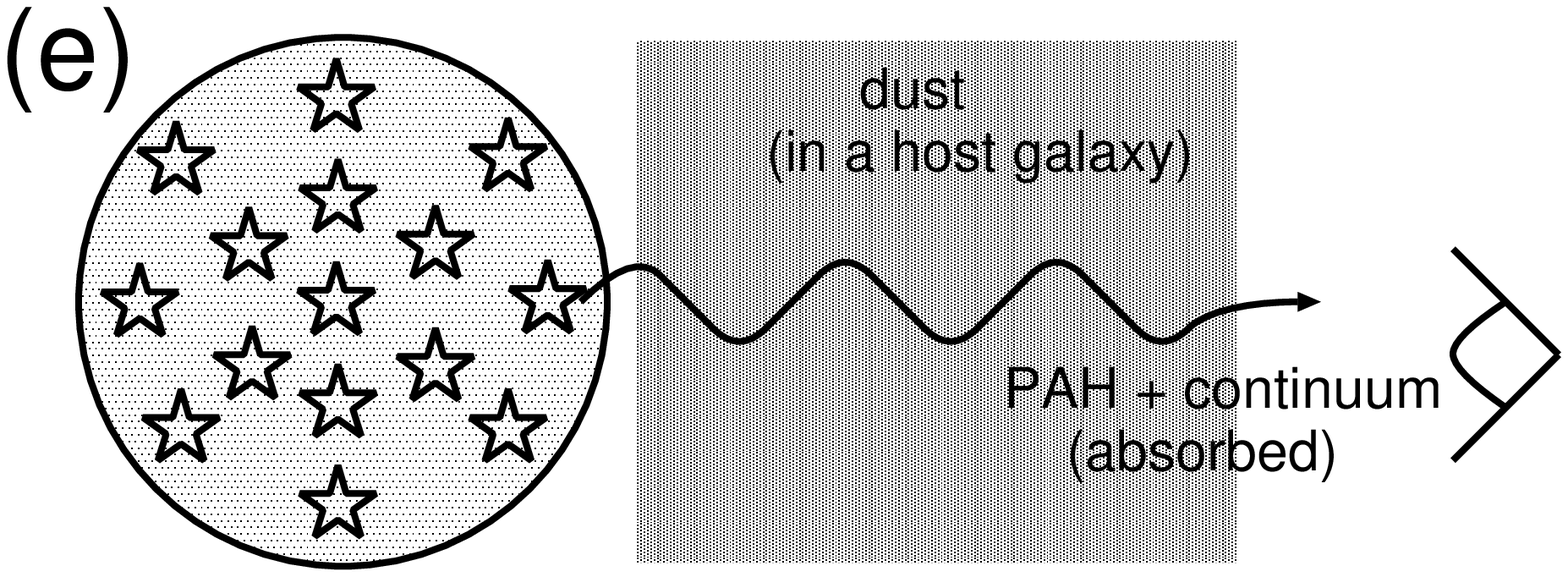}
\caption{
{\it (a)}: Geometry of energy sources and dust in a normal
starburst. The open star symbols indicate ``stars'' in a starburst. The
energy sources (stars) and dust are spatially well mixed. Large
equivalent width PAH emission is observed, regardless of the amount of
dust extinction, because both PAH and continuum emission fluxes are
similarly attenuated in this geometry. 
{\it (b)}:  Geometry of the energy source and dust in a buried AGN. The
energy source (a compact mass-accreting supermassive black hole) is more
centrally concentrated than the surrounding dust. No PAH emission is
observed, because PAHs are destroyed by strong X-ray radiation
from the AGN \citep{voi92}. 
{\it (c)}: A buried AGN and starburst composite. The starburst surrounds
the central buried AGN. The observed spectrum is a superposition of PAH
+ continuum emission from the starburst and PAH-free continuum from the
AGN. Since the buried AGN is more highly obscured than the surrounding
starbursts, the starburst emission generally makes a strong contribution
to the observed flux even if the starburst is energetically
insignificant.  
{\it (d)}: An exceptionally centrally concentrated extreme starburst
whose emitting volume is predominantly occupied by HII-regions, without
photodissociation regions and molecular gas. Such an extreme starburst
produces no PAH emission, similar to a buried AGN. 
{\it (e)}: A normal starburst nucleus (mixed dust/source geometry)
obscured by a large amount of foreground dust in an edge-on host galaxy,
exterior to the ULIRG's nuclear core. This geometry can produce a strong
dust absorption feature whose optical depth is larger than the maximum
threshold obtained in the mixed dust/source geometry.  
}
\end{figure}

%---  Figure 2 ---%
\begin{figure}
\FigureFile(80mm,80mm){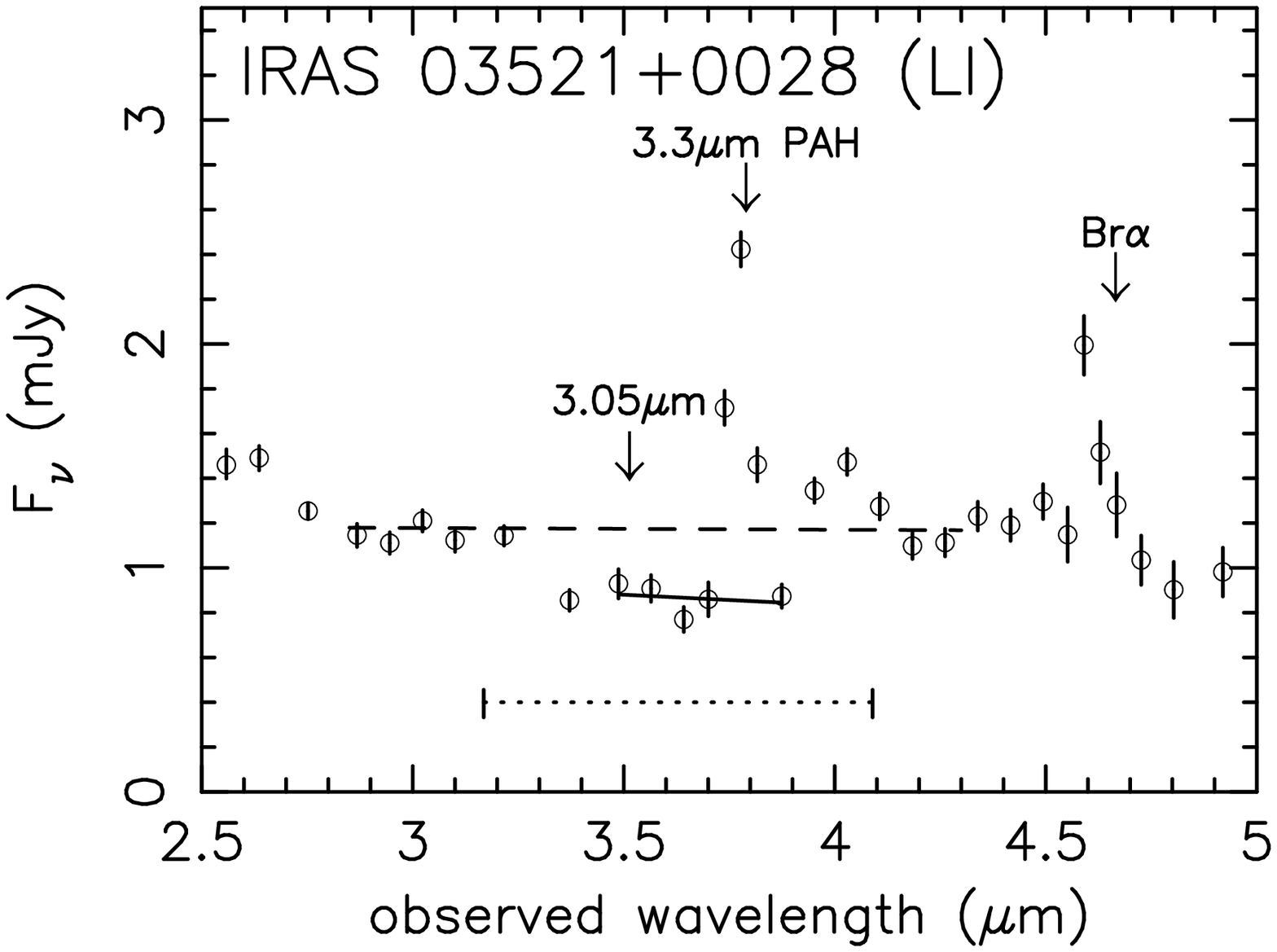}
\FigureFile(80mm,80mm){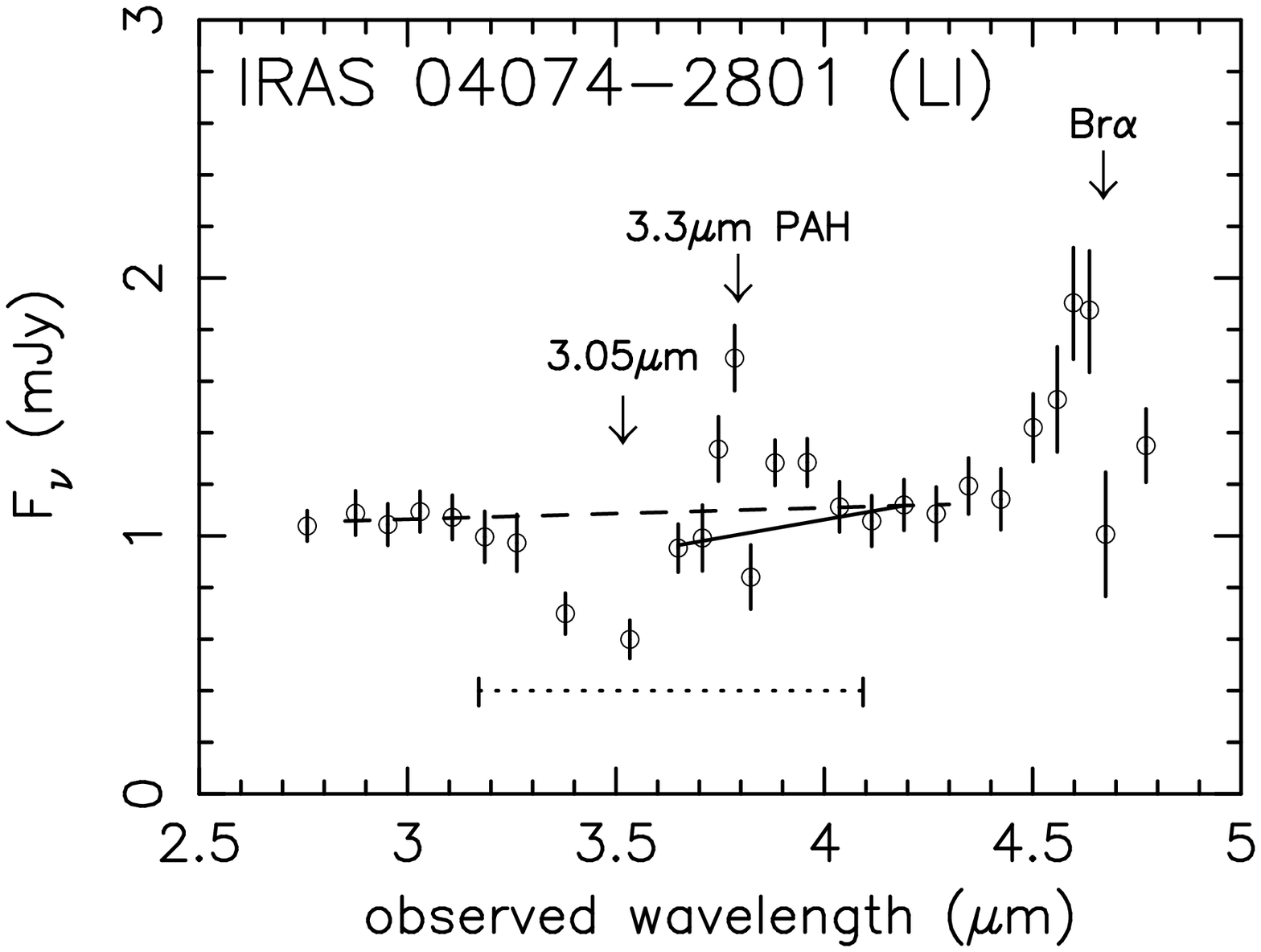} 
\FigureFile(80mm,80mm){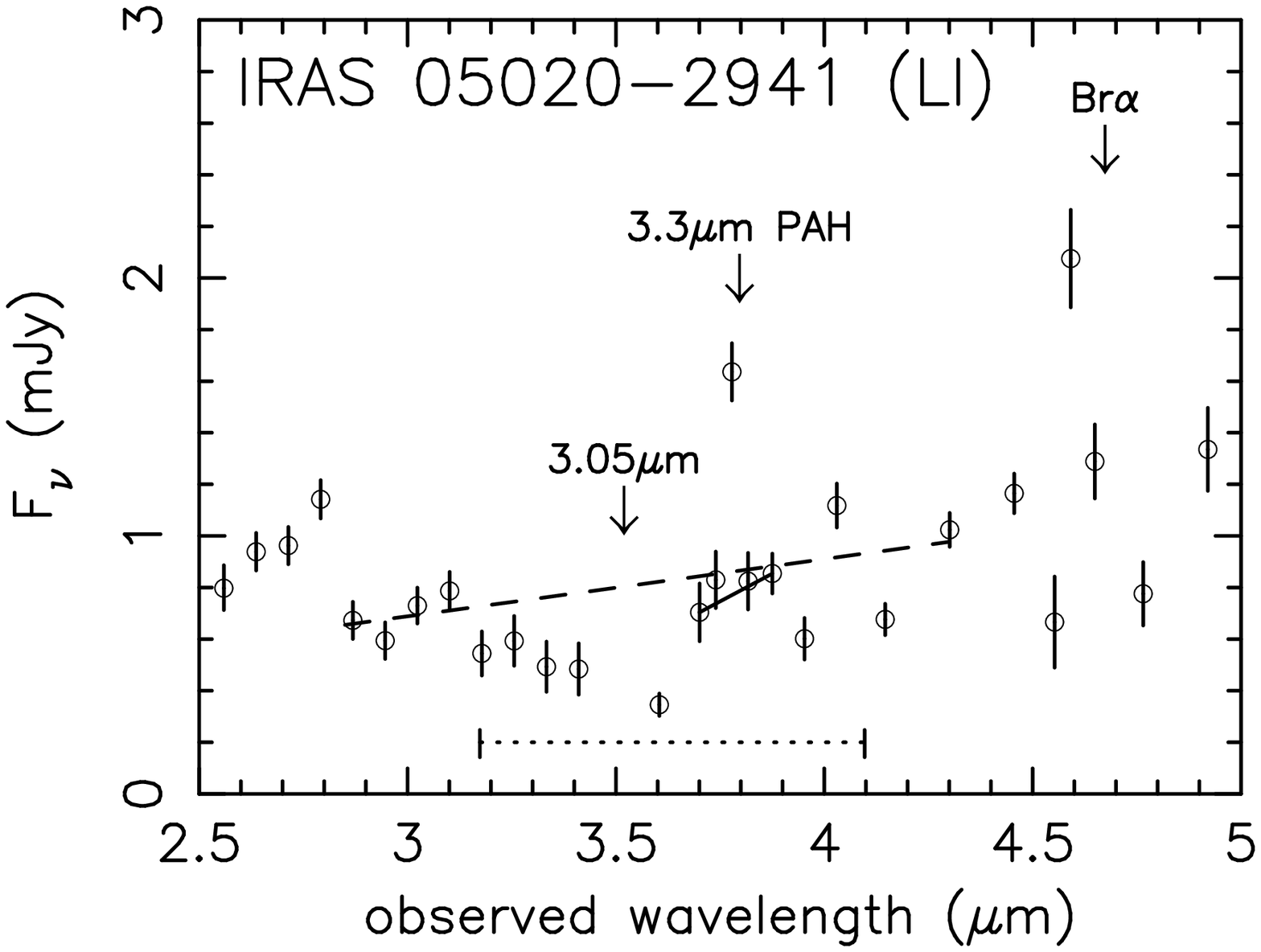}
\FigureFile(80mm,80mm){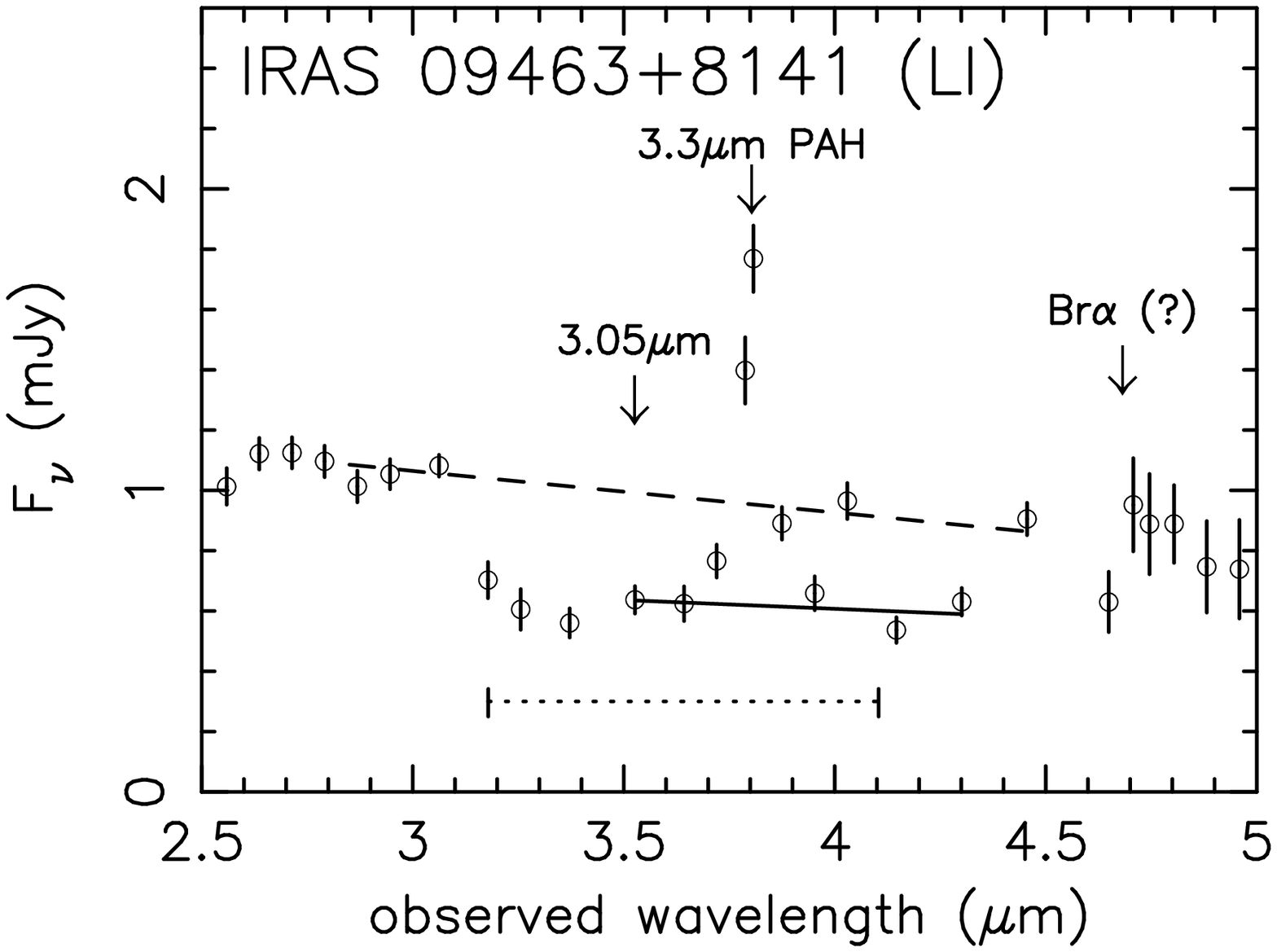} 
\FigureFile(80mm,80mm){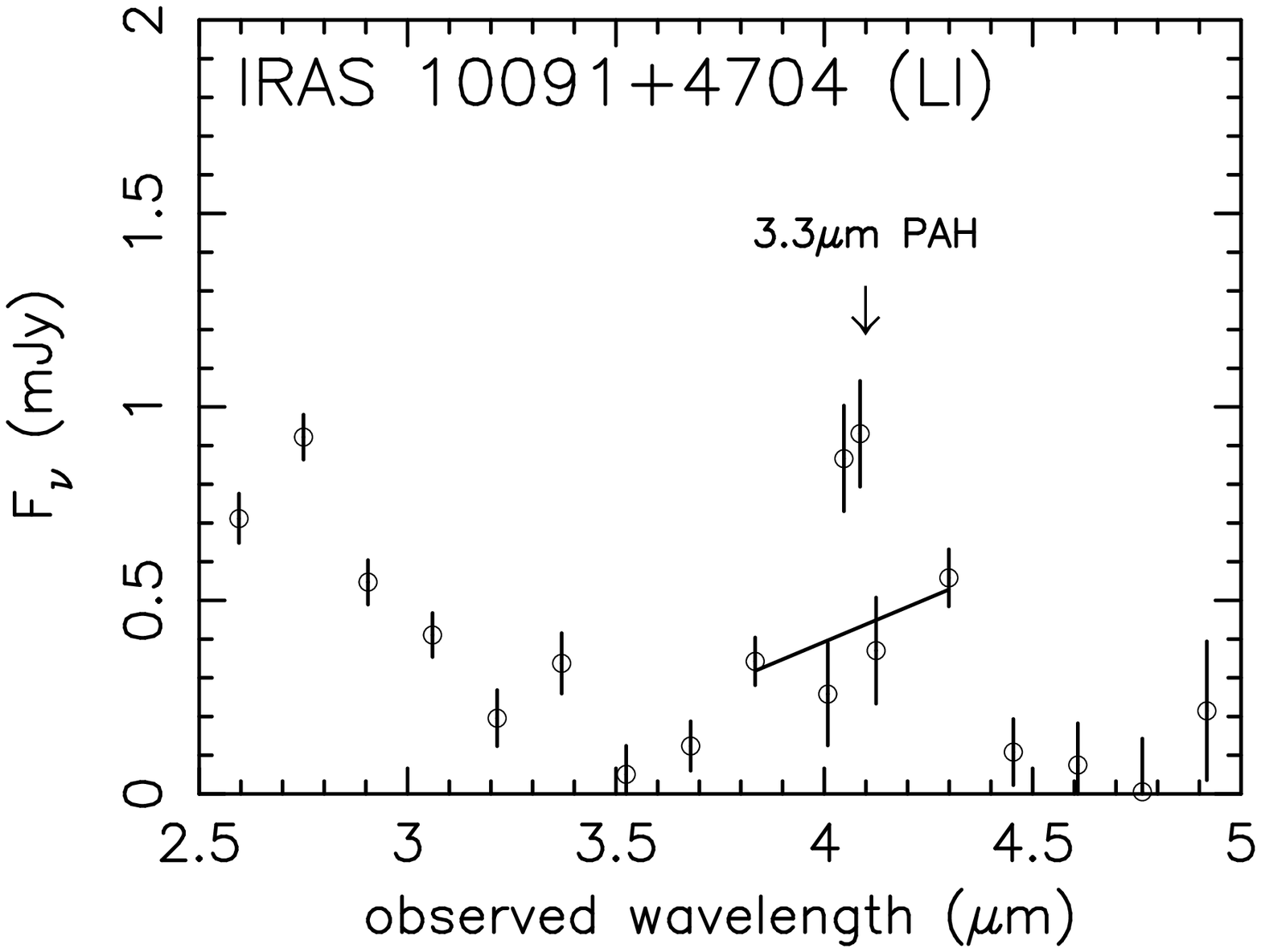}
\FigureFile(80mm,80mm){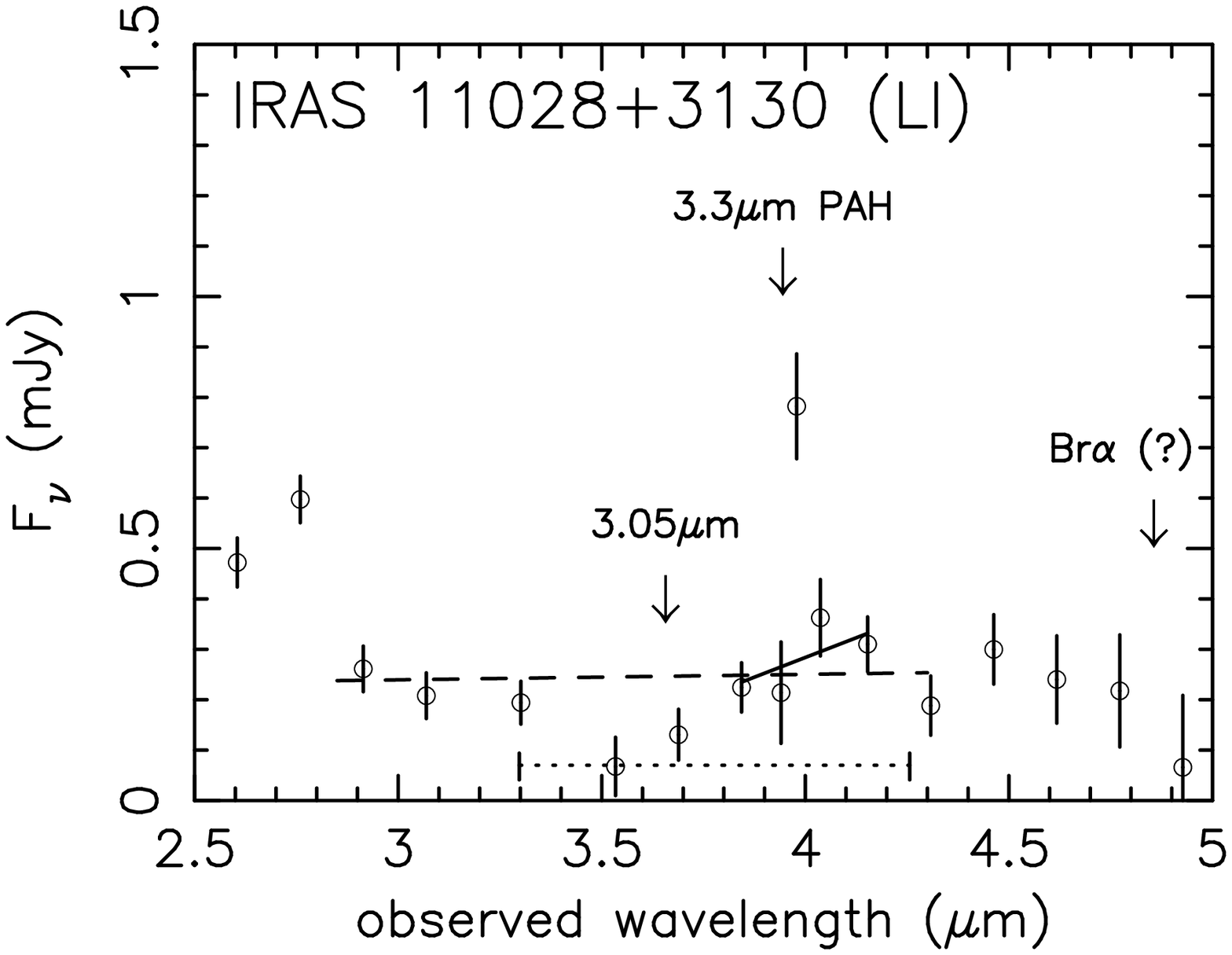} 
\end{figure}

\clearpage

\begin{figure}
\FigureFile(80mm,80mm){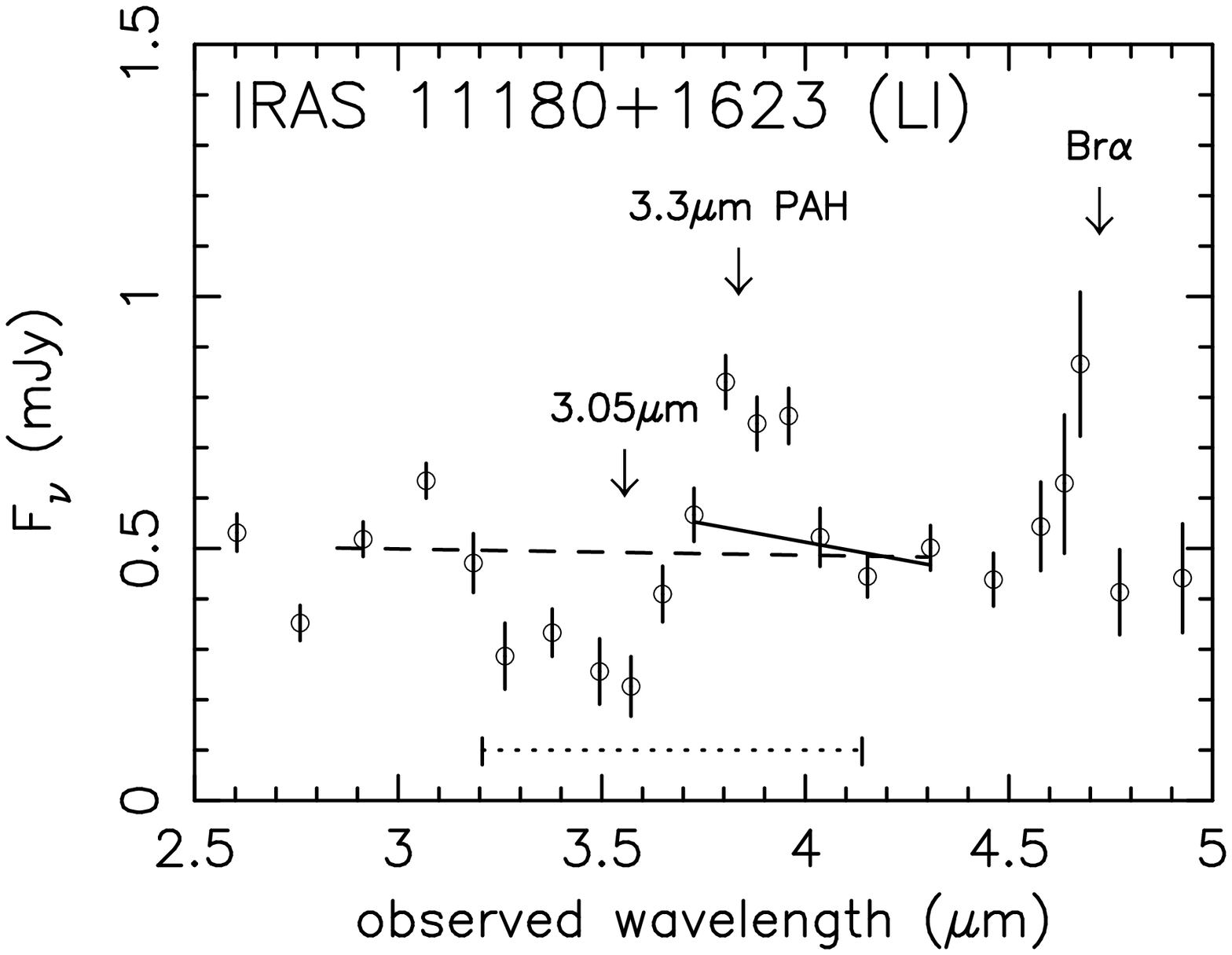}
\FigureFile(80mm,80mm){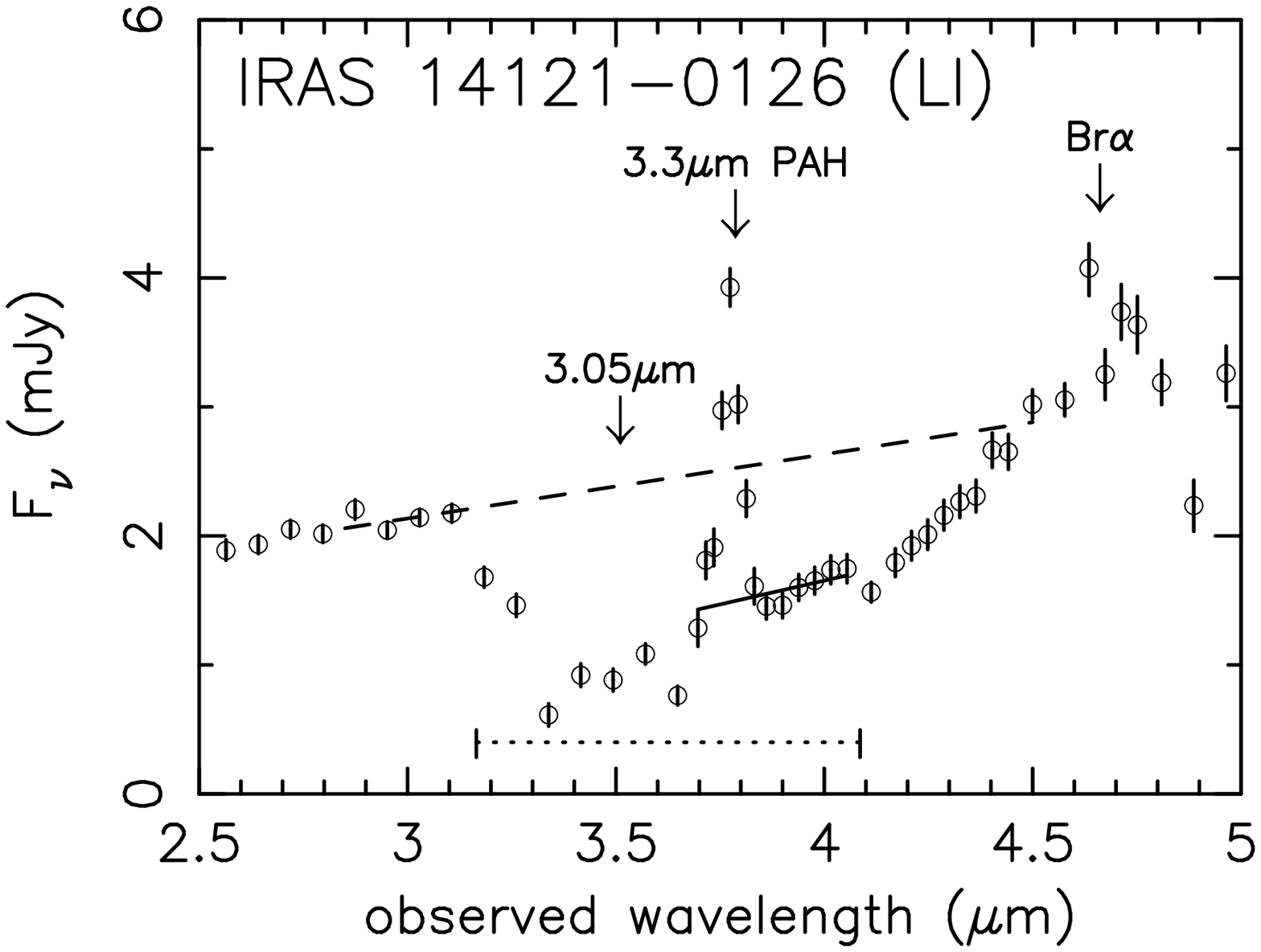}
\FigureFile(80mm,80mm){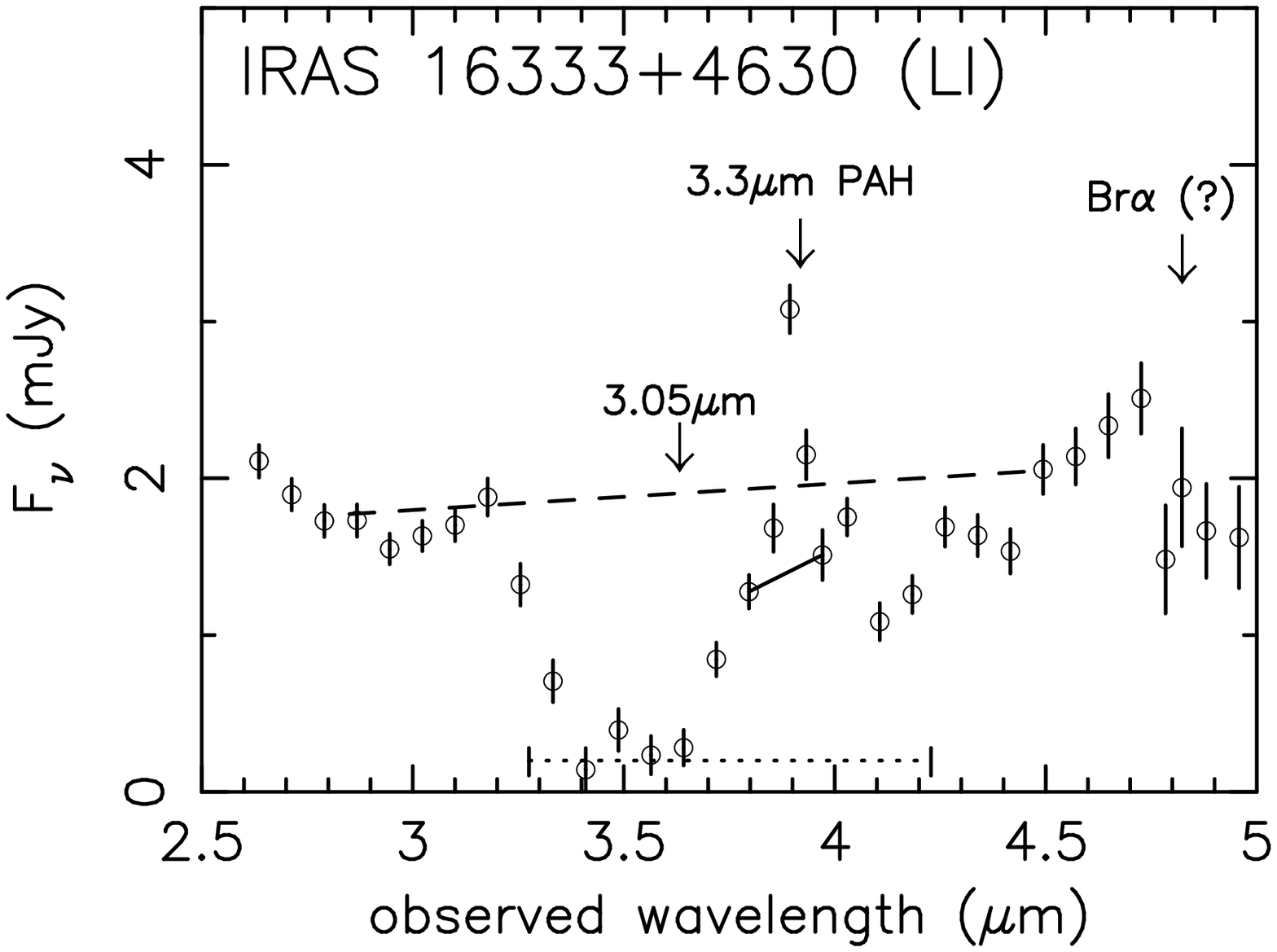}
\FigureFile(80mm,80mm){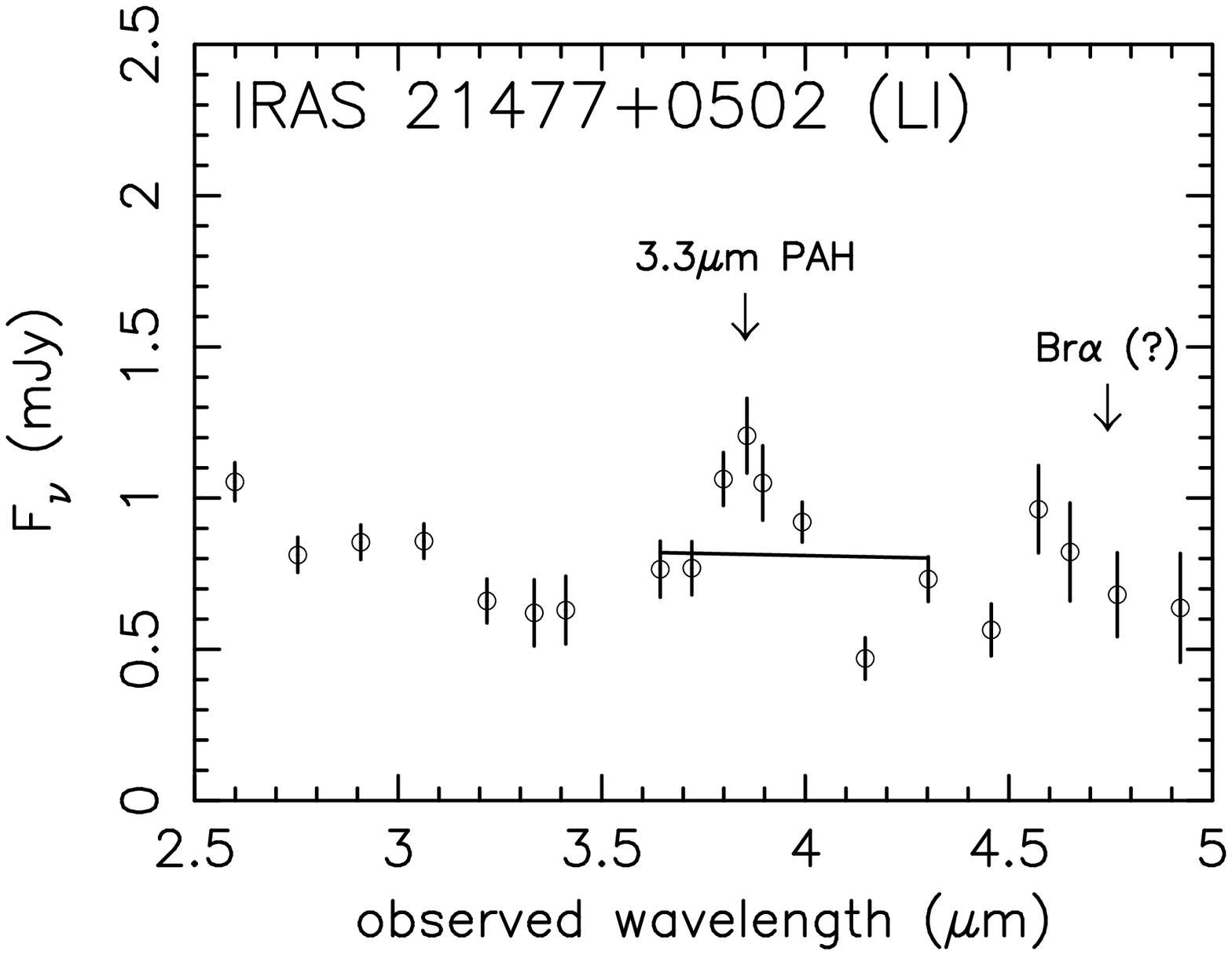}
\FigureFile(80mm,80mm){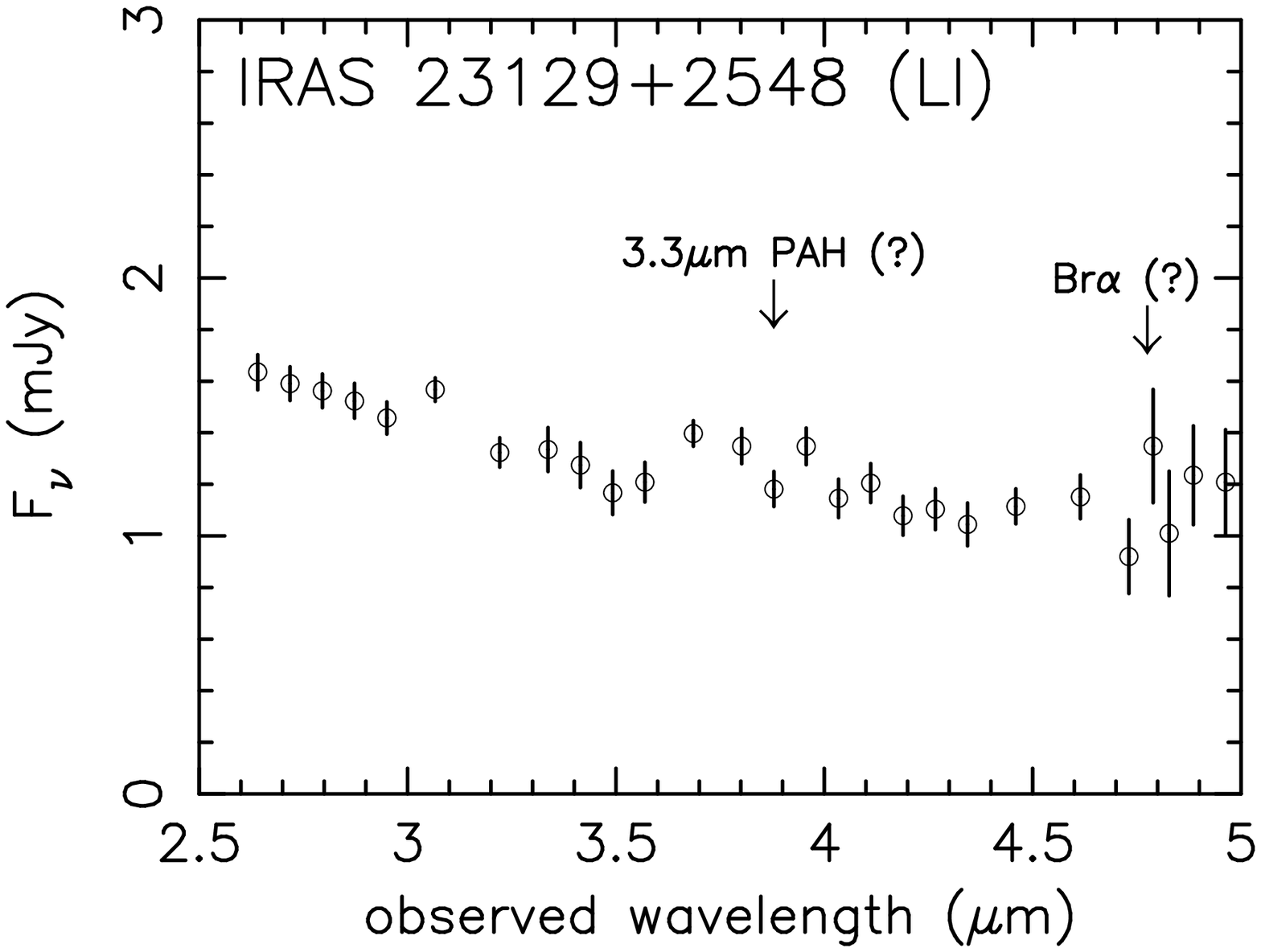}
\end{figure}

\clearpage

\begin{figure}
\FigureFile(80mm,80mm){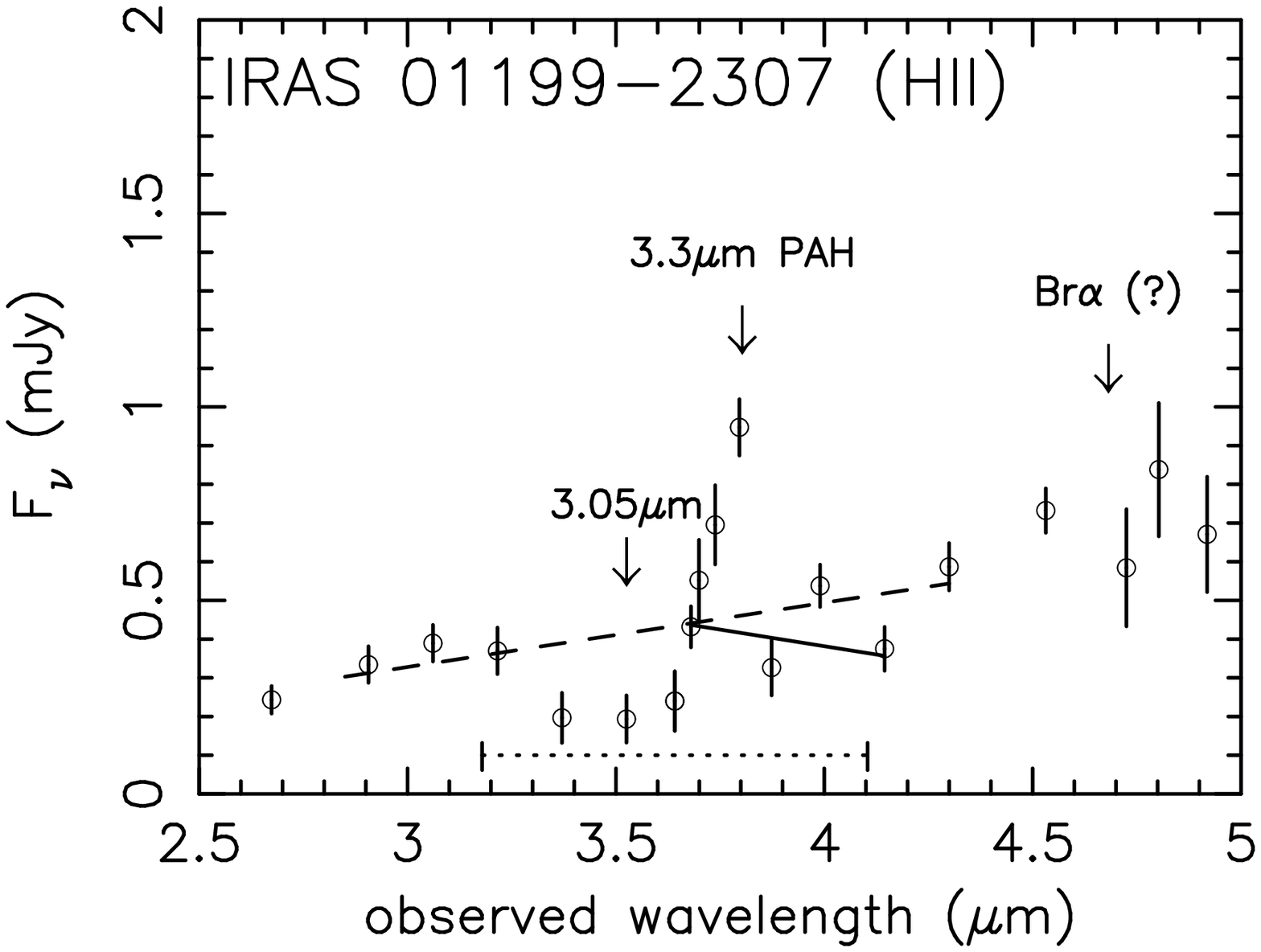}
\FigureFile(80mm,80mm){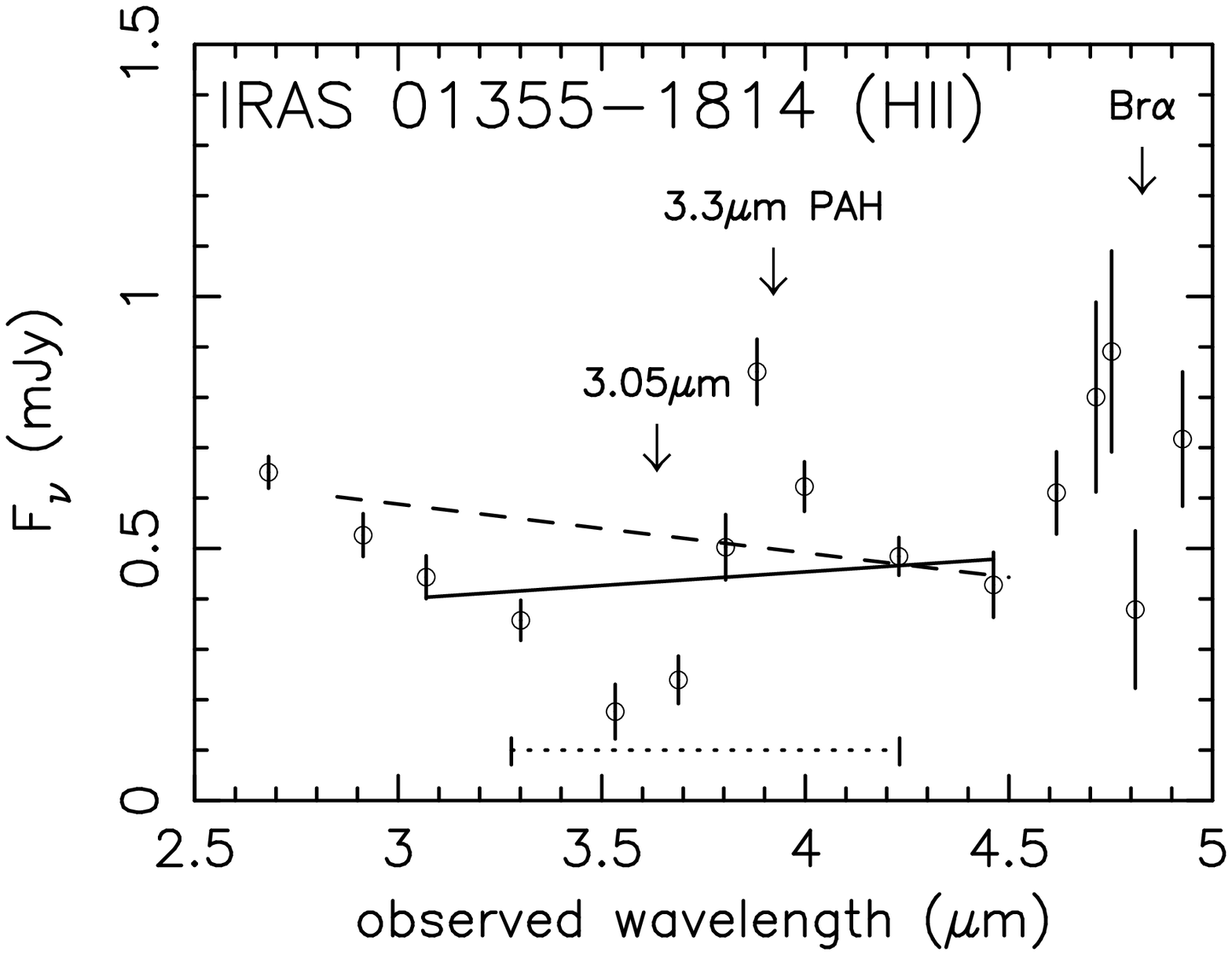}
\FigureFile(80mm,80mm){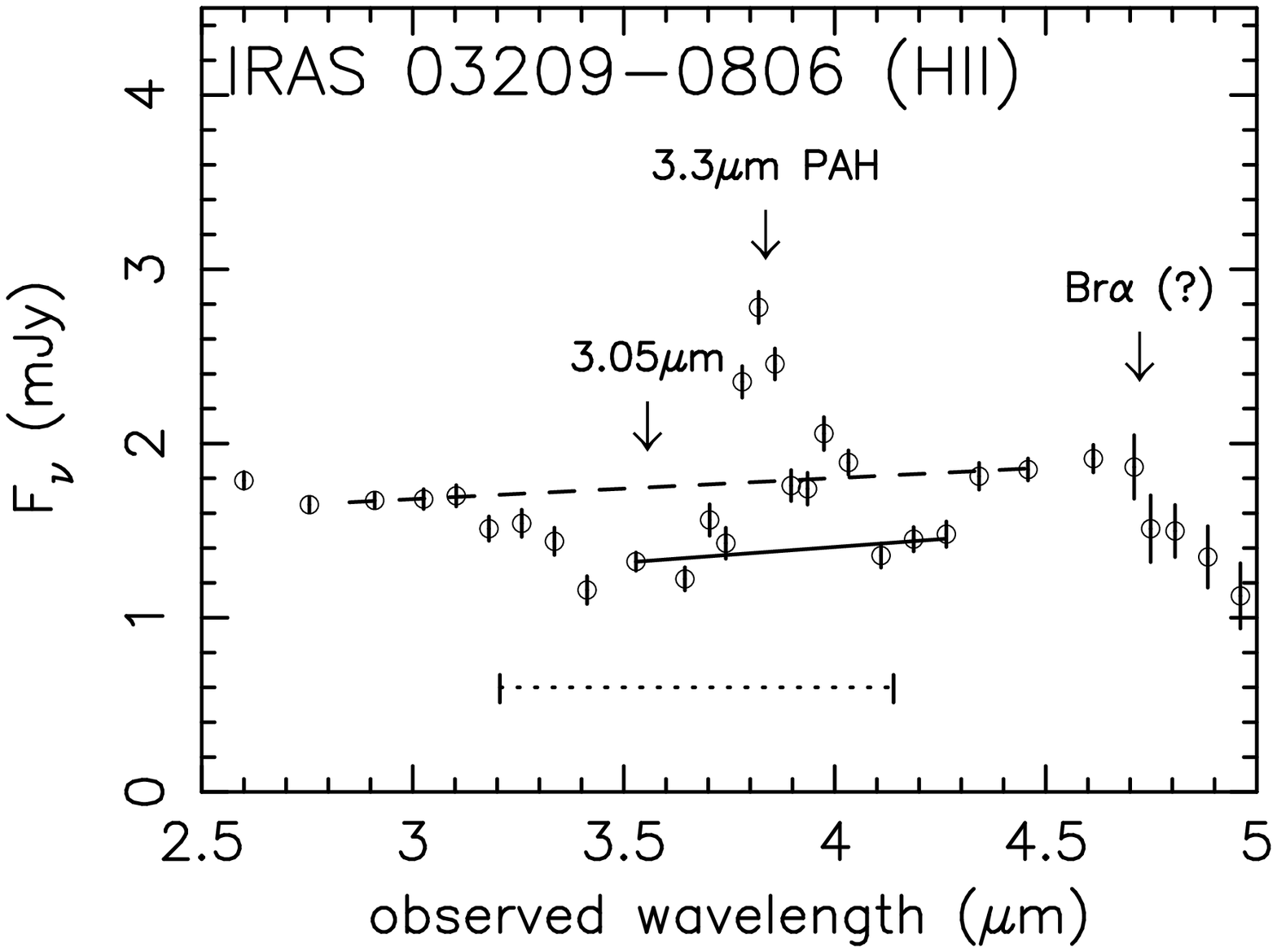}
\FigureFile(80mm,80mm){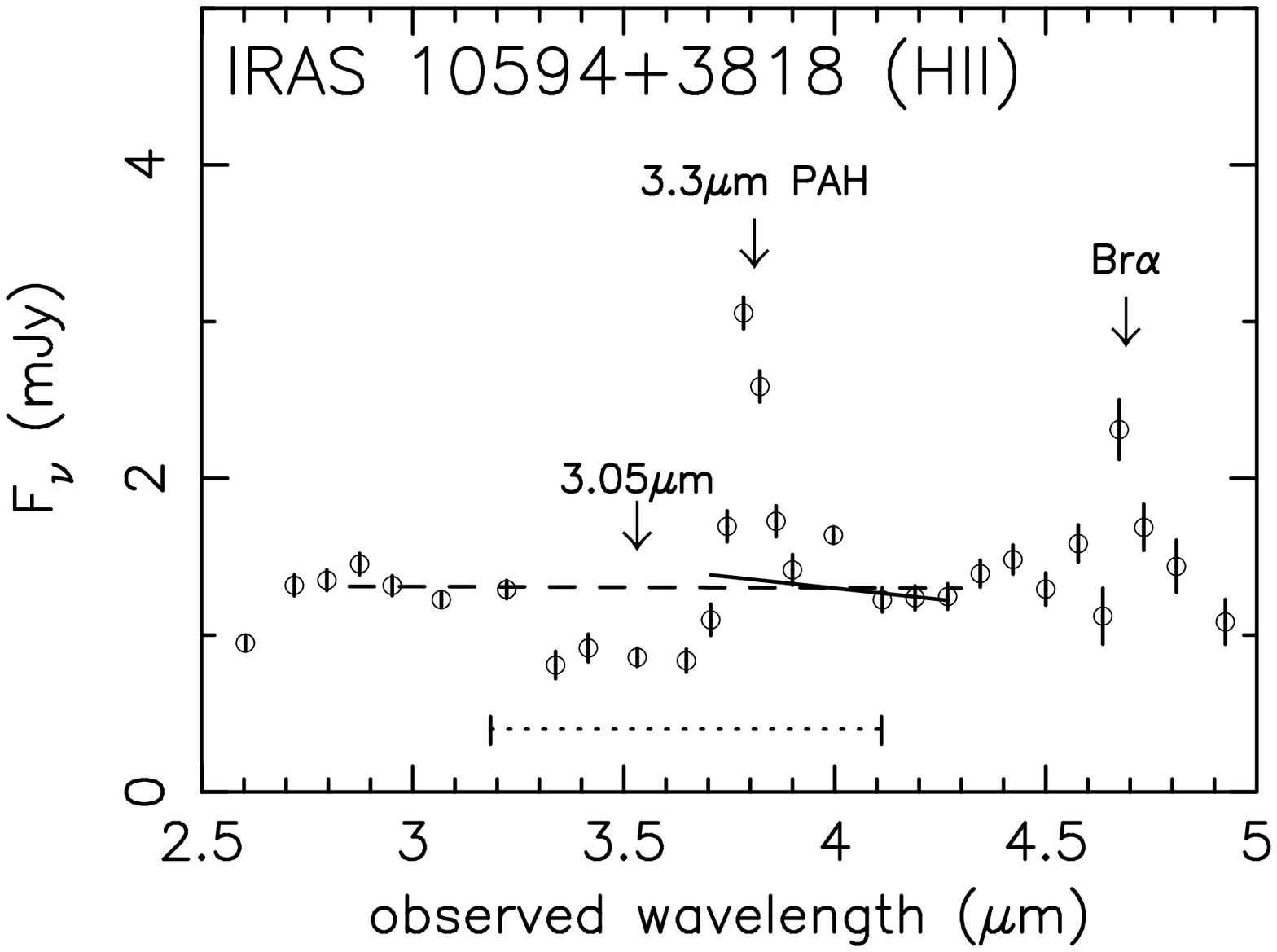}
\FigureFile(80mm,80mm){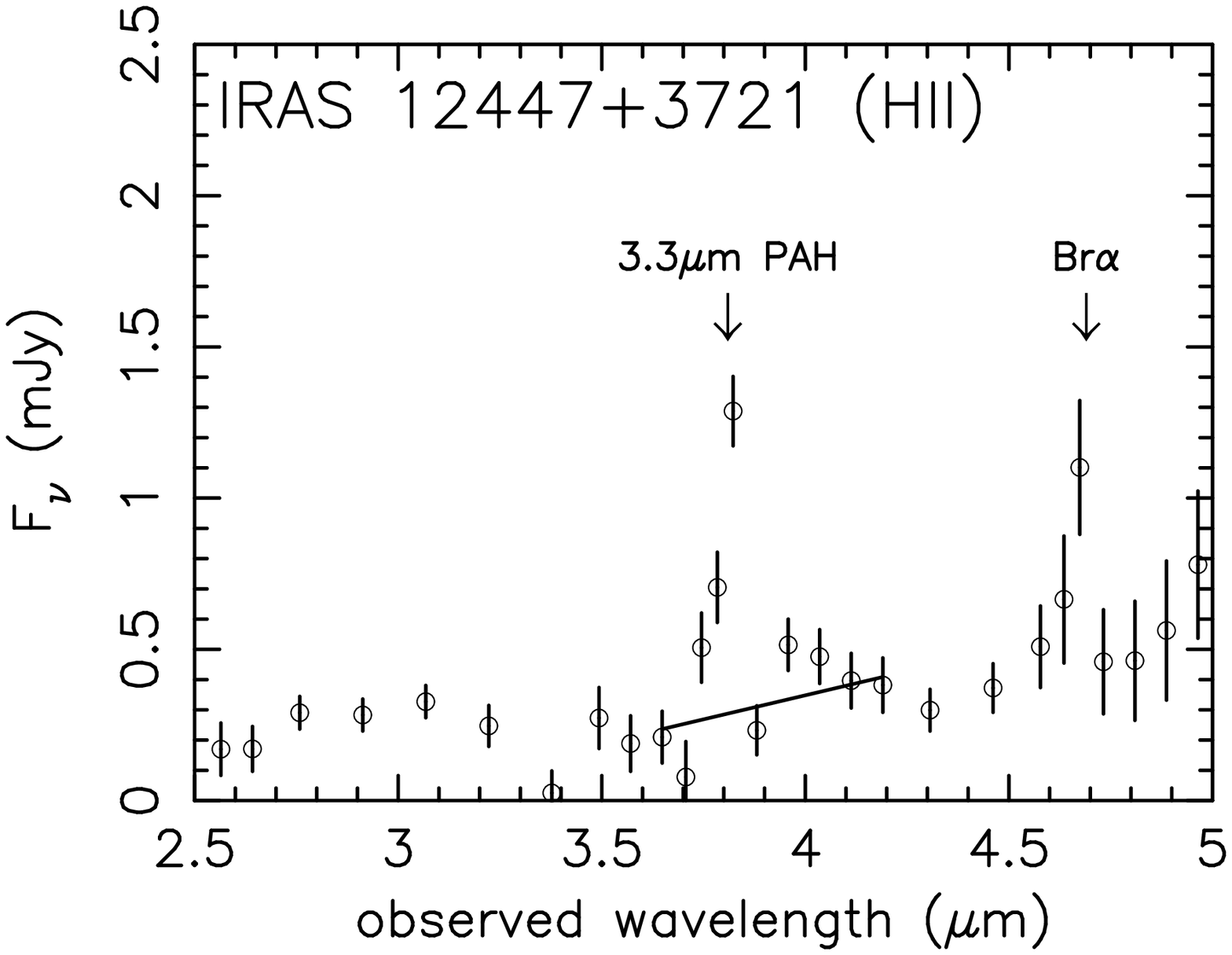}
\FigureFile(80mm,80mm){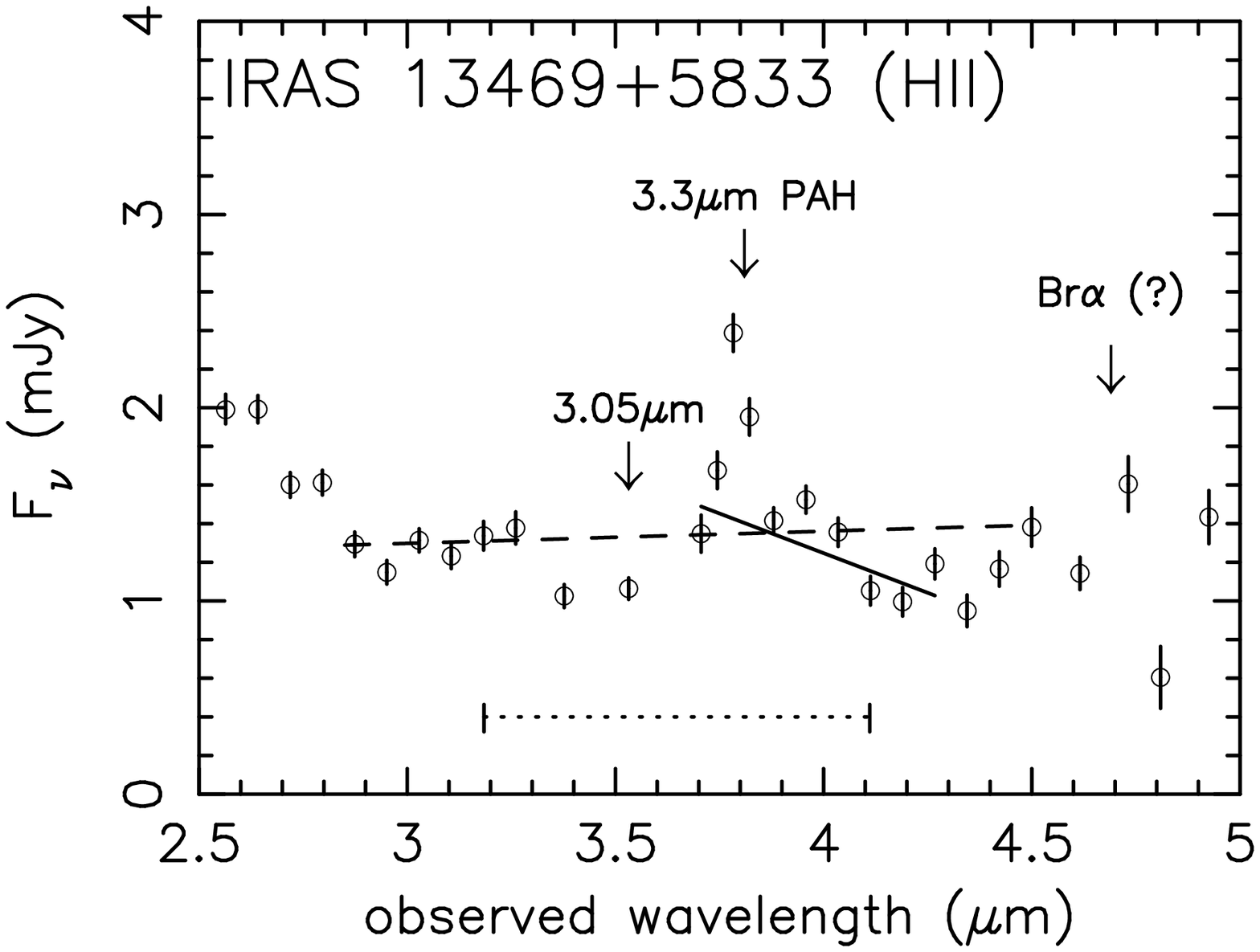}
\end{figure}

\clearpage

\begin{figure}
\FigureFile(80mm,80mm){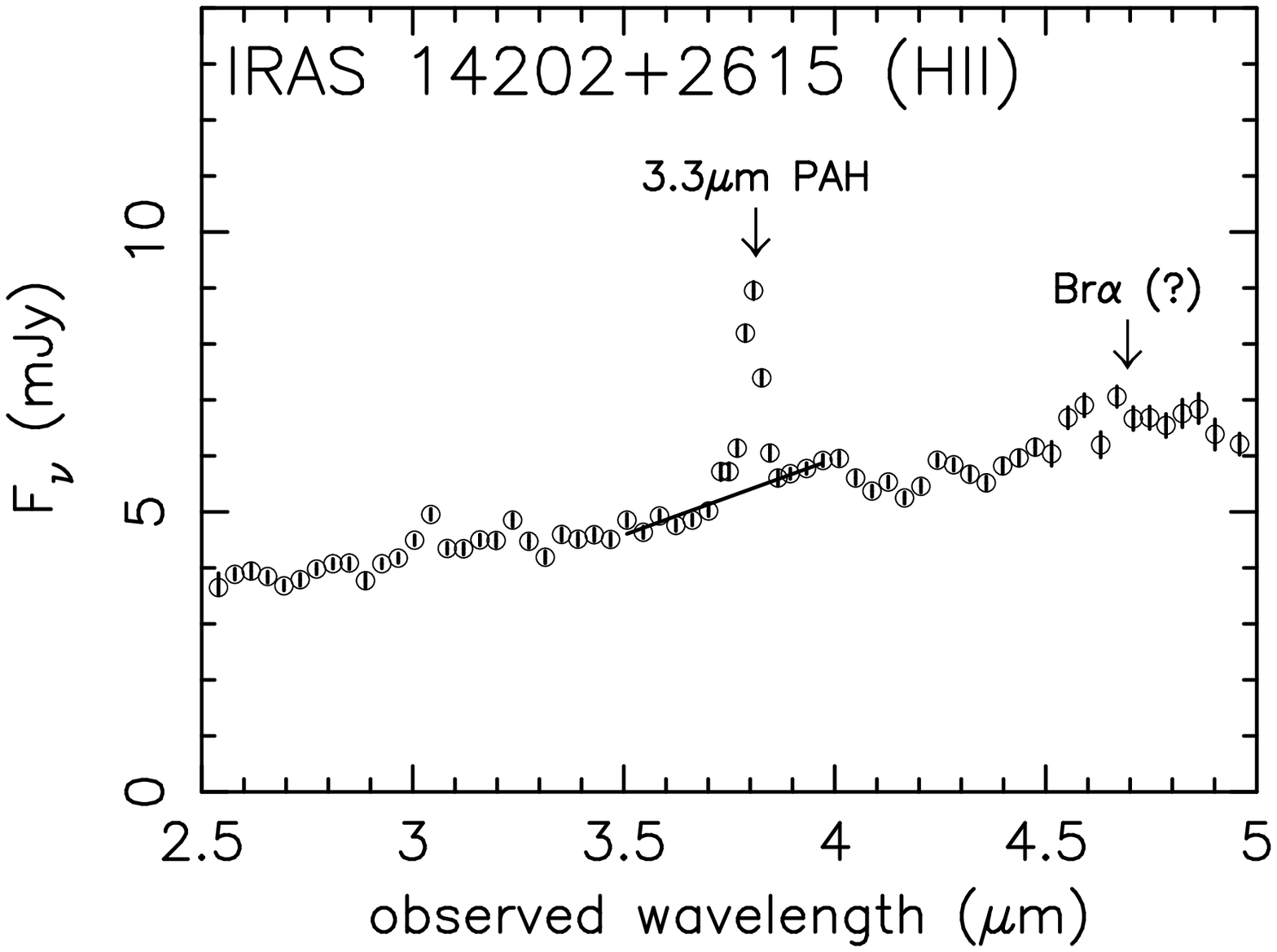}
\FigureFile(80mm,80mm){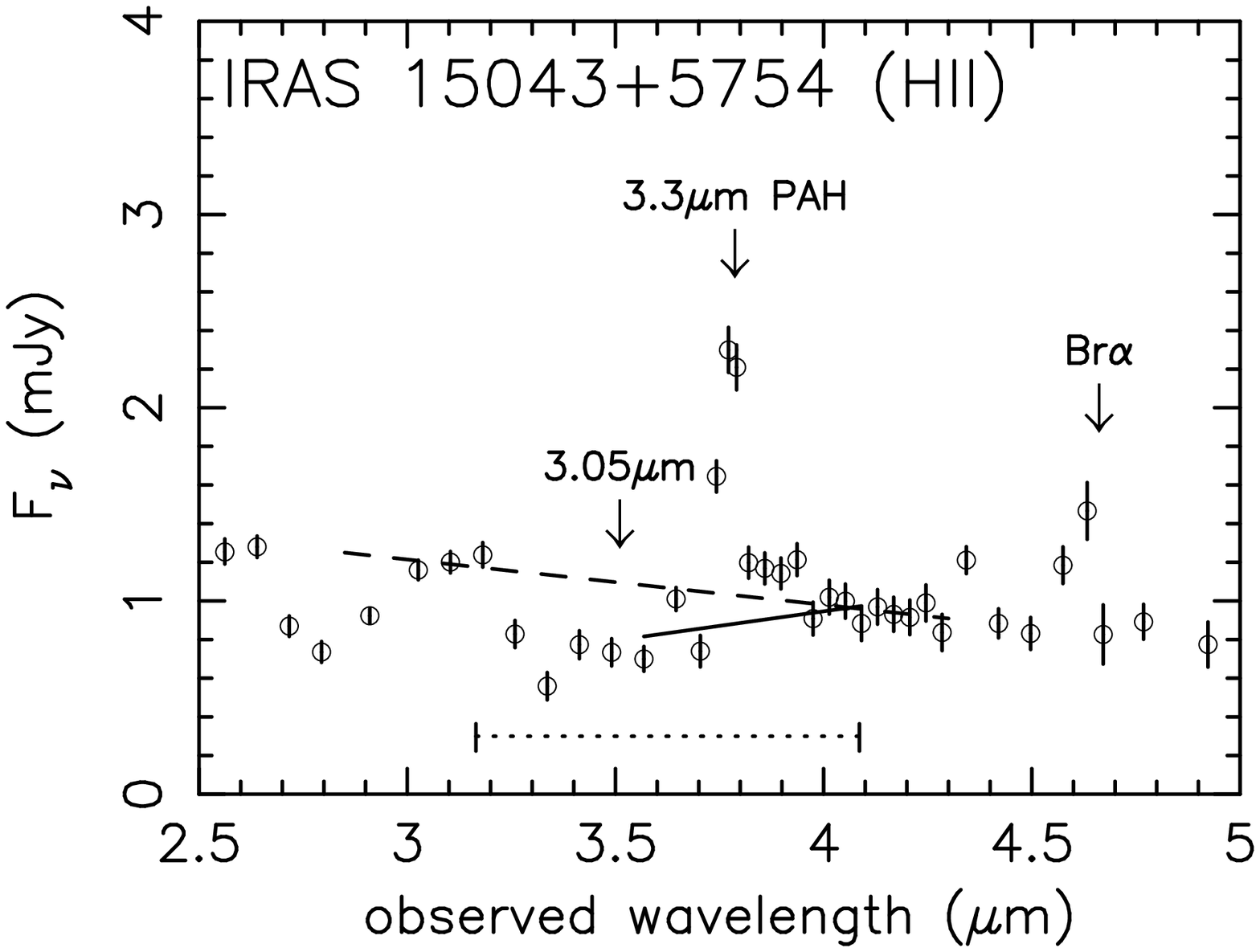}
\FigureFile(80mm,80mm){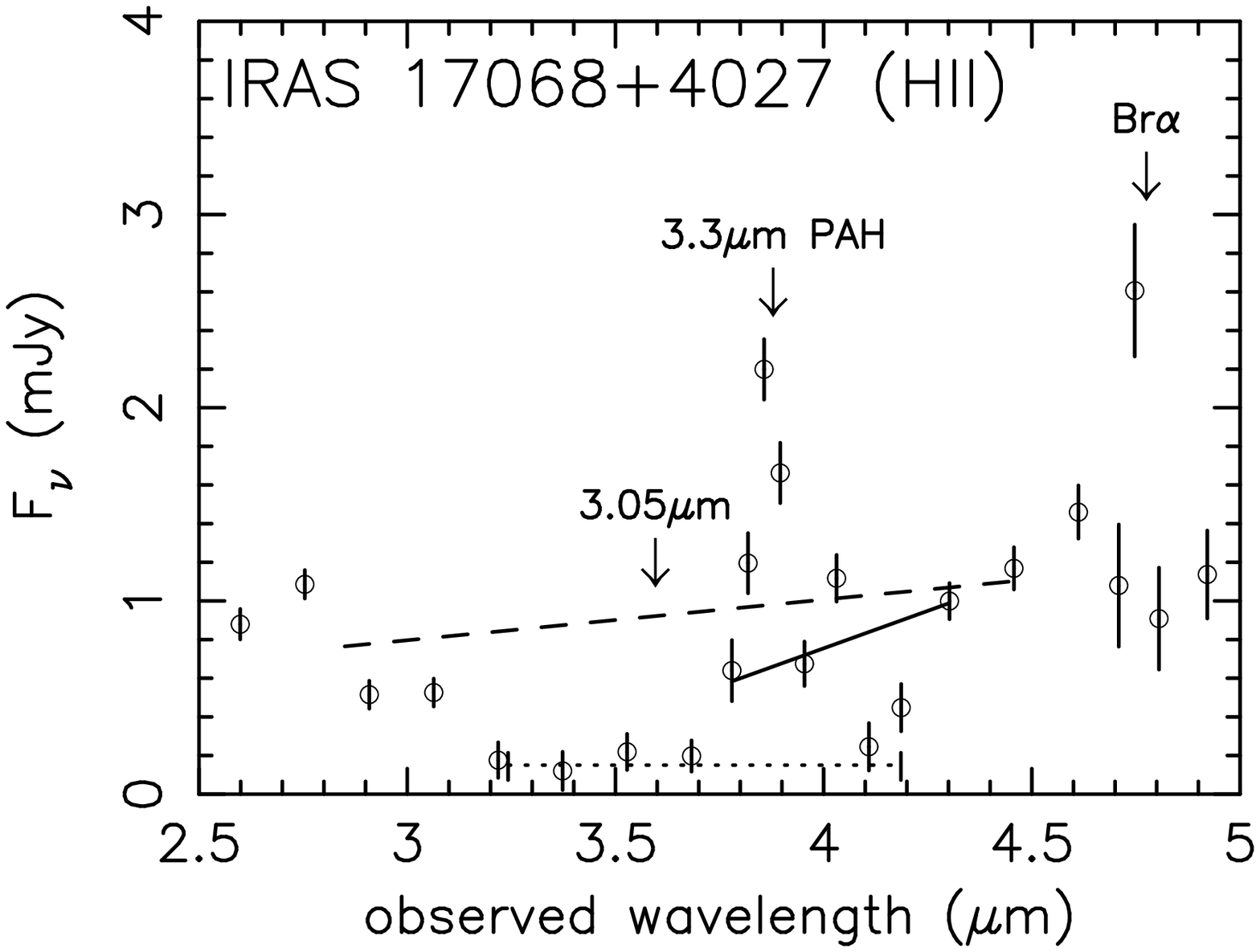}
\FigureFile(80mm,80mm){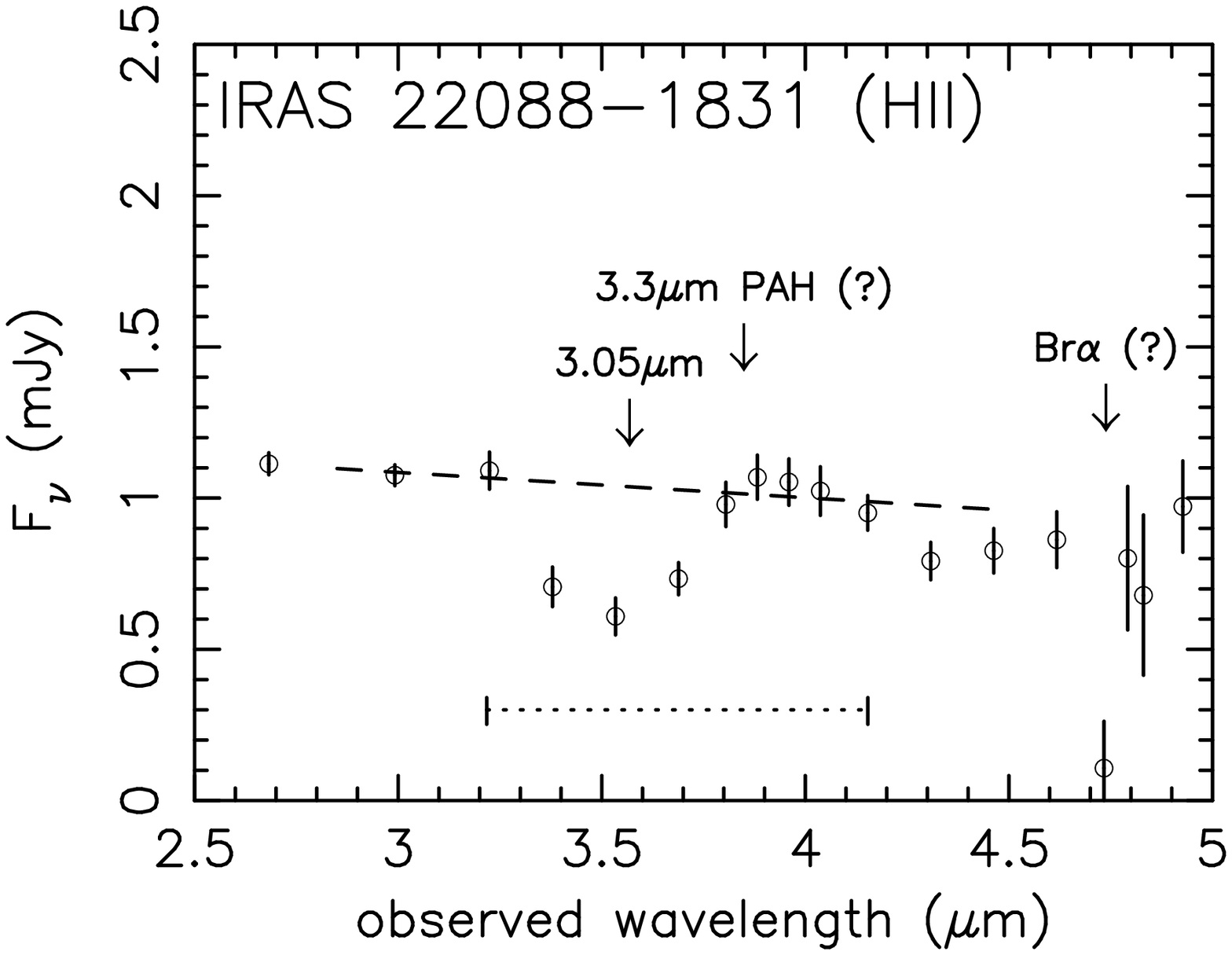}
\end{figure}

\clearpage

\begin{figure}
\FigureFile(80mm,80mm){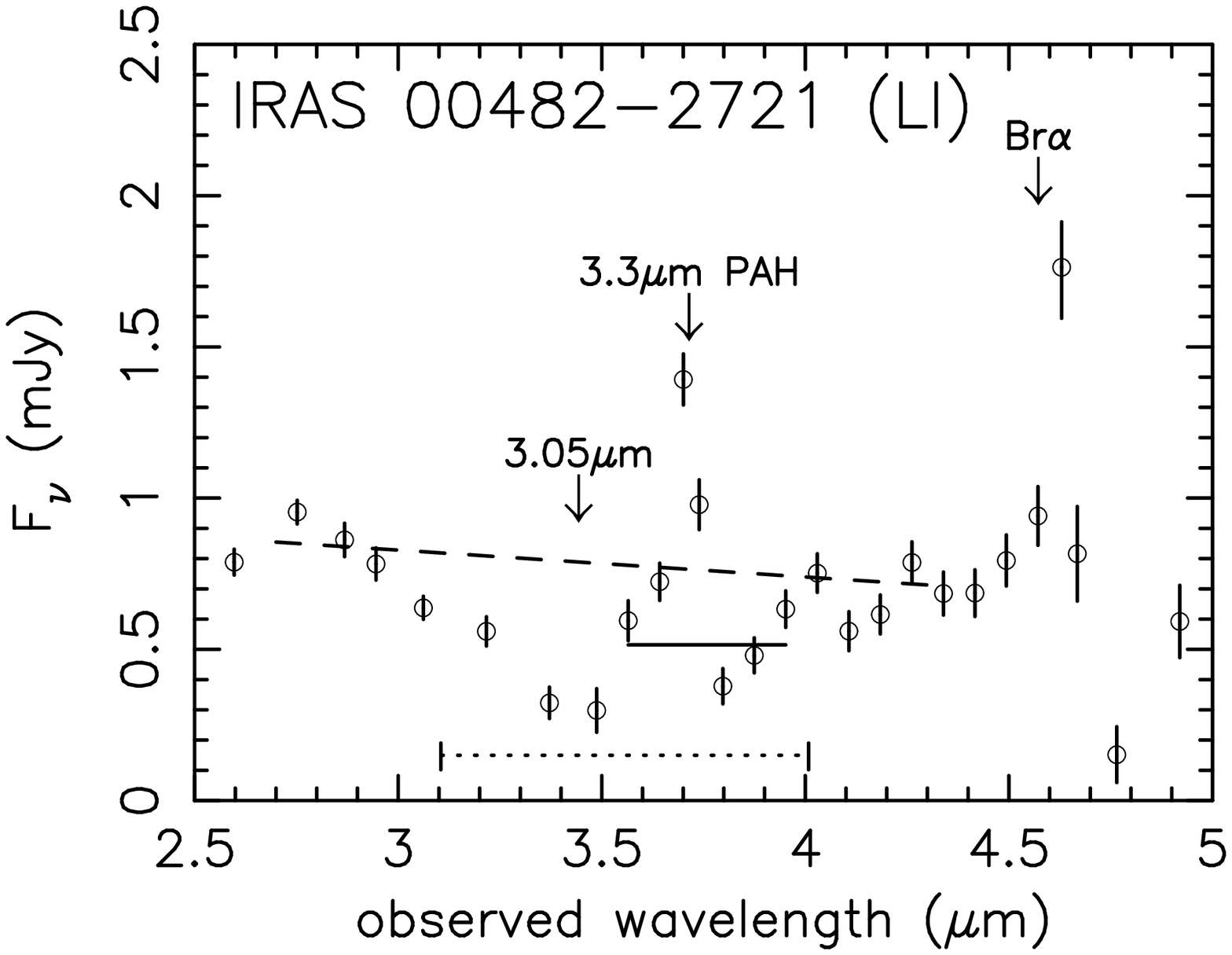}
\FigureFile(80mm,80mm){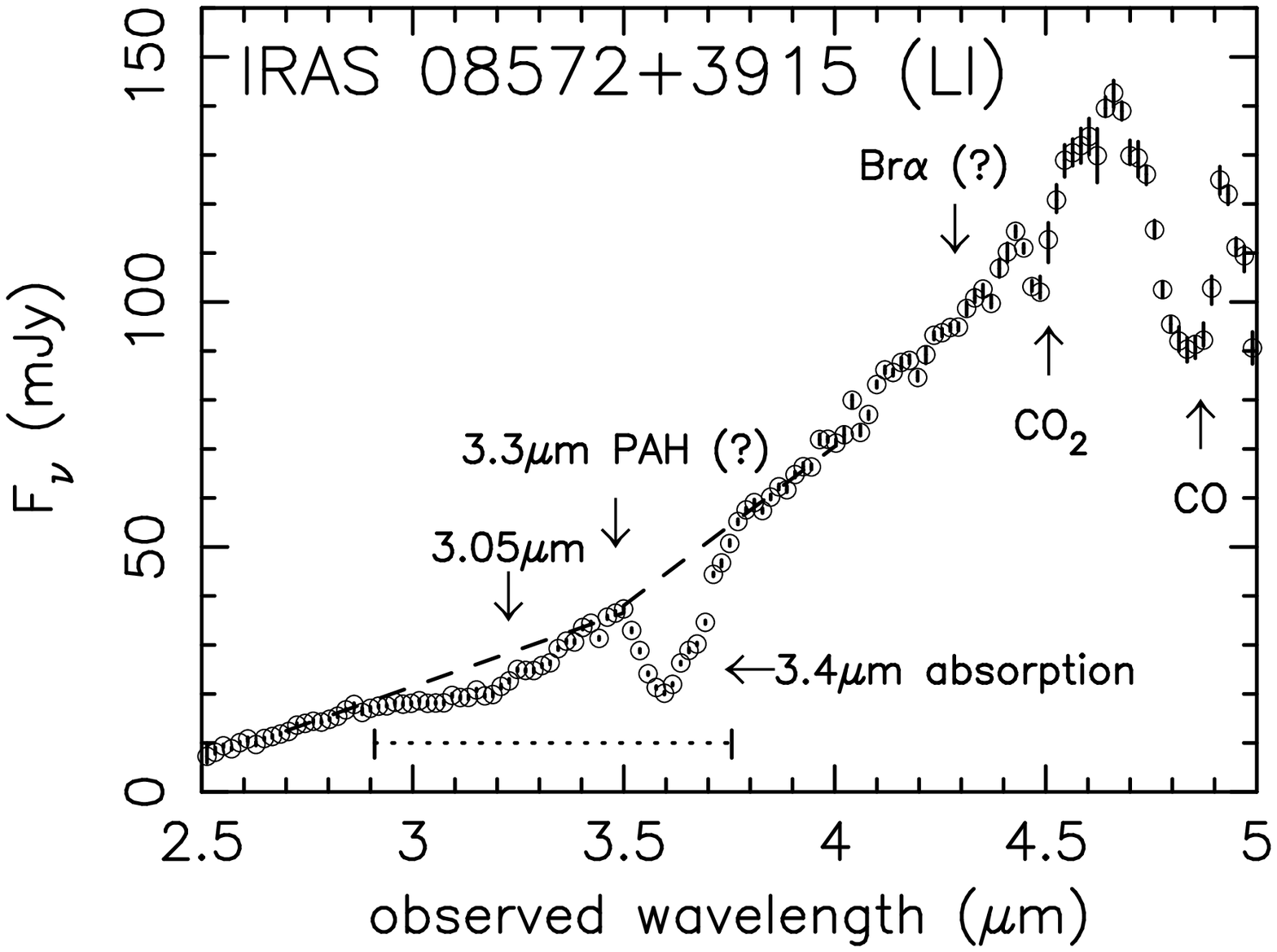}
\FigureFile(80mm,80mm){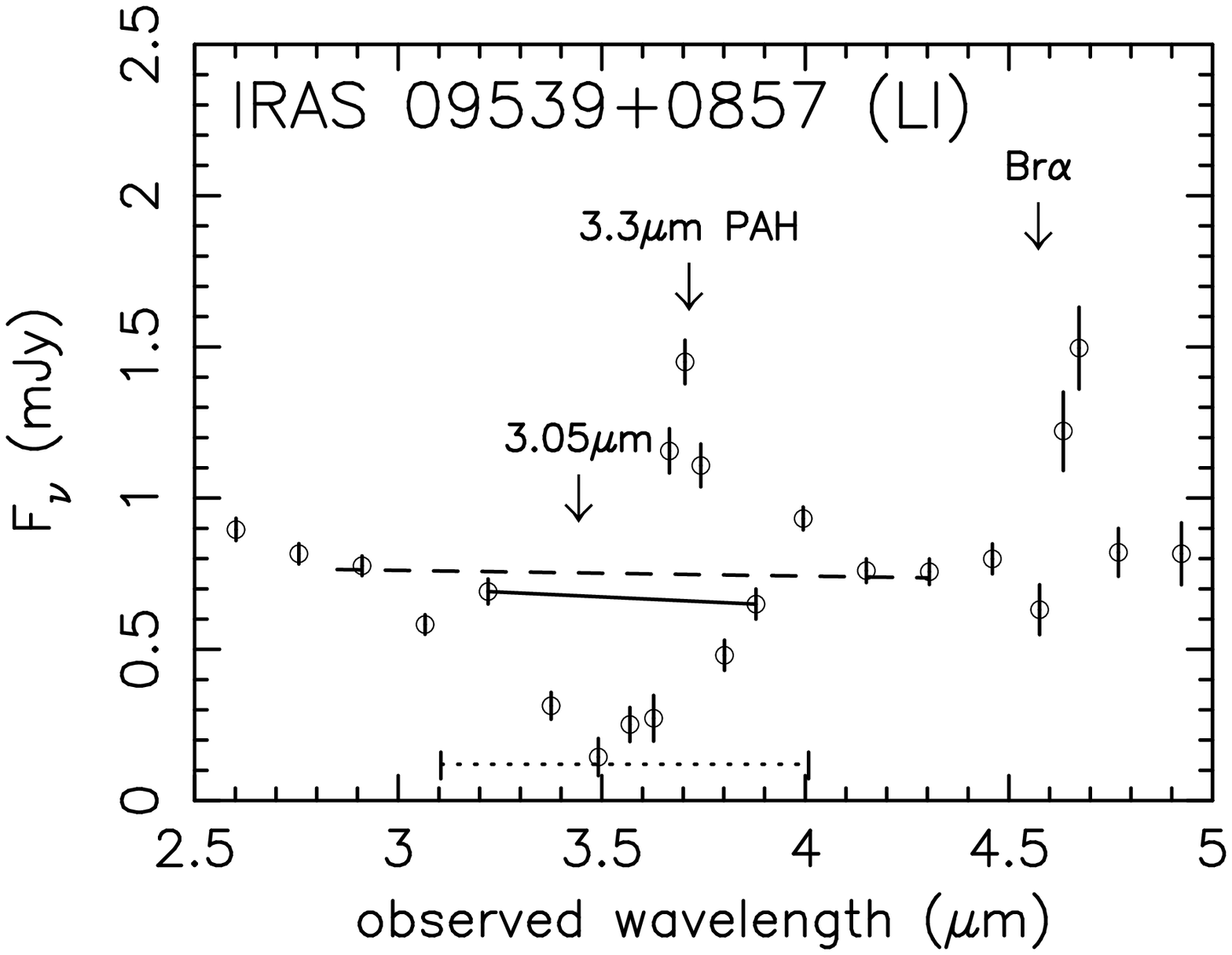}
\FigureFile(80mm,80mm){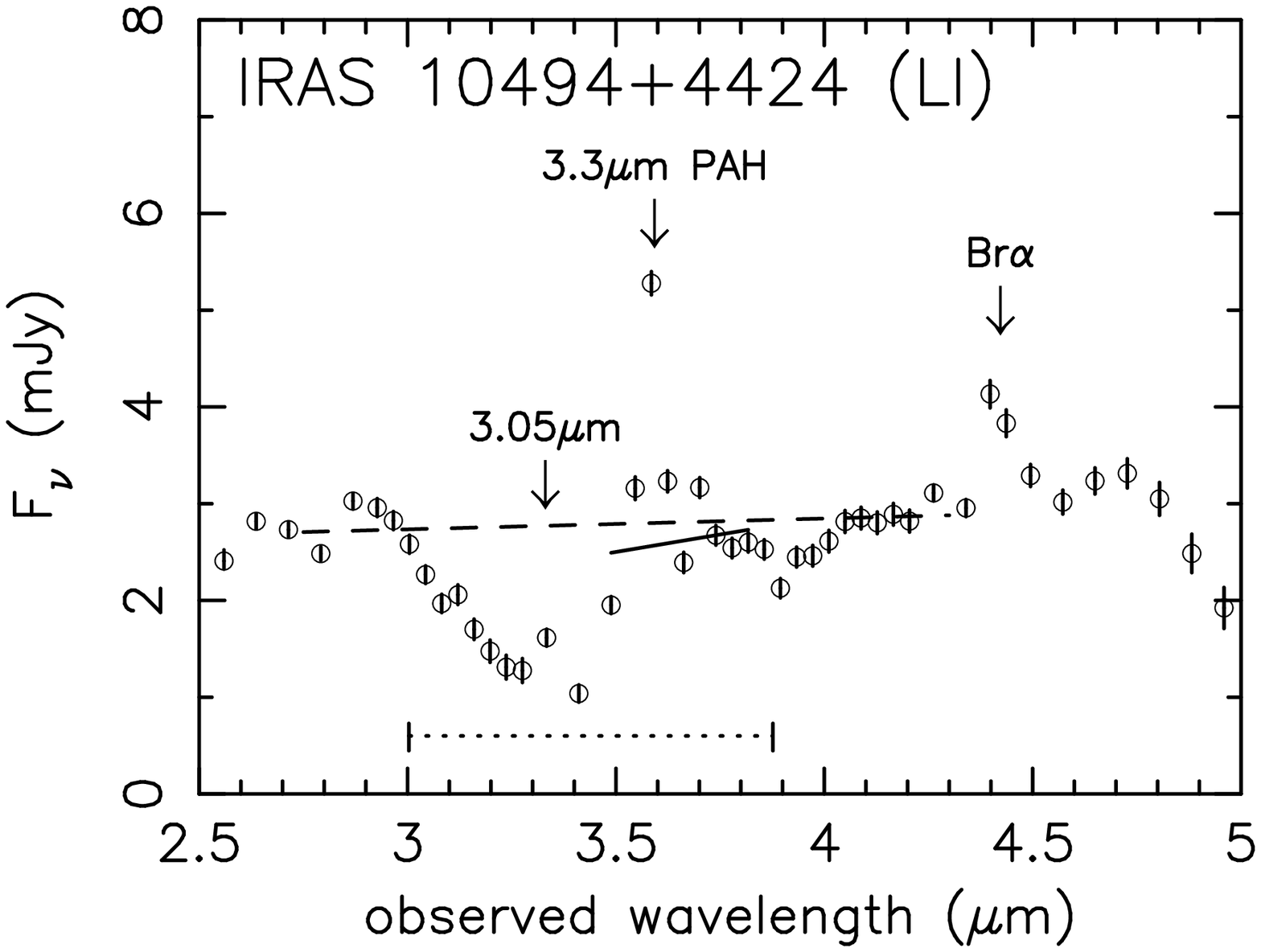}
\FigureFile(80mm,80mm){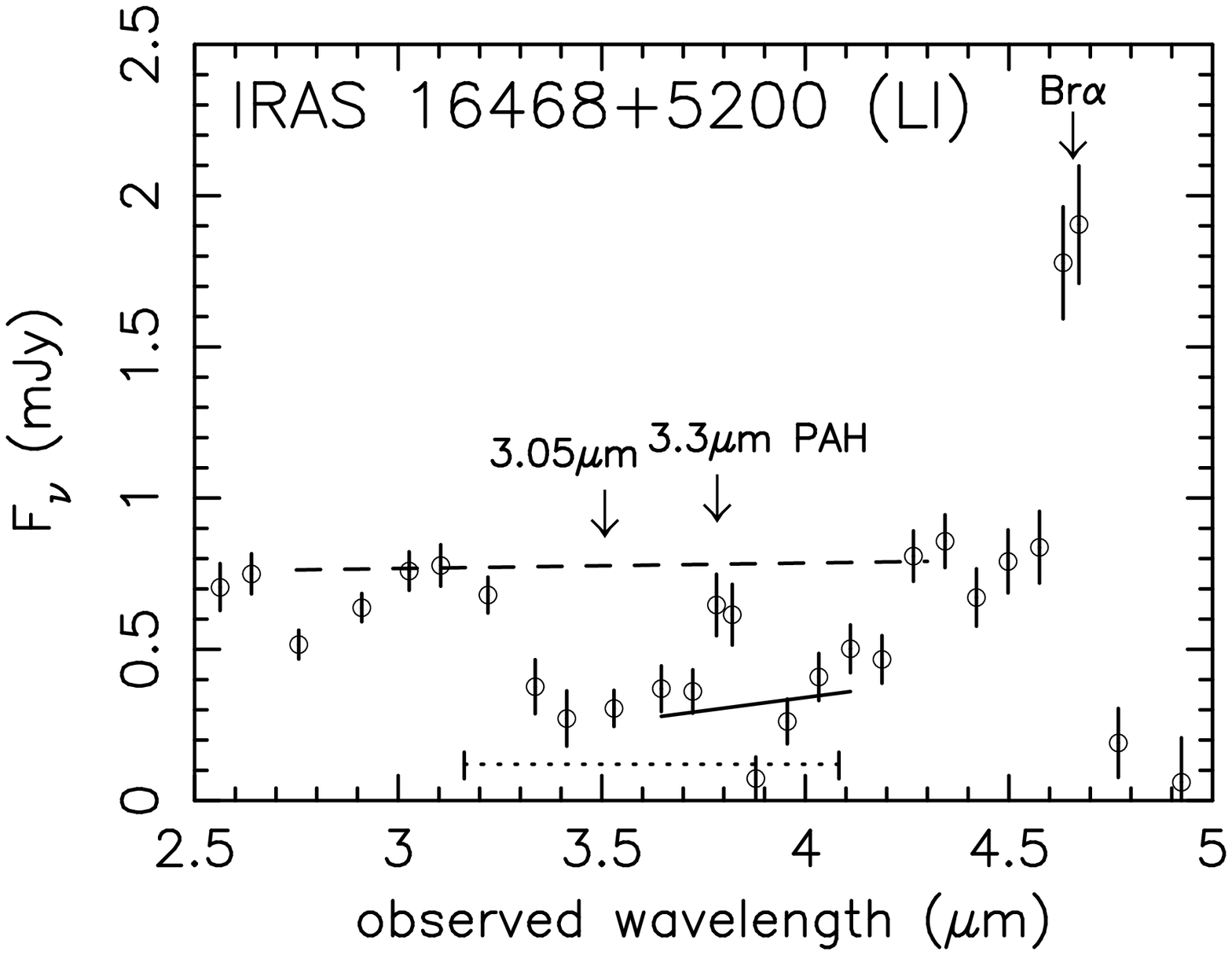}
\FigureFile(80mm,80mm){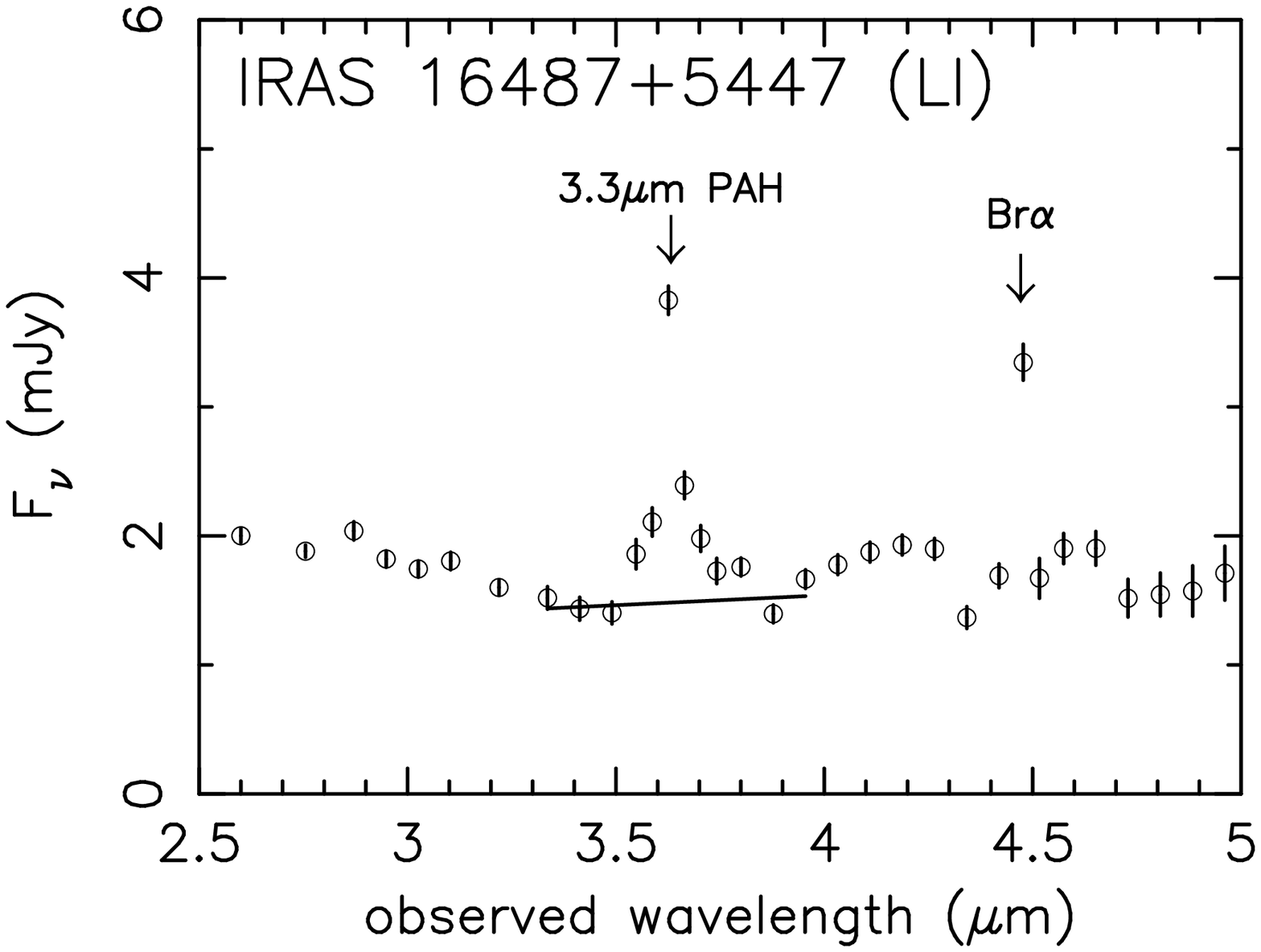}
\end{figure}

\clearpage

\begin{figure}
\FigureFile(80mm,80mm){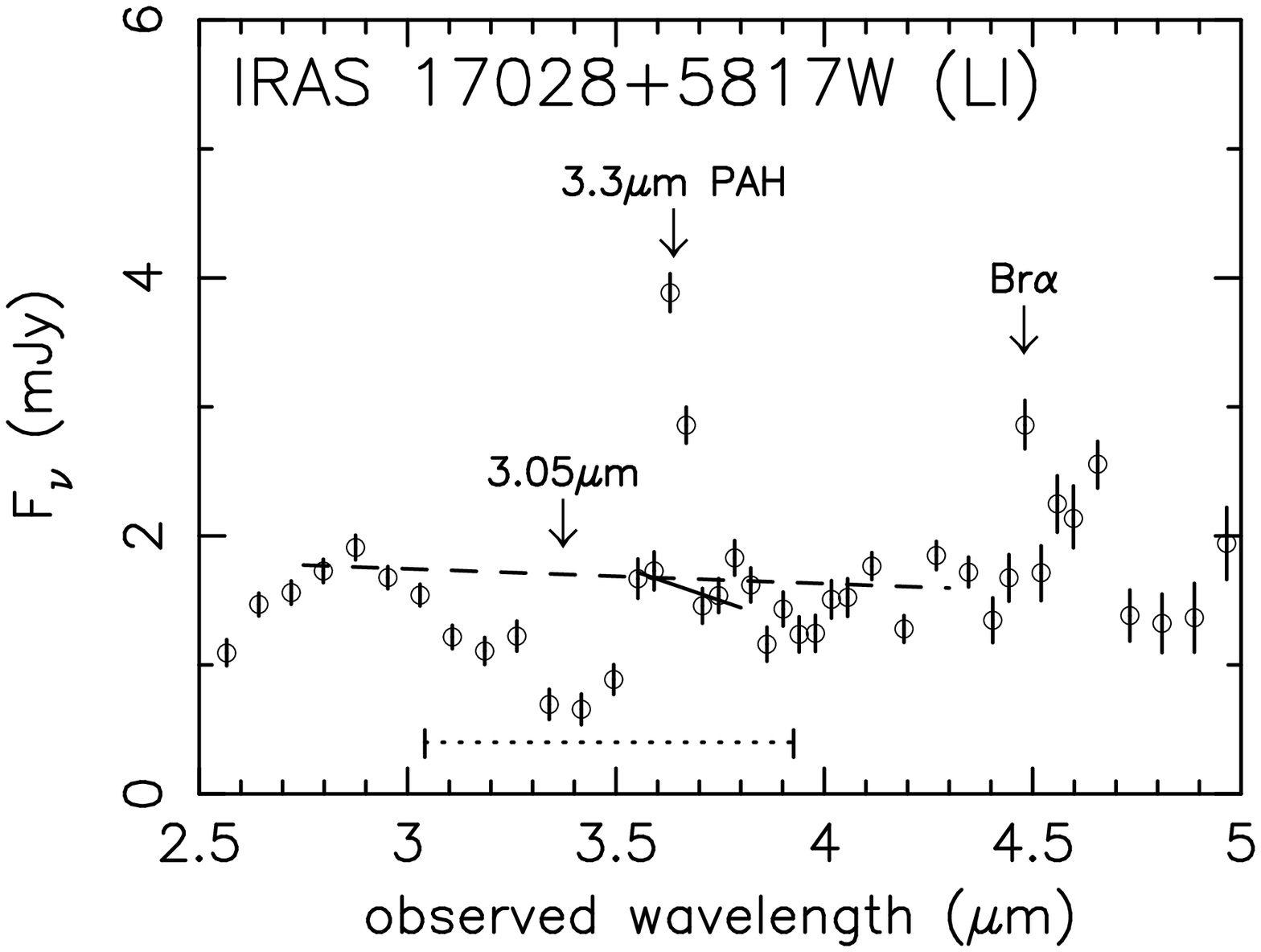}
\FigureFile(80mm,80mm){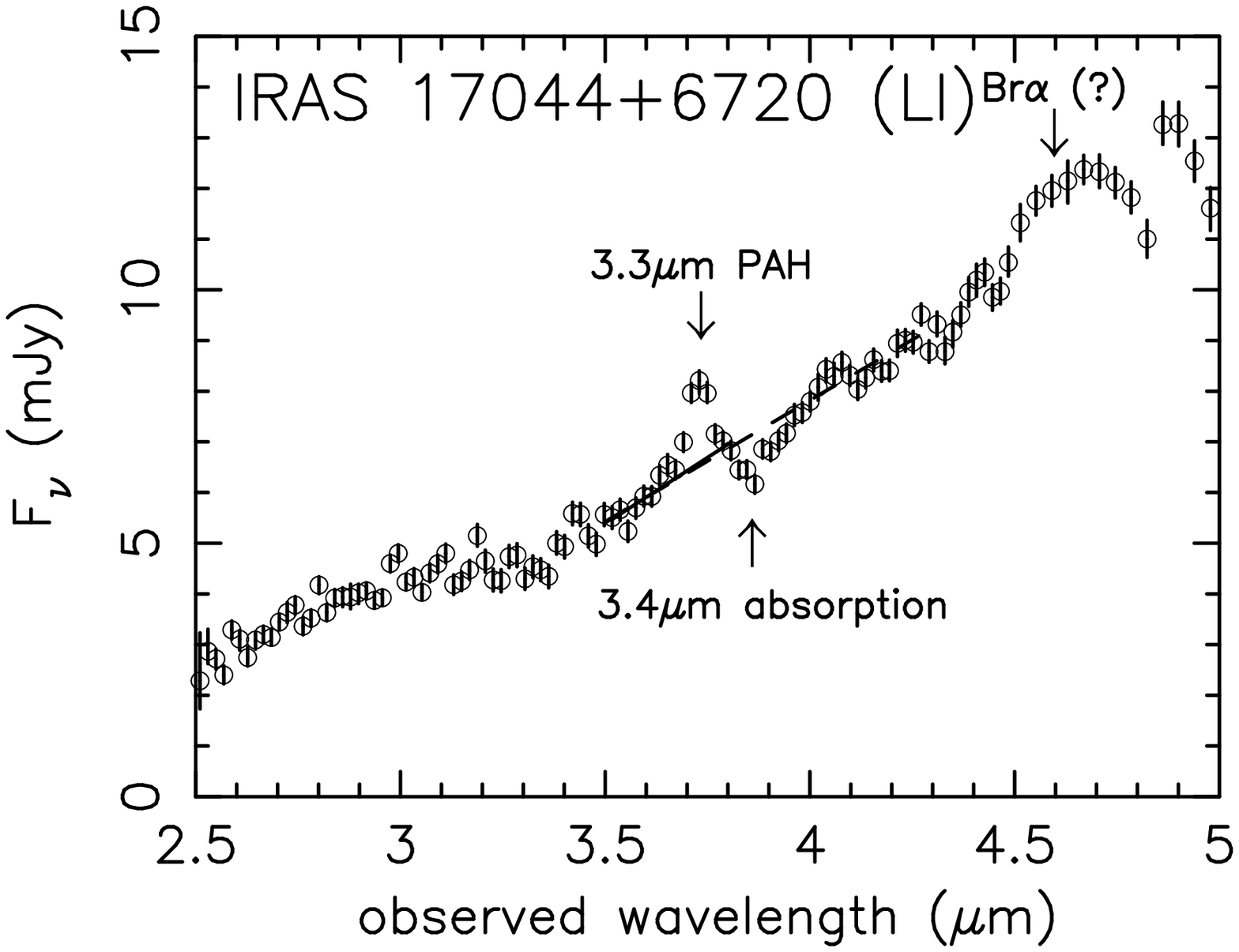}
\end{figure}

\clearpage

\begin{figure}
\FigureFile(80mm,80mm){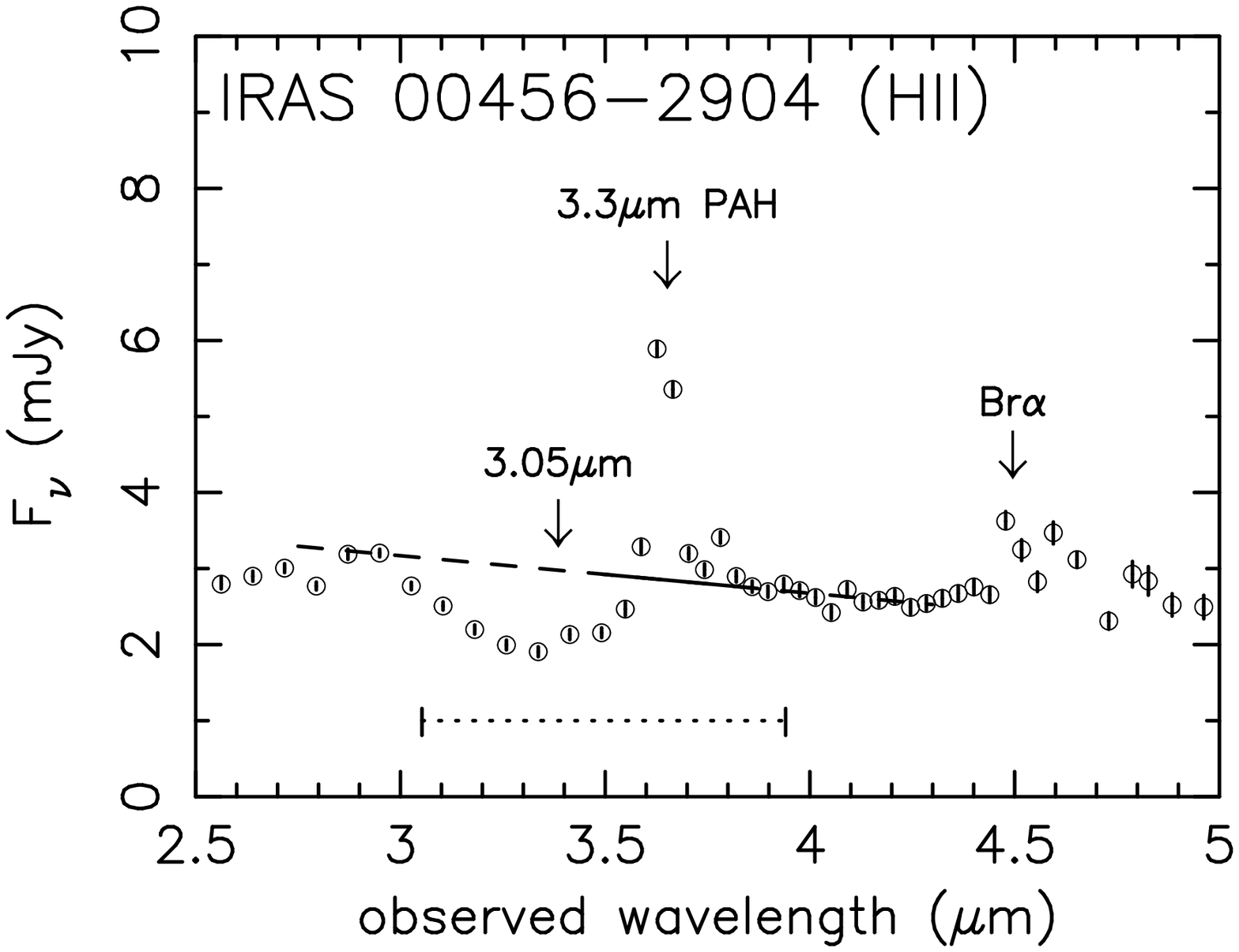}
\FigureFile(80mm,80mm){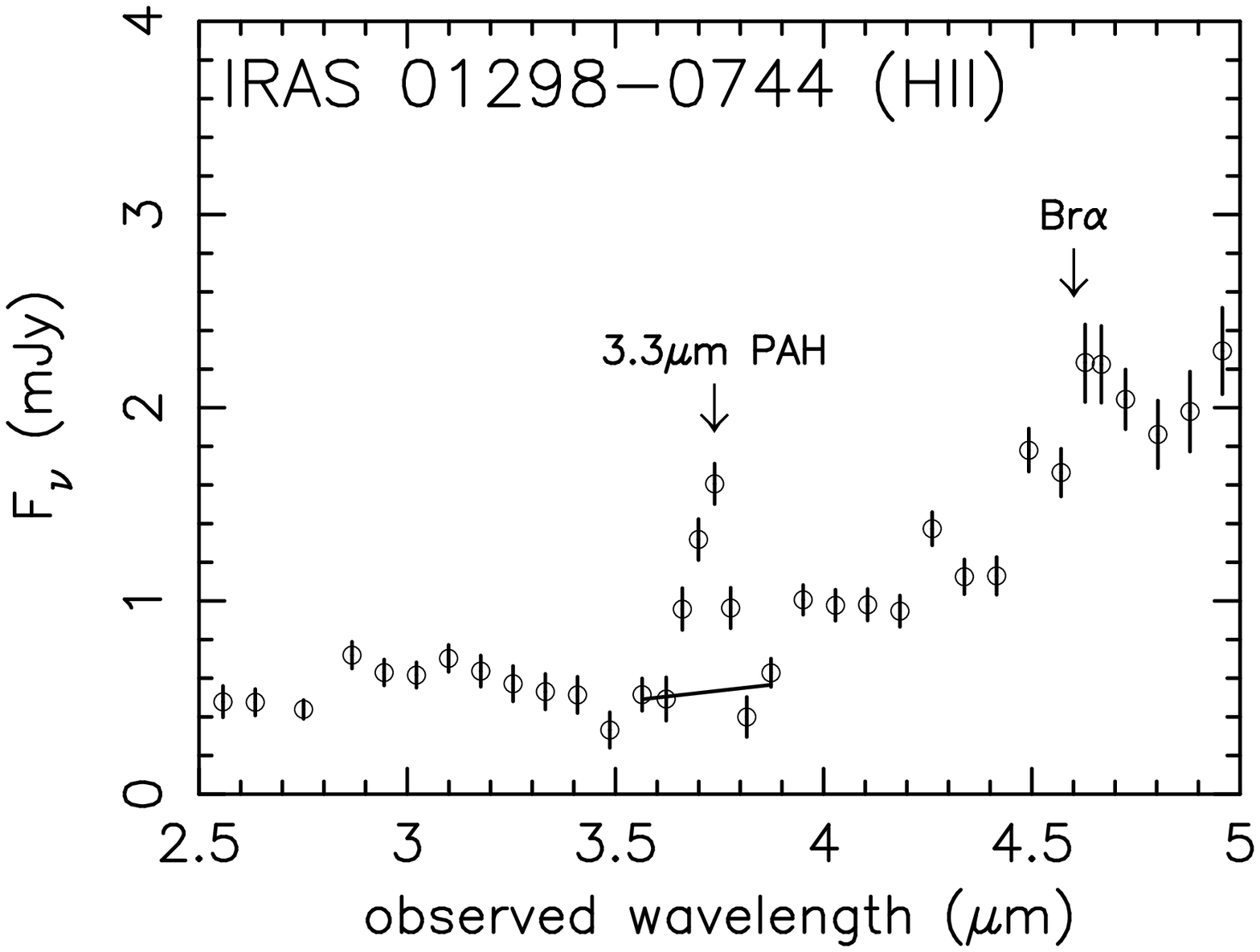}
\FigureFile(80mm,80mm){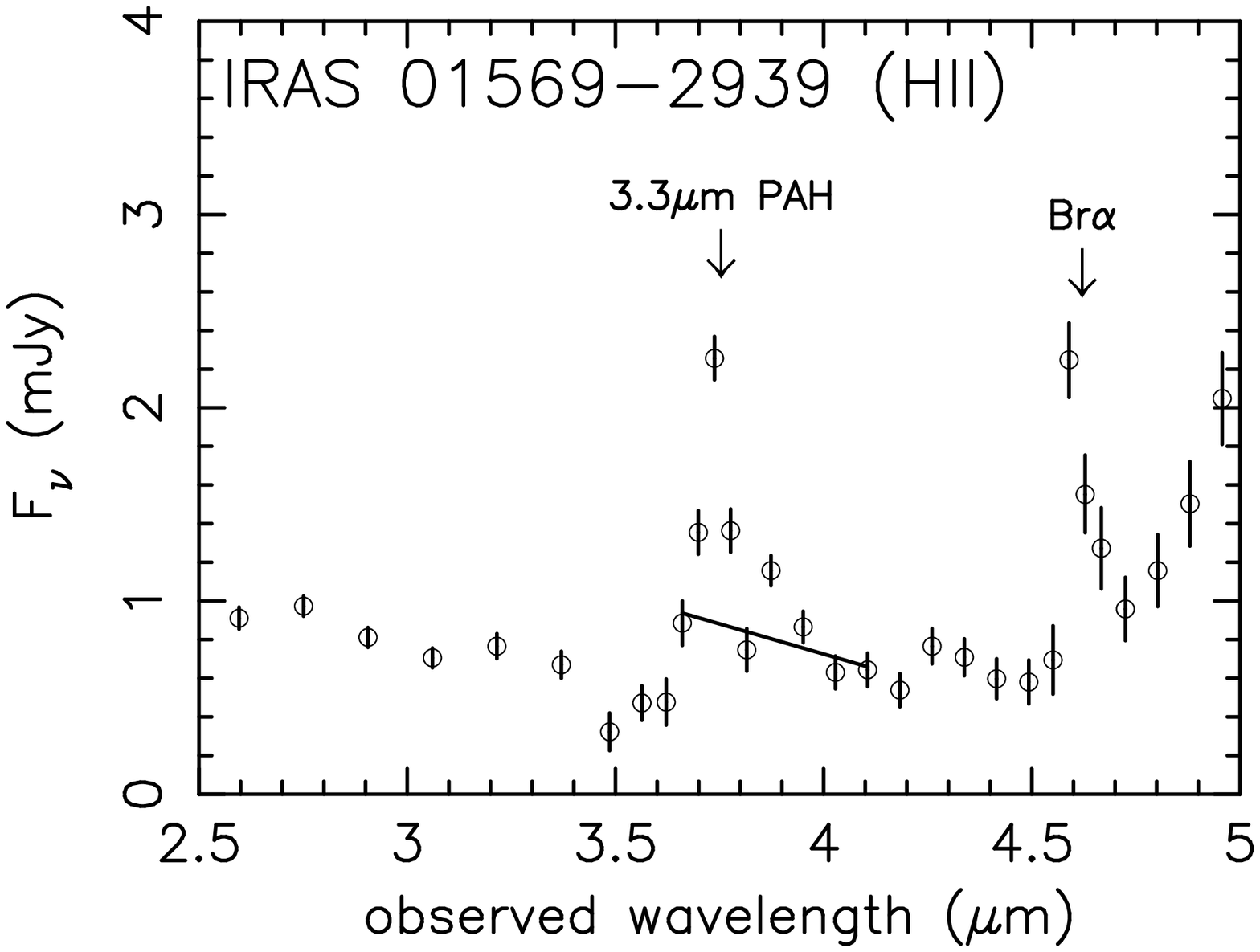}
\FigureFile(80mm,80mm){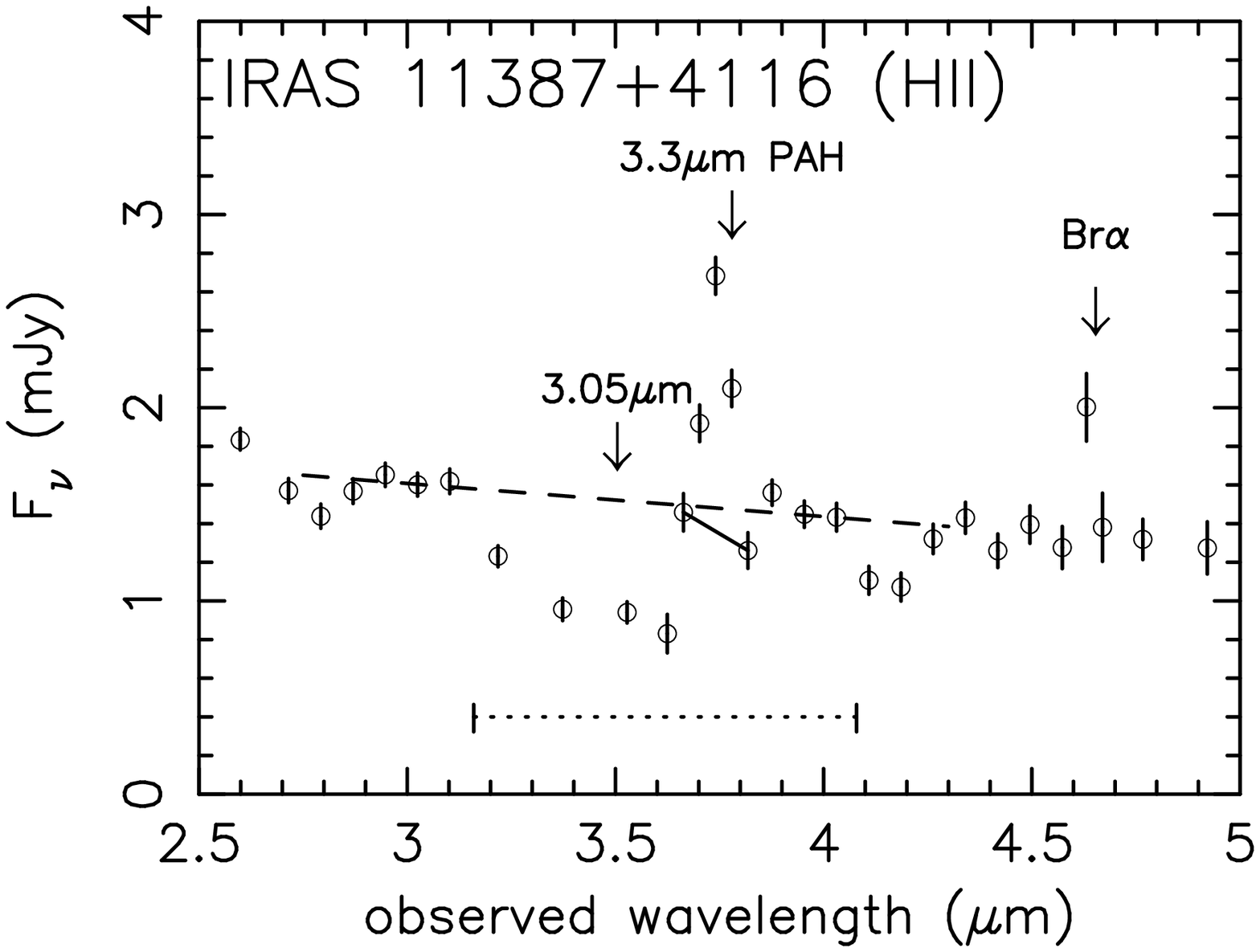}
\FigureFile(80mm,80mm){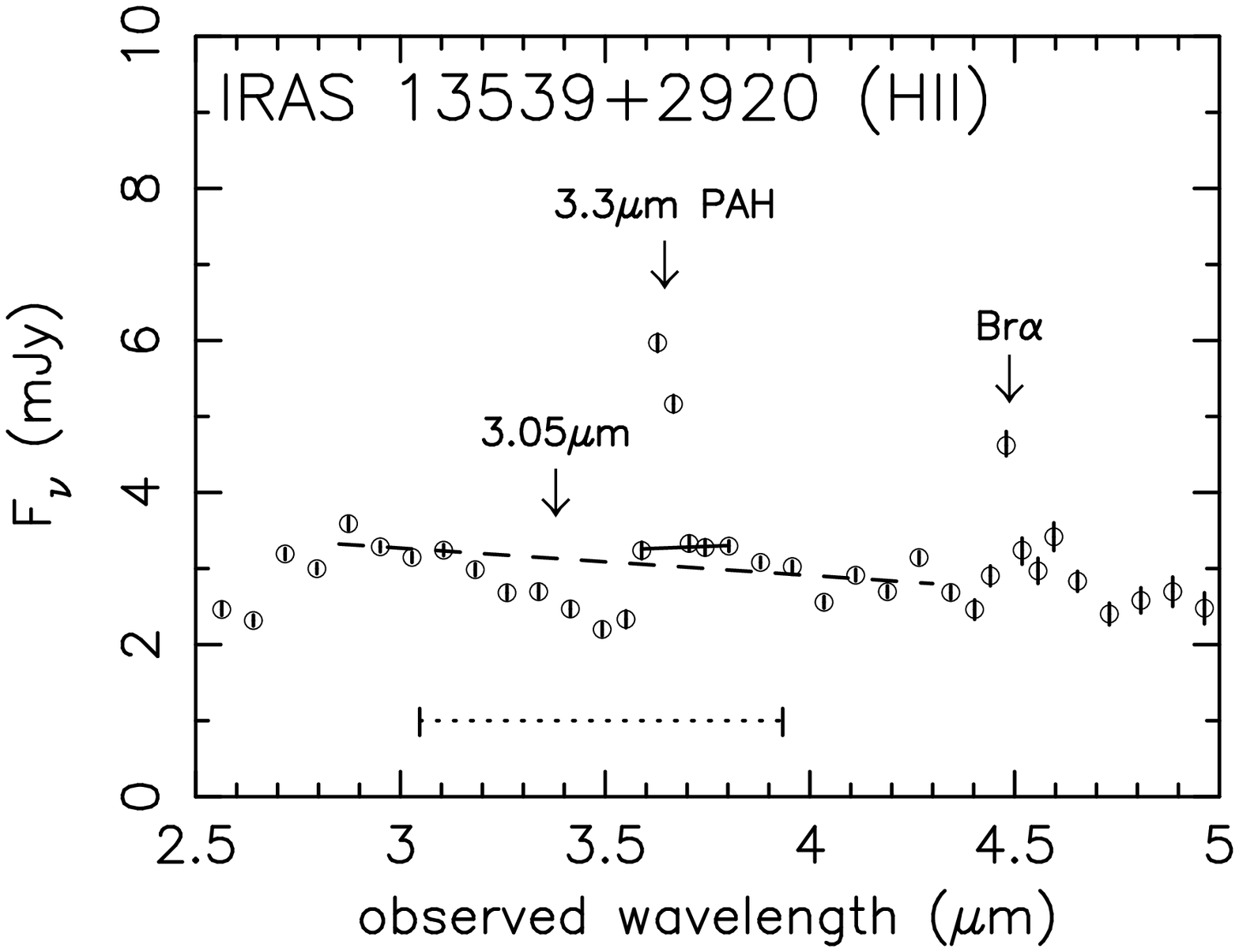}
\FigureFile(80mm,80mm){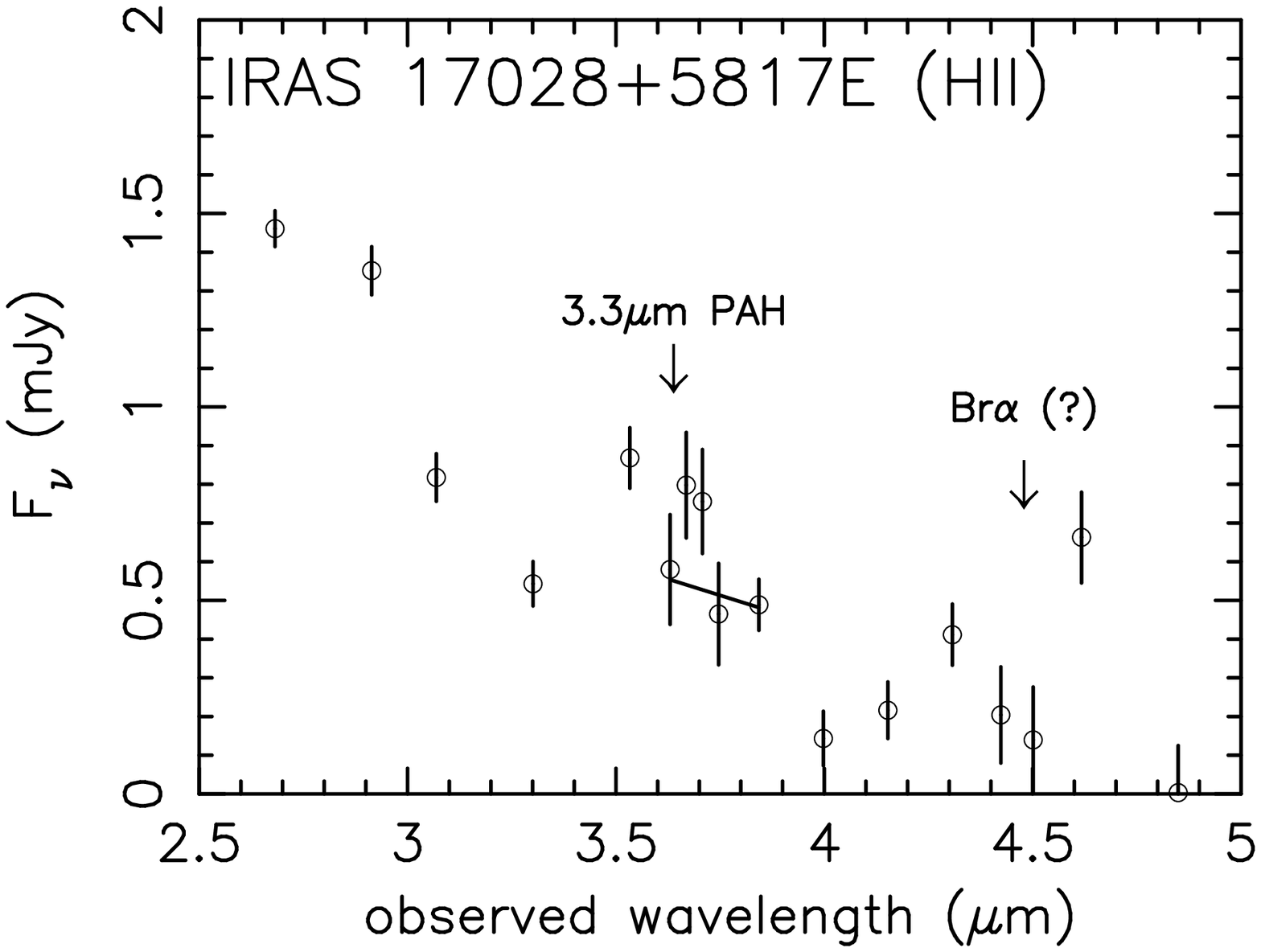}
\end{figure}

\clearpage

\begin{figure}
\FigureFile(80mm,80mm){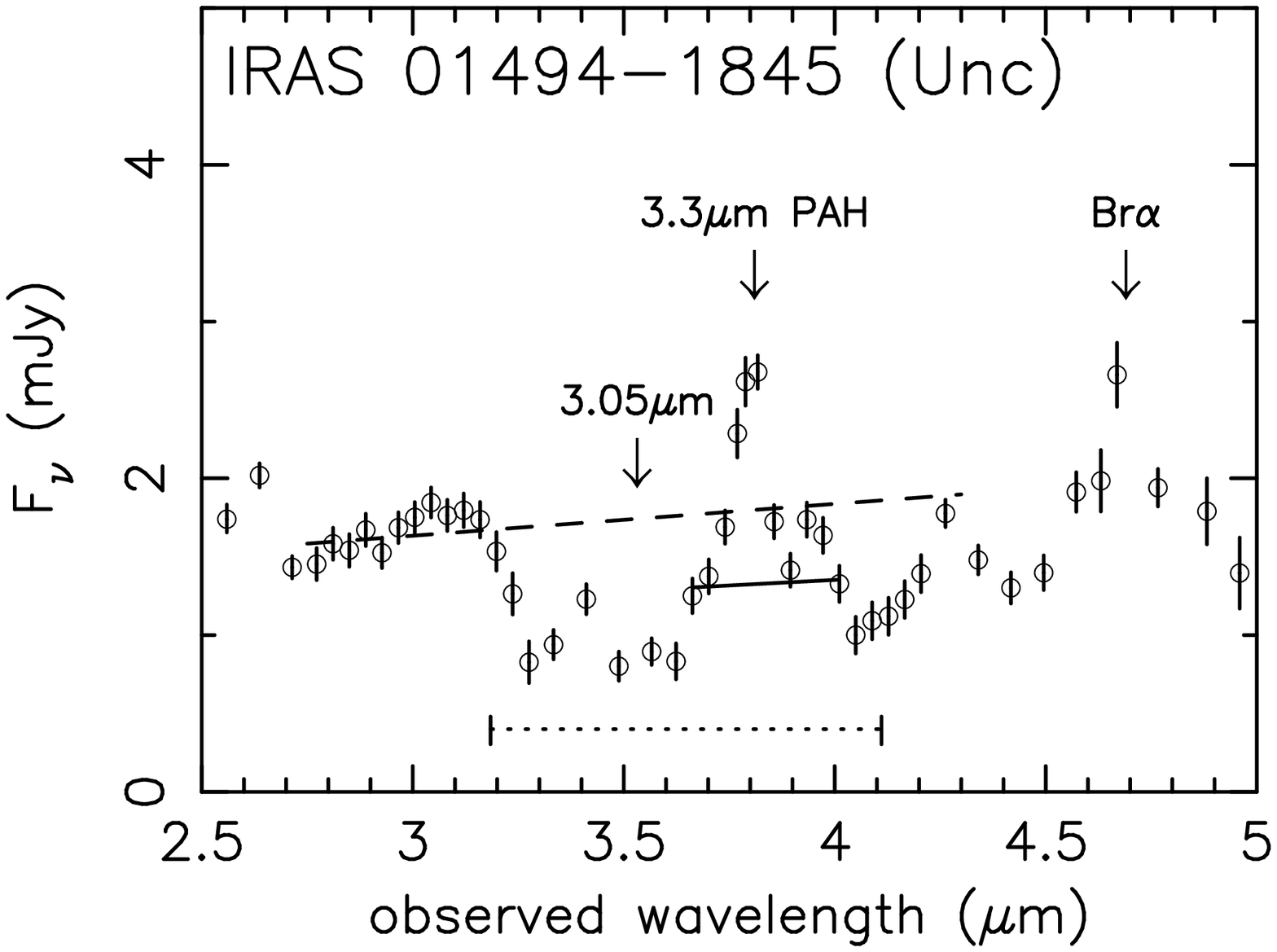}
\FigureFile(80mm,80mm){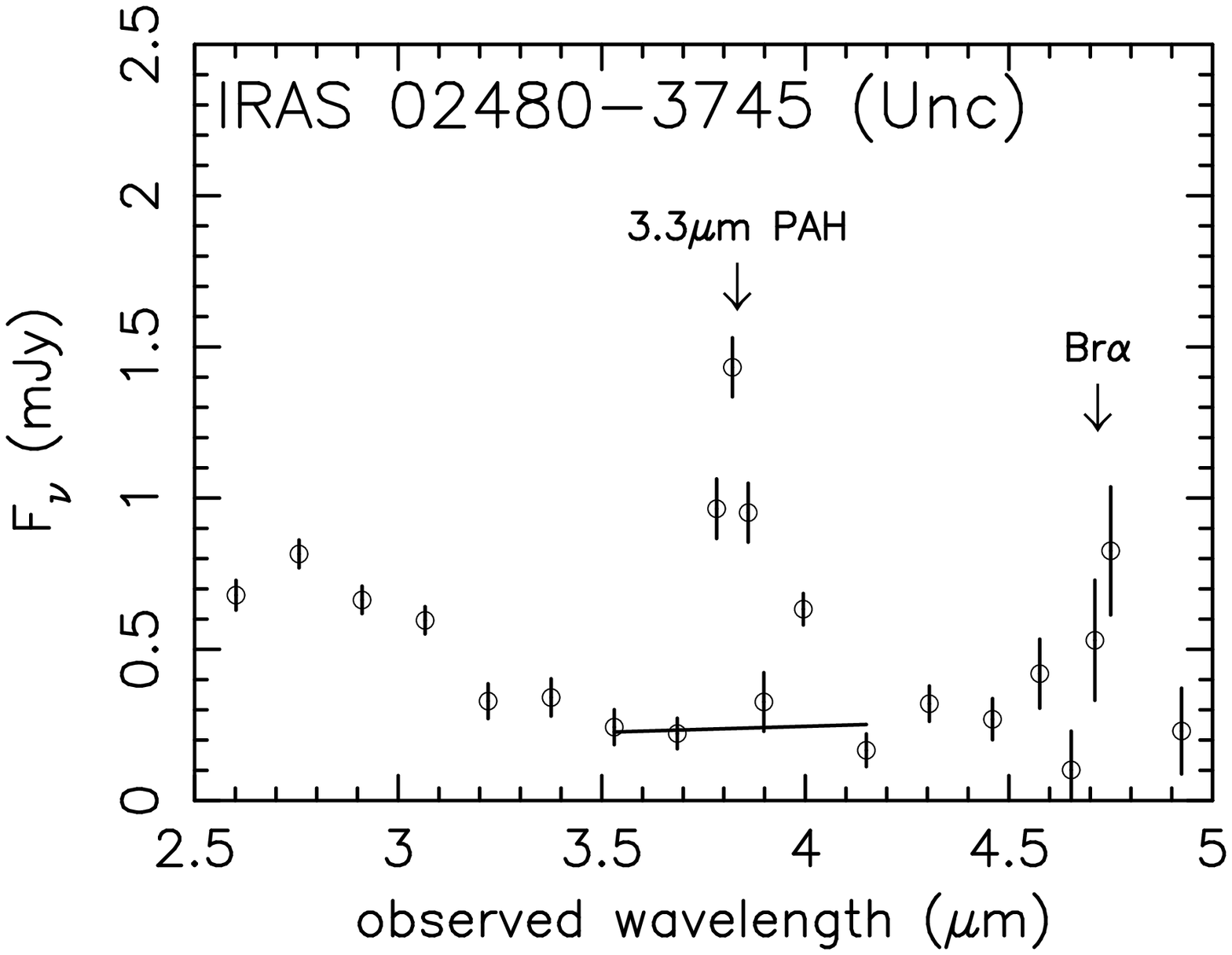}
\FigureFile(80mm,80mm){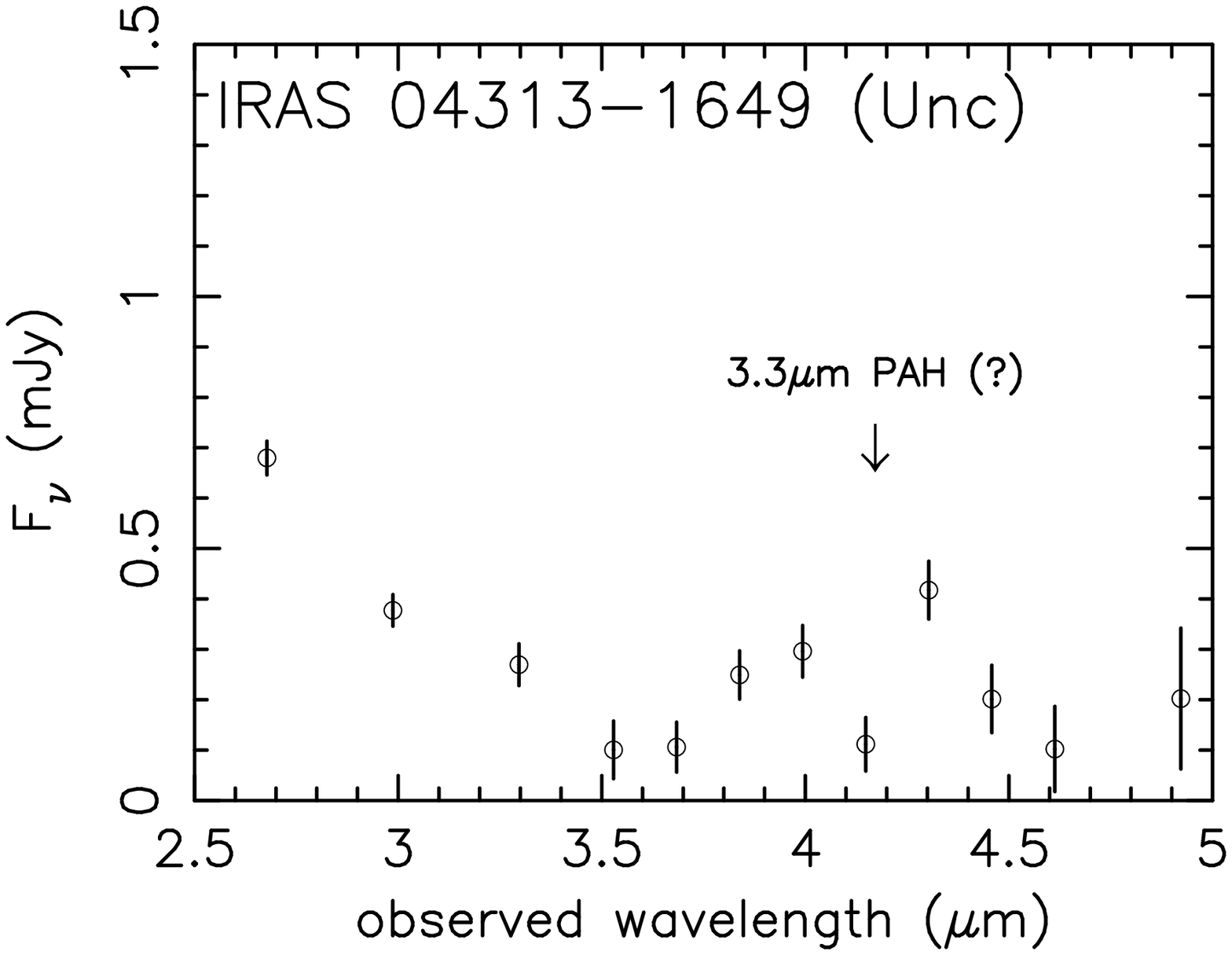}
\FigureFile(80mm,80mm){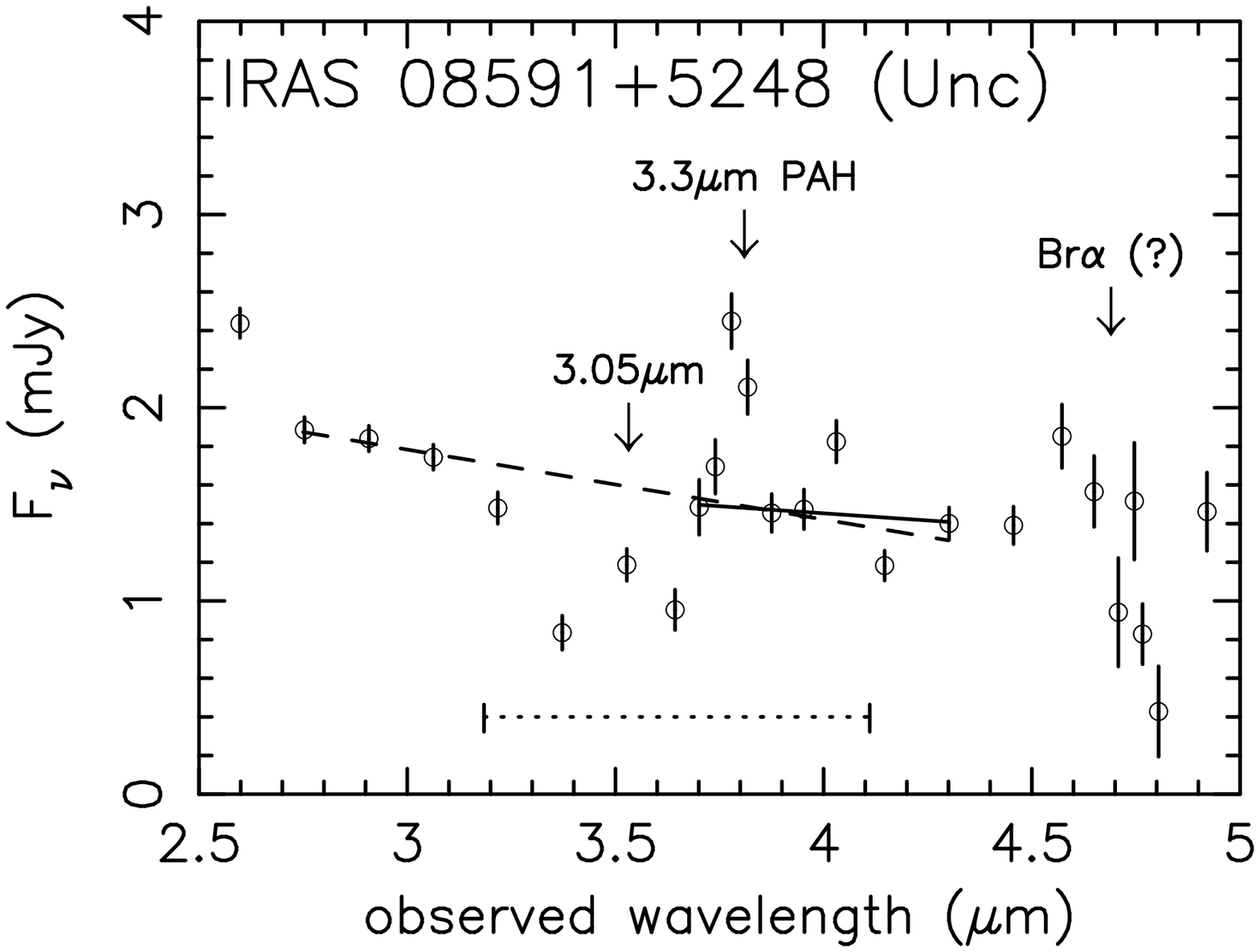}
\FigureFile(80mm,80mm){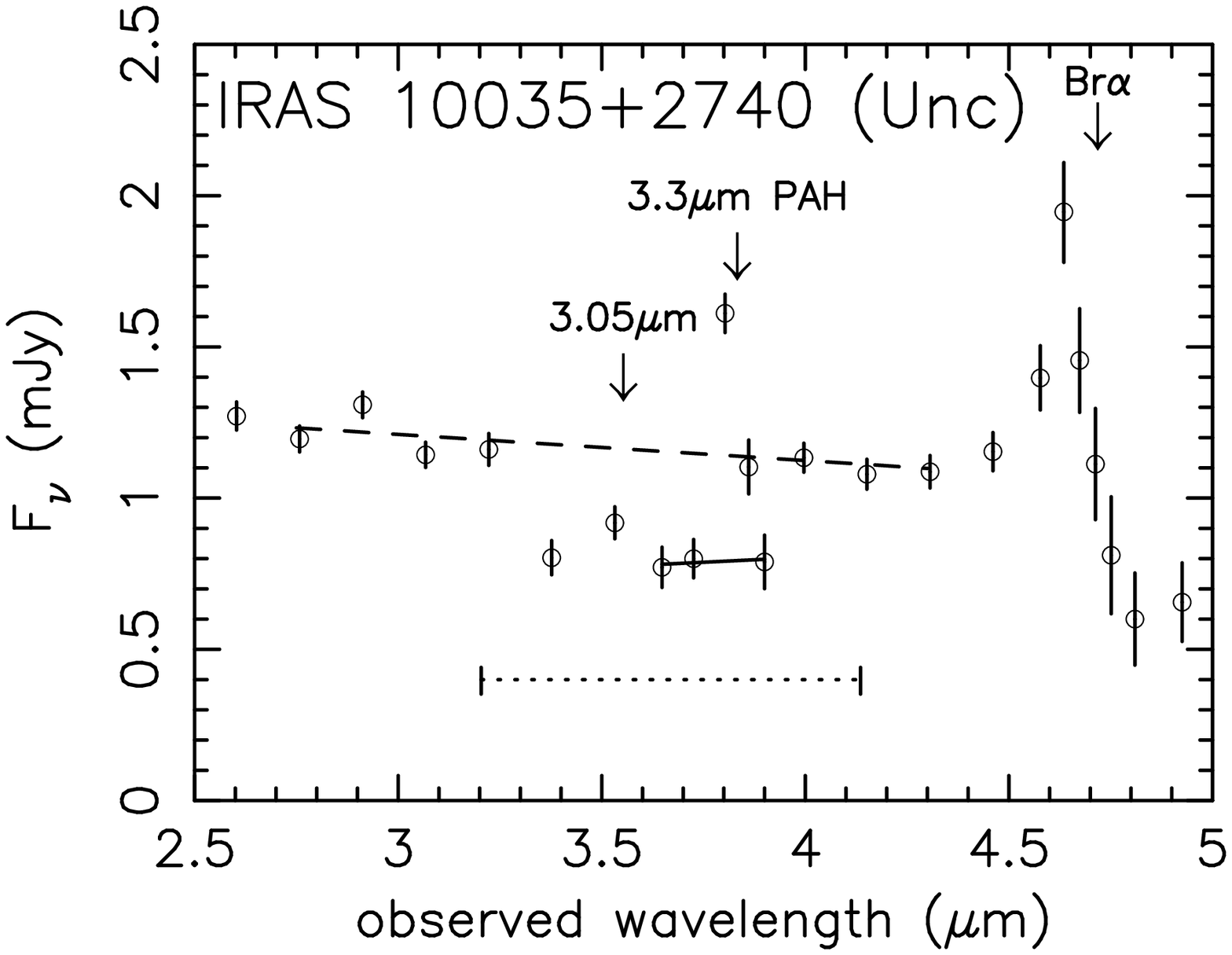}
\end{figure}

\clearpage

\begin{figure}
\FigureFile(80mm,80mm){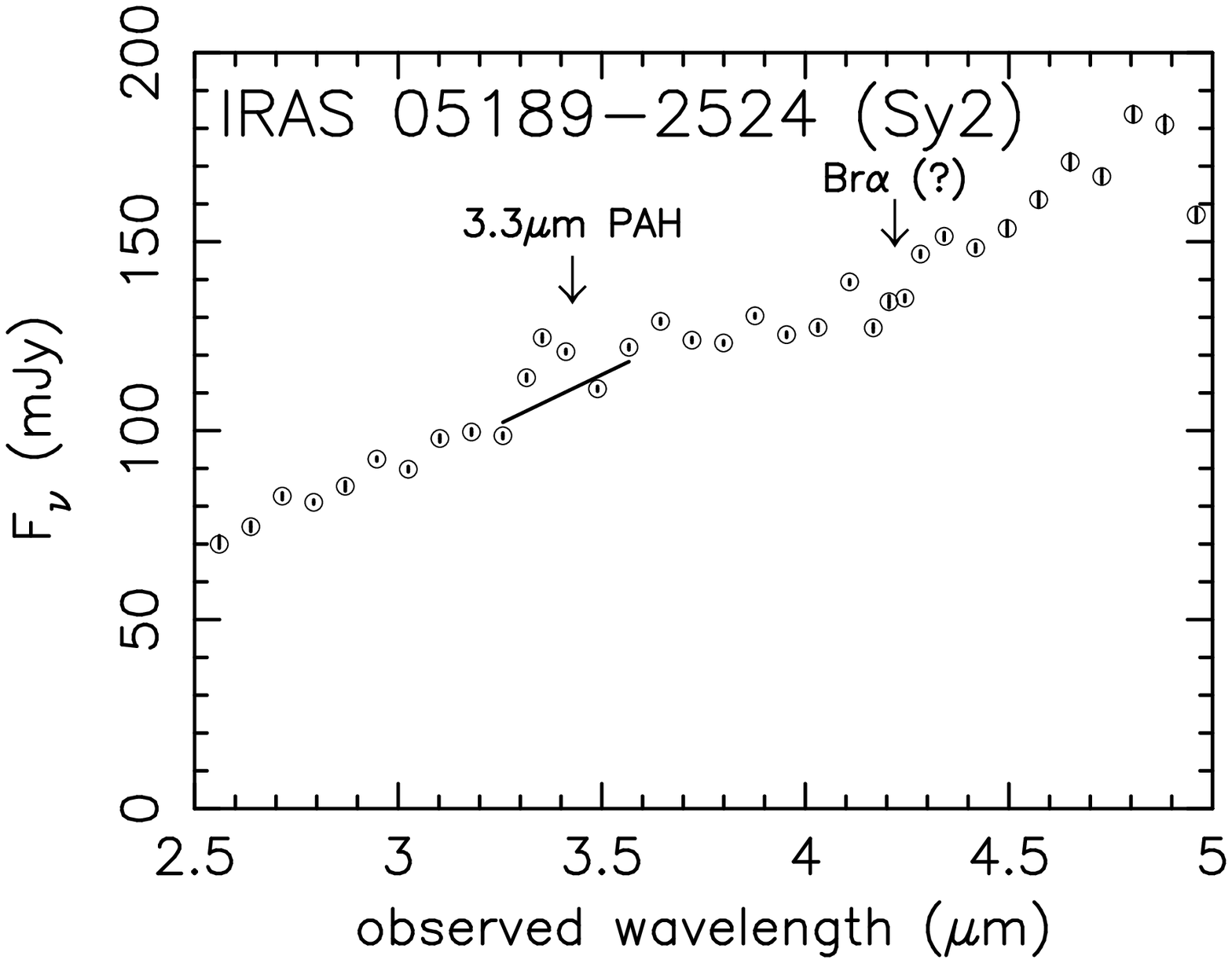}
\FigureFile(80mm,80mm){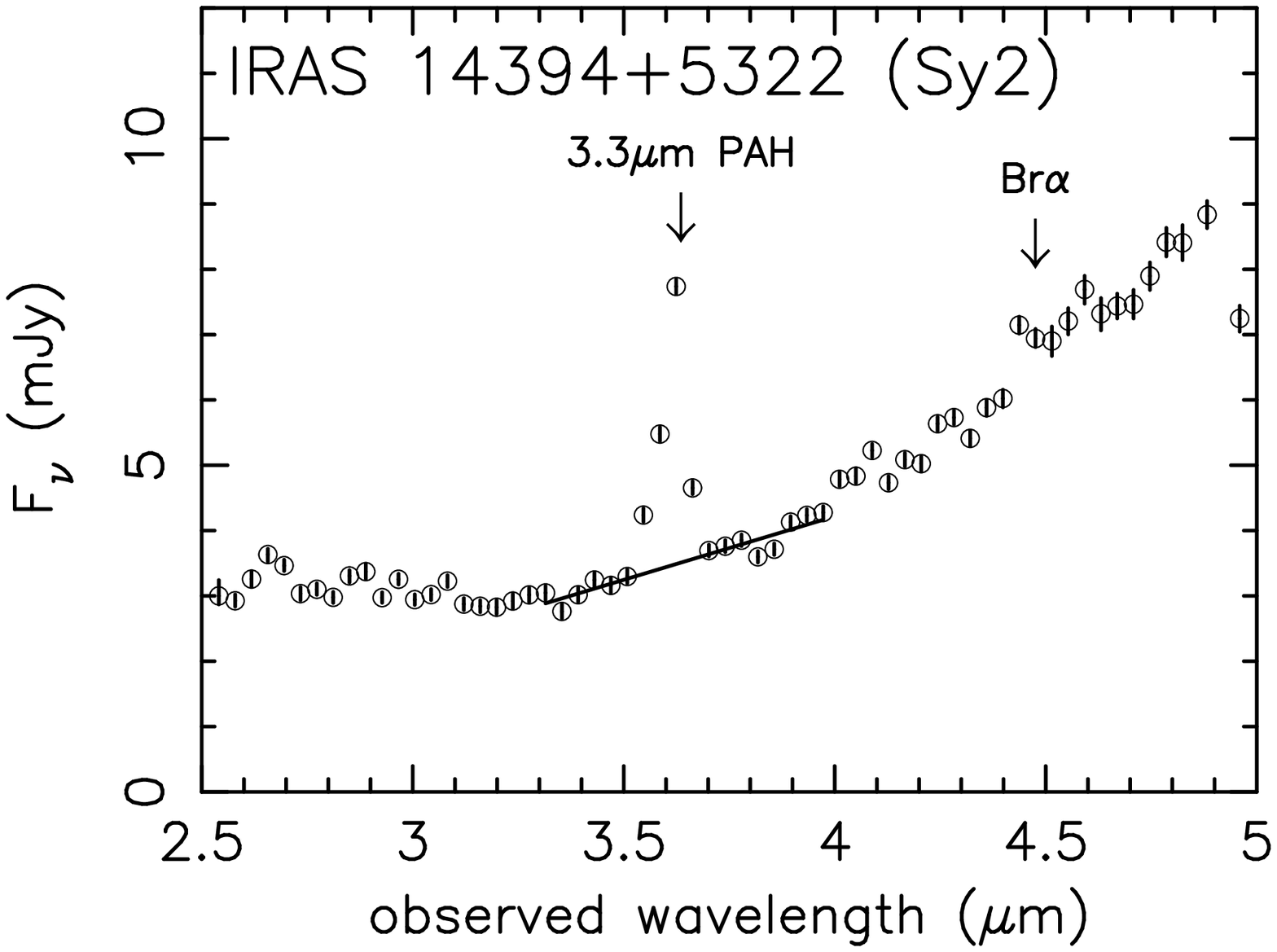}
\FigureFile(80mm,80mm){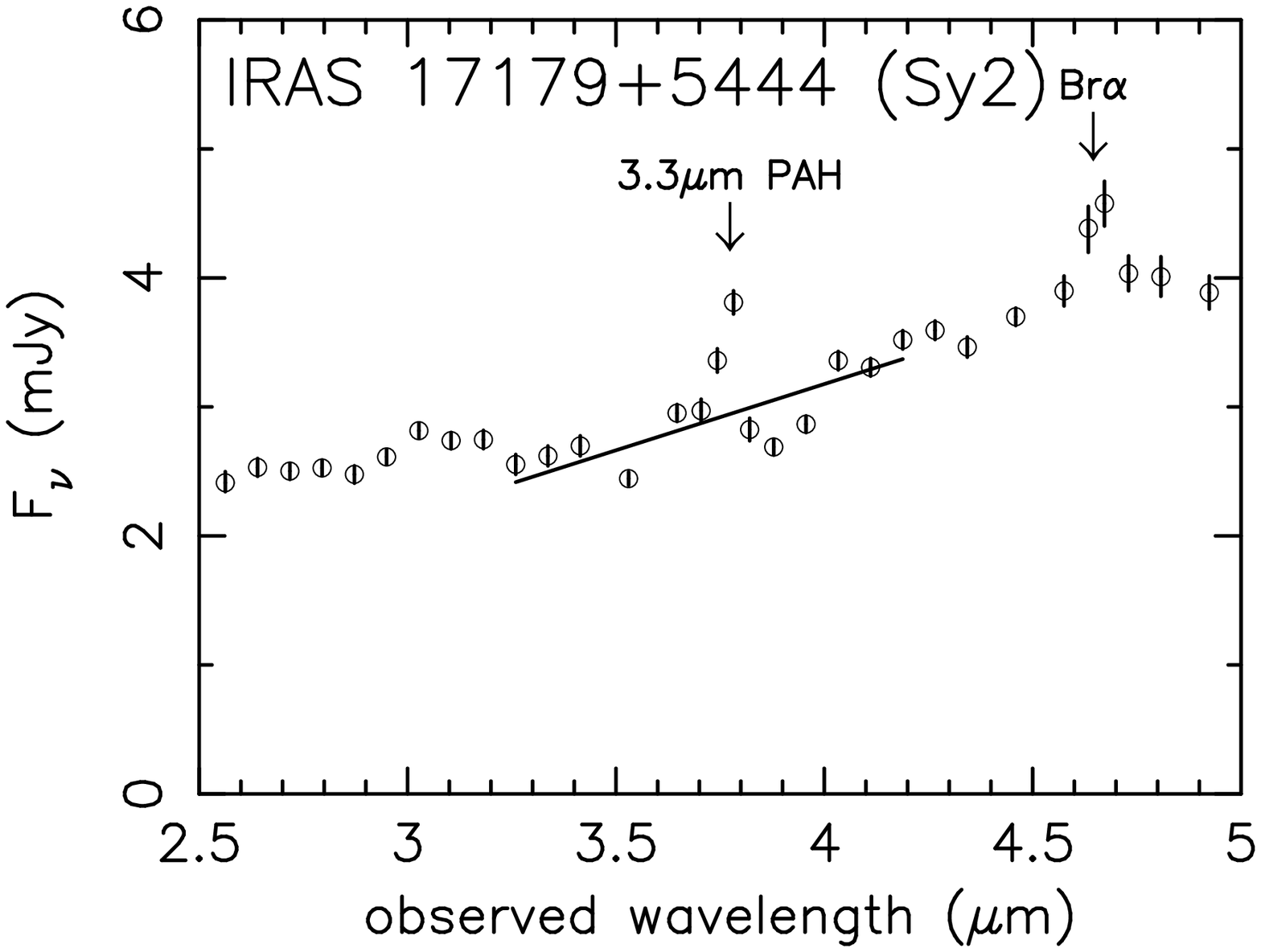}
\FigureFile(80mm,80mm){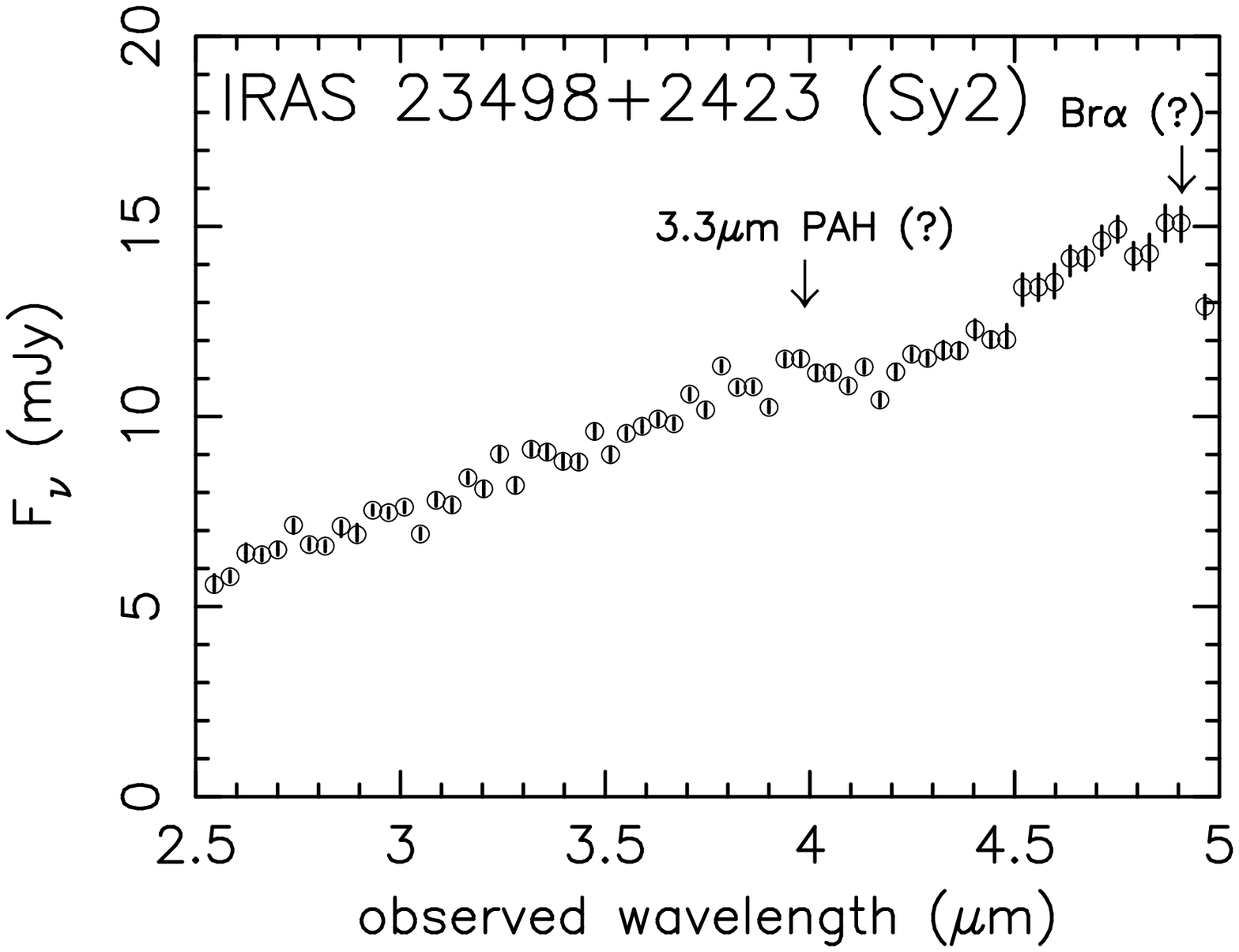}
\end{figure}

\clearpage

\begin{figure}
\FigureFile(80mm,80mm){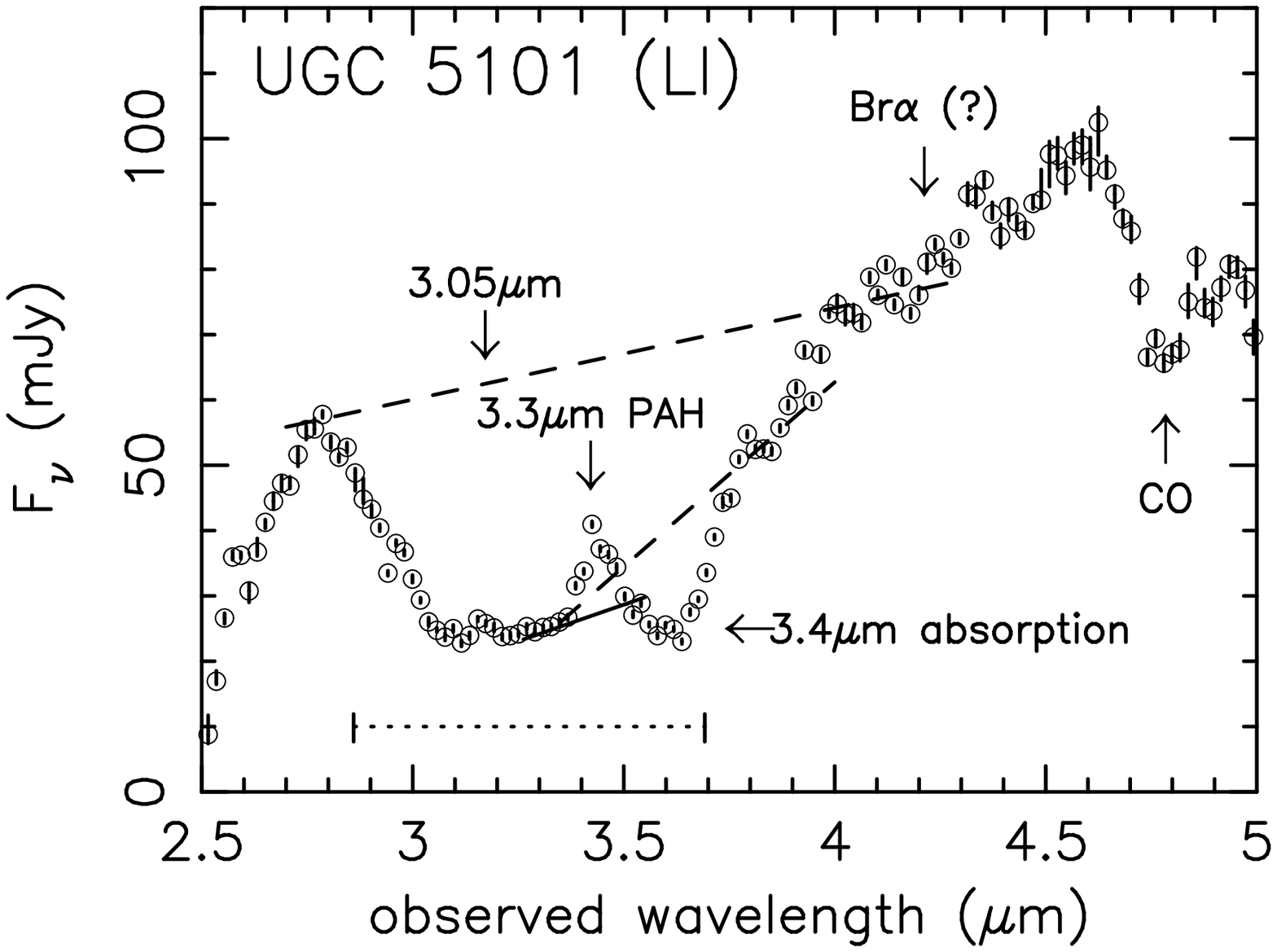}
\FigureFile(80mm,80mm){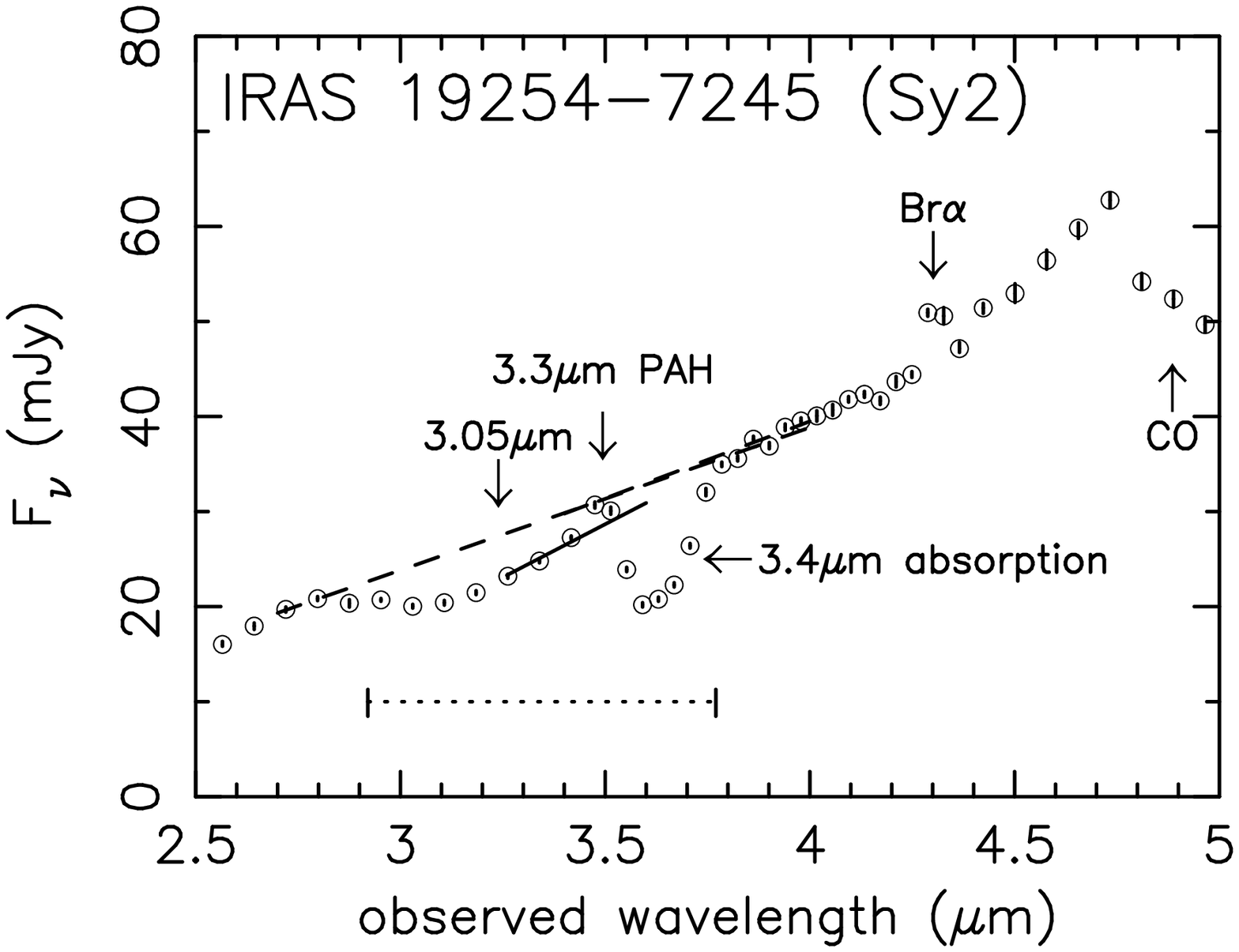}
\caption{
AKARI IRC 2.5--5.0 $\mu$m spectra of ULIRGs. The order of
objects is the same as shown in Table 1, except IRAS 17028+5817E, whose
spectrum is shown in the group of HII-region ULIRGs. ``LI'', ``HII'',
``Unc'', and ``Sy2'' indicate optically LINER, HII-region, unclassified,
and Seyfert 2 ULIRGs, respectively.  
The abscissa and ordinate are the observed wavelength in $\mu$m and
F$_{\nu}$ in mJy, respectively. The lower arrows with ``3.3 $\mu$m PAH''
indicate the expected wavelength of the 3.3 $\mu$m PAH emission
($\lambda_{\rm rest}$ = 3.29 $\mu$m). The solid lines are the adopted
continua used to estimate the 3.3 $\mu$m PAH emission fluxes for
PAH-detected sources. The arrows
with ``3.05 $\mu$m'' or ``3.4 $\mu$m'' denote the 3.1 $\mu$m ice-covered
and 3.4 $\mu$m bare carbonaceous dust absorption features,
respectively. Regarding ULIRGs with clear signatures for these
absorption features, dashed straight lines are plotted to show the
adopted continuum levels used to estimate the absorption optical
depths. The dotted lines indicate the wavelength range in which effects
of the broad 3.1 $\mu$m ice absorption feature can be significant
($\lambda_{\rm rest}$ = 2.75--3.55 $\mu$m, adopted from the spectrum of
Elias 16; Smith et al. 1989). The ice absorption feature is usually very
strong at $\lambda_{\rm rest}$ = 2.9--3.2 $\mu$m, but the profiles and
wavelength range of weaker absorption wings are found to vary among
different Galactic objects (Smith et al. 1989). Br$\alpha$ emission is
also indicated. ``CO'' or ``CO$_{2}$'' denote CO and CO$_{2}$ absorption
features, respectively.     
}
\end{figure}

\clearpage

%---  Figure 3 ---%
\begin{figure}
\begin{center}
\FigureFile(120mm,120mm){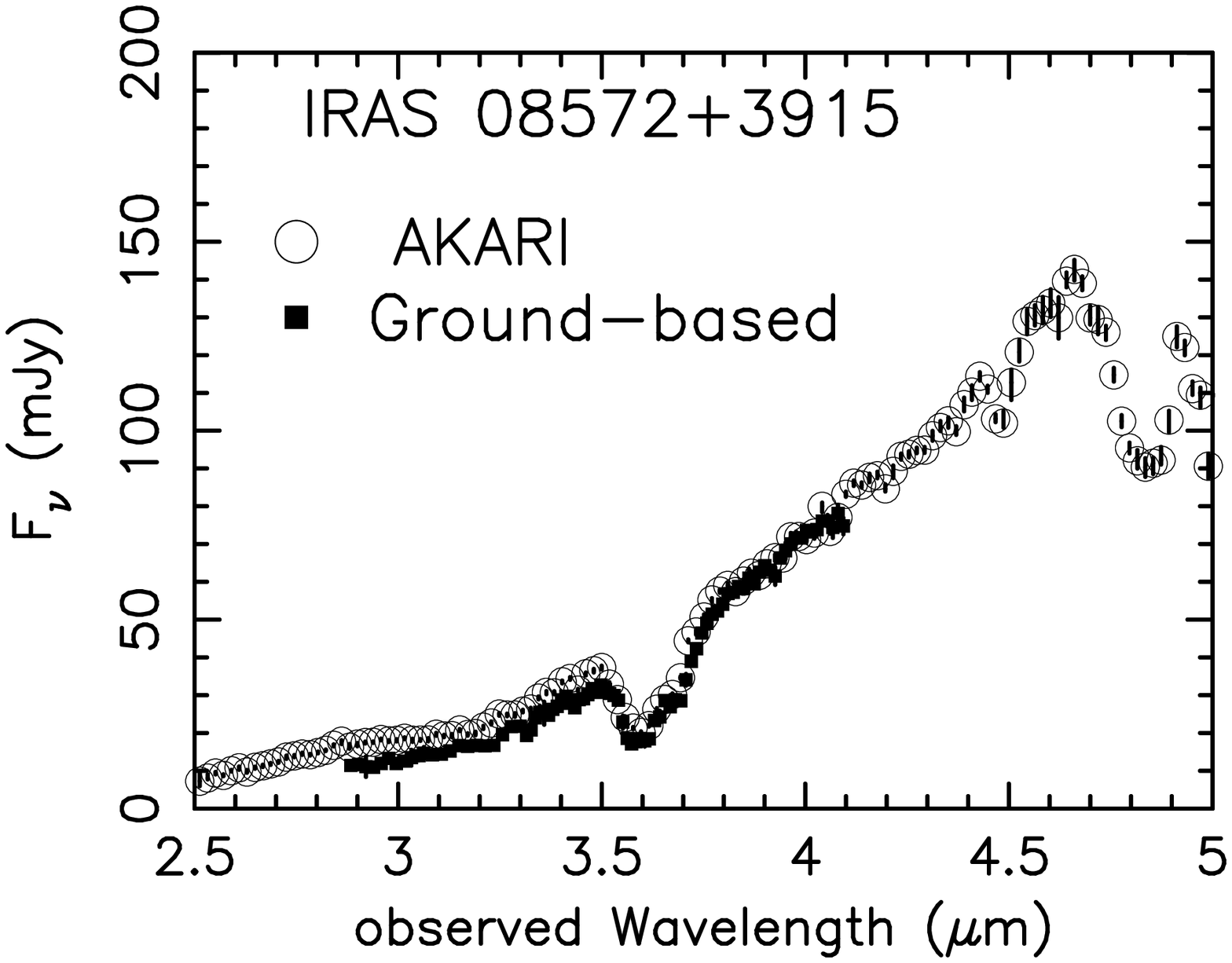}
\caption{
Comparison of the AKARI IRC slitless spectrum of IRAS
08572+3915 (open circles) with ground-based 1$\farcs$6-wide slit
spectrum taken with IRTF SpeX (filled squares) \citep{idm06}. The seeing
size at the time of the IRTF observations was 0$\farcs$6--0$\farcs$7 in
the $K$-band (2--2.5 $\mu$m). The ordinate of the IRTF SpeX spectrum is
converted from F$_{\lambda}$ in W m$^{-2}$ $\mu$m$^{-1}$ to F$_{\nu}$ in
mJy. Both spectra have similar spectral shapes and flux
levels. Statistical 1$\sigma$ errors are shown as vertical lines, but
are too small to discern at $\lambda_{\rm obs}$ $<$ 4.5 $\mu$m. A slight
discrepancy at the shorter wavelength could be explained by a
non-negligible contribution from extended stellar emission in the host
galaxy (see $\S$4 for more details). 
}
\end{center}
\end{figure}

%--- Figure 4 ---%
\begin{figure}
\begin{center}
\FigureFile(80mm,80mm){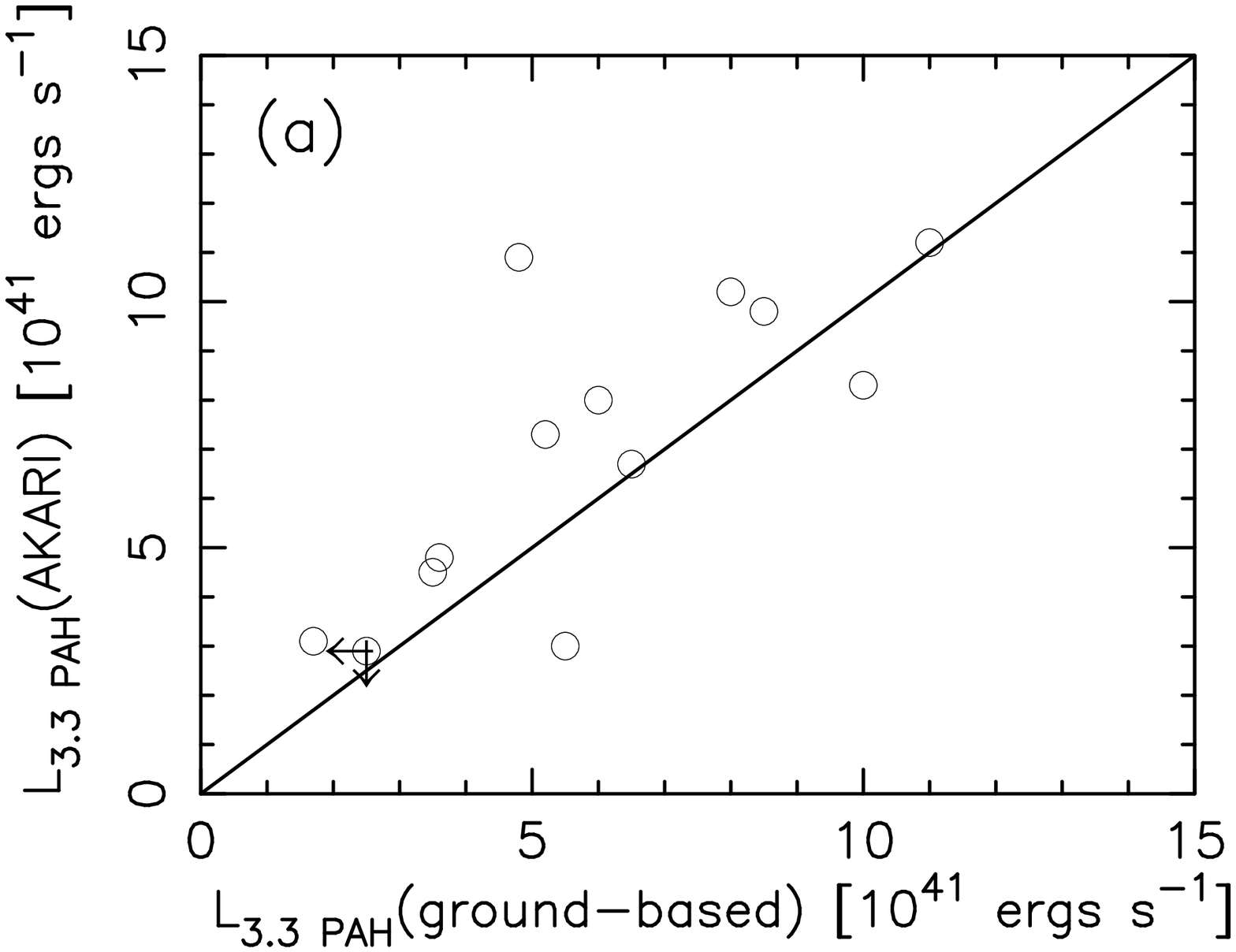}
\FigureFile(80mm,80mm){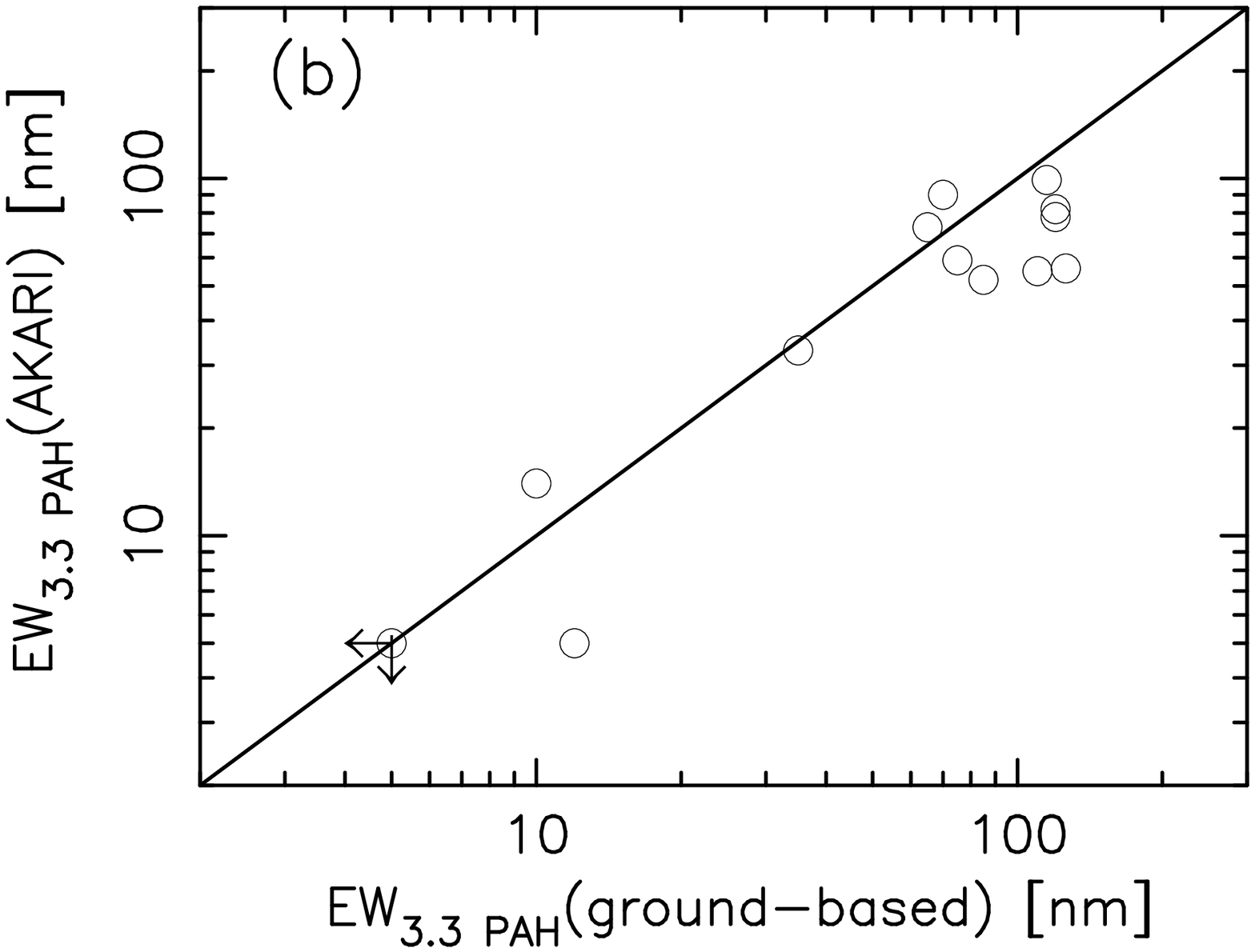}
\caption{
{\it (a)} : Comparison of the 3.3 $\mu$m PAH luminosity
measured with ground-based slit spectra (abscissa) and AKARI IRC
slitless spectra (ordinate) for ULIRGs for which both AKARI and
ground-based spectra at 3--4 $\mu$m are available. These ULIRGs are IRAS
00482$-$2721, 08572+3915, 09539+0857, 10494+4424, 16468+5200,  
16487+5447, 17028+5817W, 17044+6720, 00456$-$2904, 11387+4116, 
13539+2920, UGC 5101, and IRAS 19254$-$7245 (Imanishi \& Maloney 2003;
Risaliti et al. 2003; Imanishi et al. 2006a; this paper). All ULIRGs are
at $z <$ 0.15.  
{\it (b)} : Comparison of rest-frame 3.3 $\mu$m PAH equivalent width
measured with ground-based slit spectra (abscissa) and AKARI IRC
slitless spectra (ordinate) for the same ULIRGs.    
}
\end{center}
\end{figure}

%--- Figure 5 ---%
\begin{figure}
\begin{center}
\FigureFile(80mm,80mm){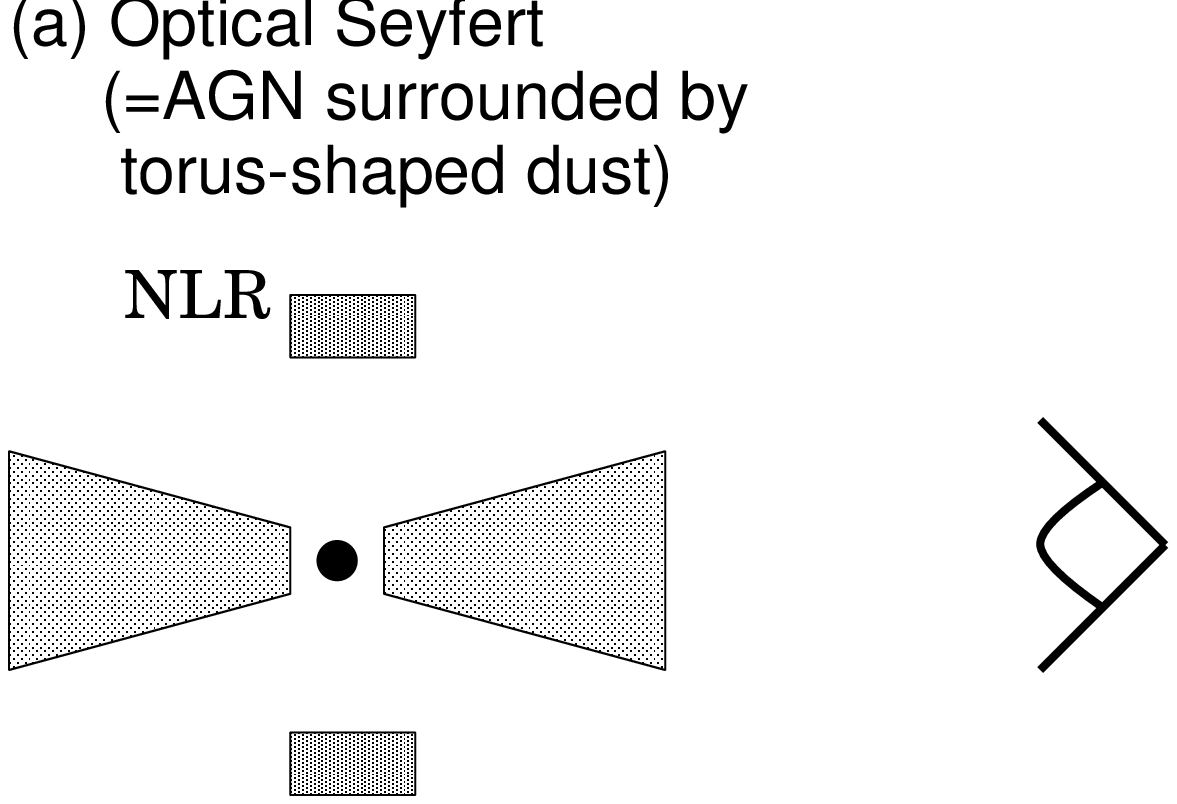}
\FigureFile(120mm,120mm){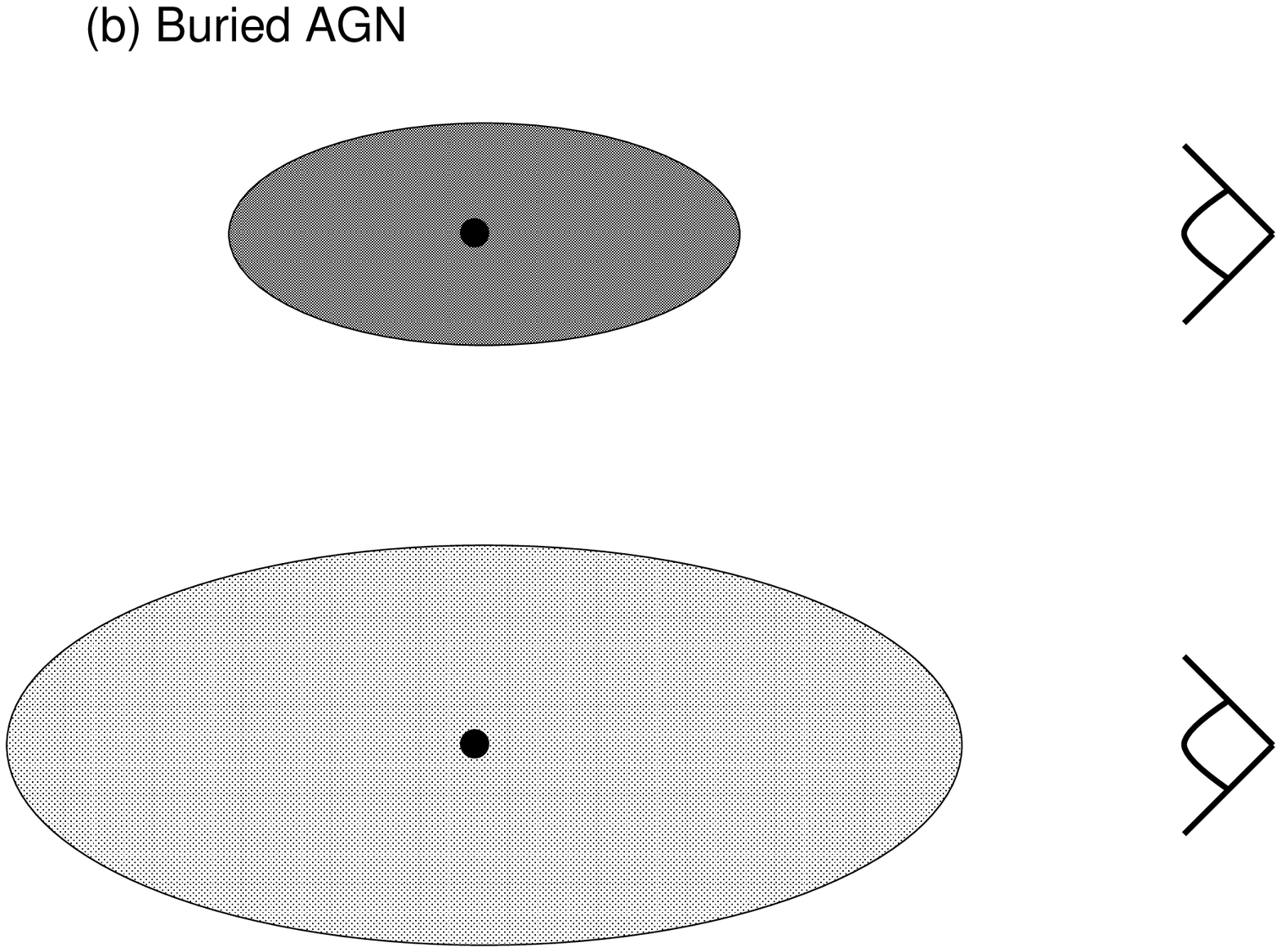}
\caption{
Schematic diagram of nuclear dust distribution.
{\it (a): } An AGN classified optically as a Seyfert. The small filled
circle at the center represents the central mass-accreting supermassive
black hole. When we look from the right side, the AGN is classified as a
Seyfert 2. As the dust has angular momentum, dust distribution is
axisymmetrical. When the total amount of nuclear dust is modest, dust
along the direction of the lowest dust column density can be transparent
to the AGN's ionizing radiation, producing the so-called narrow line
regions (NLRs), photoionized by the AGN's radiation, above a torus scale
height. In this case, dust distribution can be approximated as
torus-shaped. Optical emission line ratios from NLRs are classified as
Seyfert 2s, which are different from normal starburst galaxies.  
{\it (b): } A buried AGN classified optically as a non-Seyfert. Dust
surrounds the central AGN along virtually all lines-of-sight. Since the
total amount of nuclear dust is larger than that of a Seyfert-type AGN,
even the direction of the lowest dust column density can be opaque to
the AGN's radiation, and dust column density along our line-of-sight is
generally larger than a Seyfert 2 AGN (see $\S$5.7). The darker shade
indicates higher dust volume density. The larger amount of total nuclear
dust in a buried AGN compared to a Seyfert 2 AGN can originate either in
a higher dust volume density with a similar volume (upper case), a
larger volume of nuclear obscuring dust with similar volume density
(lower case), or both larger volume and higher volume density of nuclear
dust.  
}
\end{center}
\end{figure}

%--- Figure 6 ---%
\begin{figure}
\begin{center}
\FigureFile(80mm,80mm){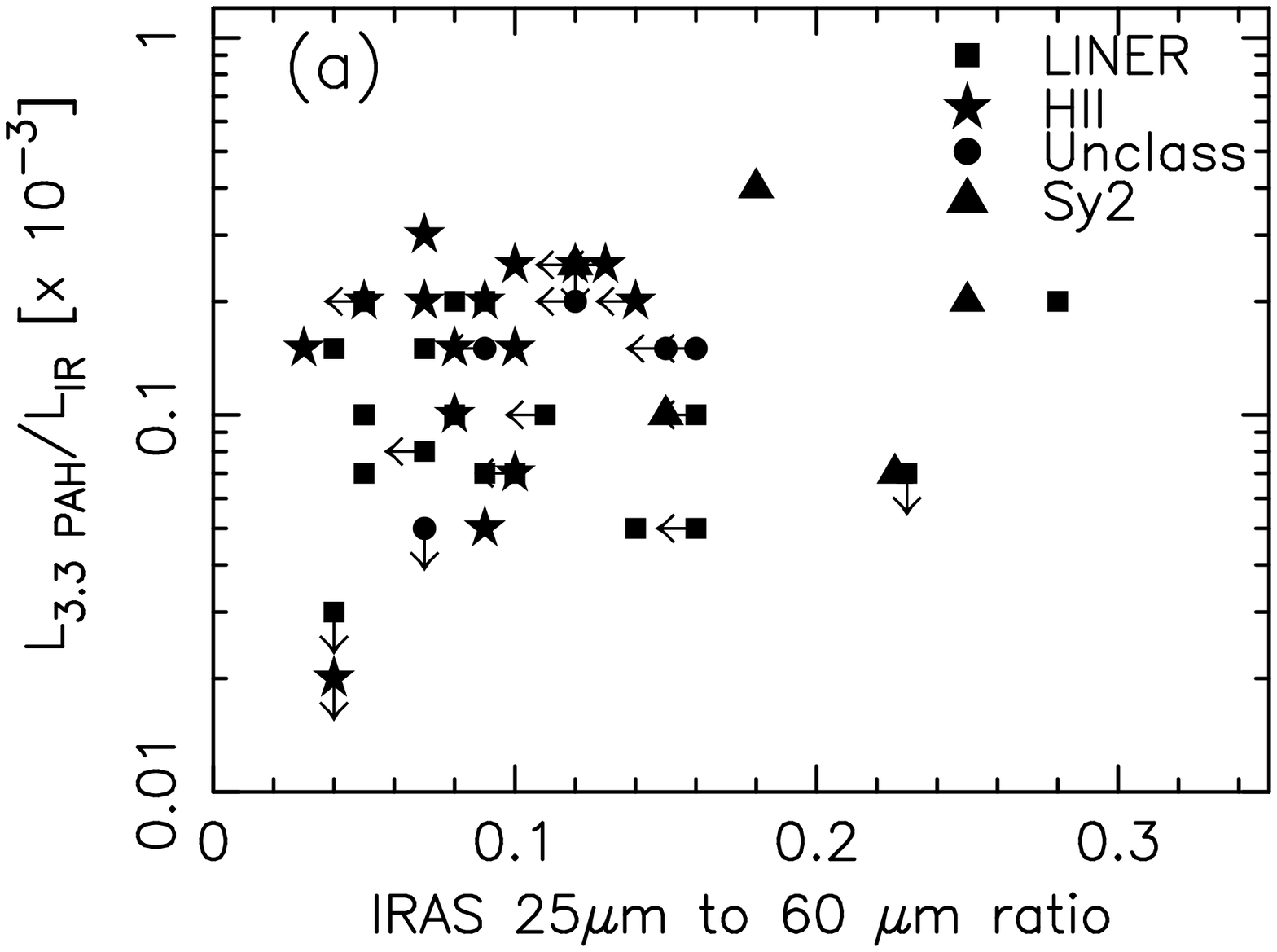}
\FigureFile(80mm,80mm){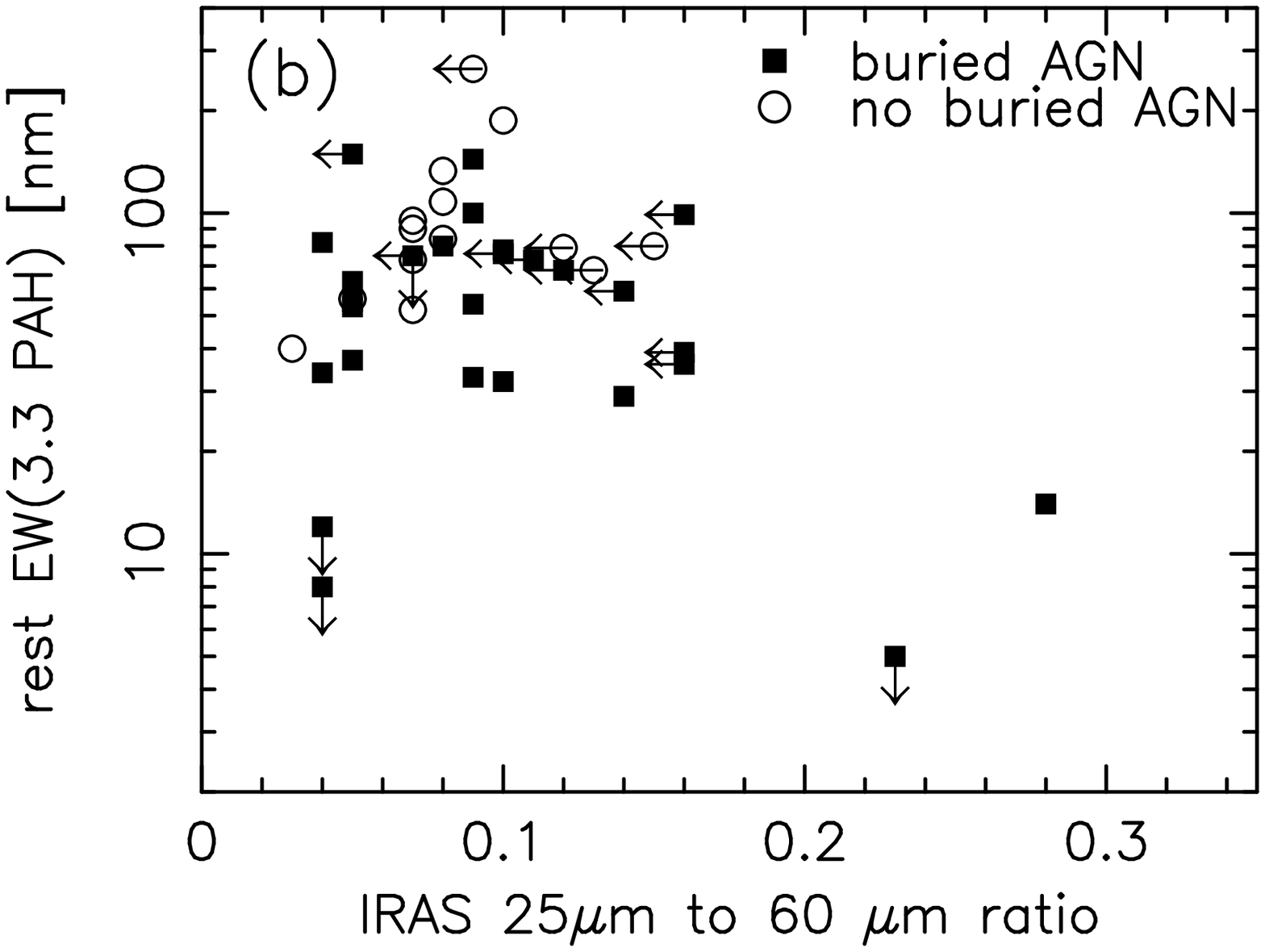}
\caption{
{\it (a)}: {\it IRAS} 25 $\mu$m-to-60 $\mu$m flux ratio (abscissa) and
observed 3.3 $\mu$m PAH to infrared luminosity ratio (ordinate).
Filled squares: LINER ULIRGs. 
Filled stars: HII-region ULIRGs.
Filled circles: Optically unclassified ULIRGs.
Filled triangles: Seyfert 2 ULIRGs.
The two ULIRGs of interest, UGC 5101 and IRAS 19254$-$7245, are
classified as LINER and Seyfert 2, respectively. For IRAS 17028+5817,
emission from double nuclei is resolved in AKARI IRC spectra, but not in
the {\it IRAS} data.  For this source, we assume that both nuclei have
the same far-infrared colors as measured with {\it IRAS}. 
{\it (b)}: {\it IRAS} 25 $\mu$m-to-60 $\mu$m flux ratio (abscissa) and
rest frame equivalent widths of the 3.3 $\mu$m PAH emission (ordinate)
for non-Seyfert ULIRGs. 
Filled squares: ULIRGs with buried AGN signatures. 
Open circles: ULIRGs with no obvious buried AGN signatures in the AKARI
IRC 2.5--5 $\mu$m spectra. 
}
\end{center}
\end{figure}

%--- Figure 7 ---%
\begin{figure}
\FigureFile(80mm,80mm){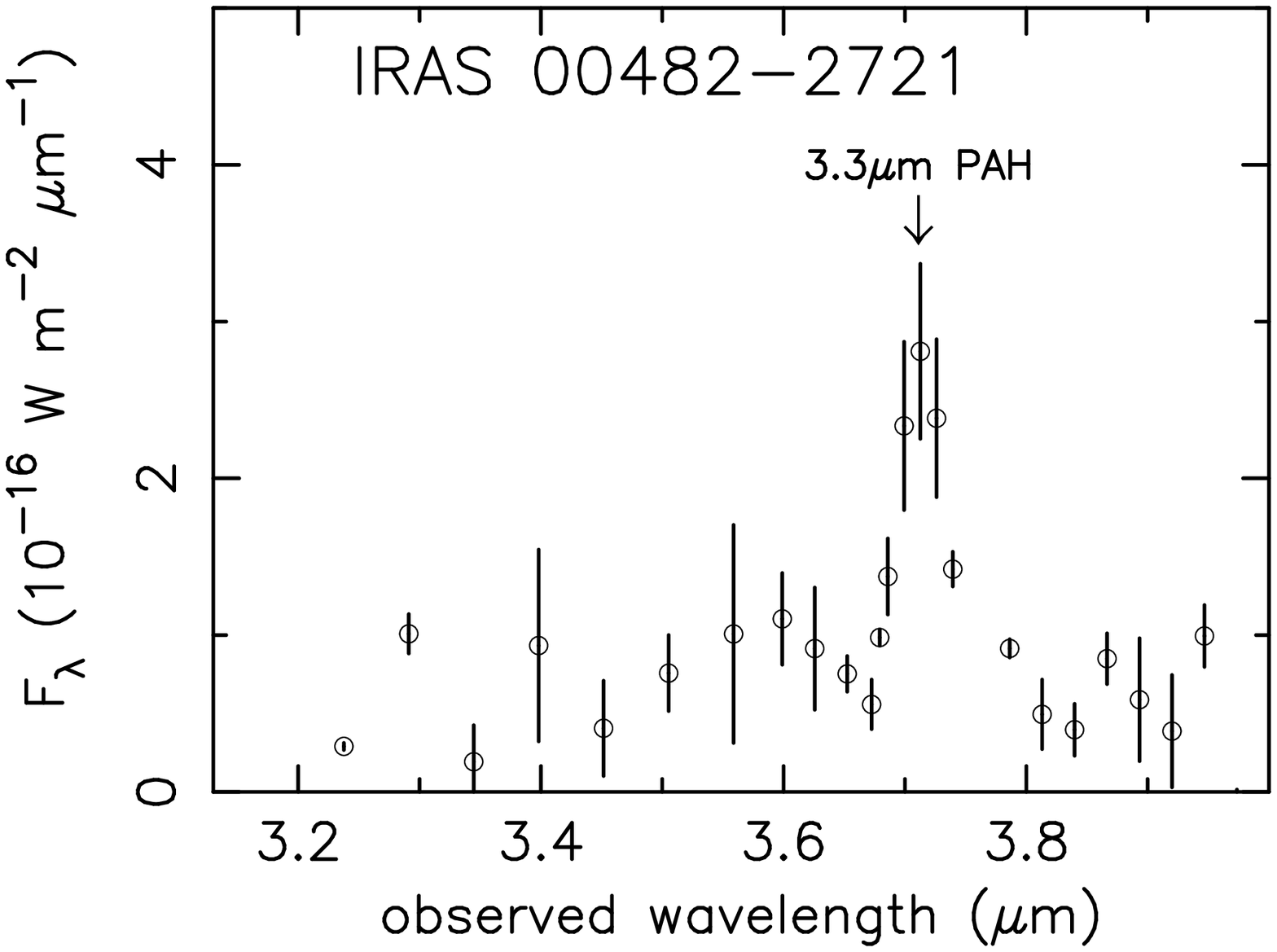}
\FigureFile(80mm,80mm){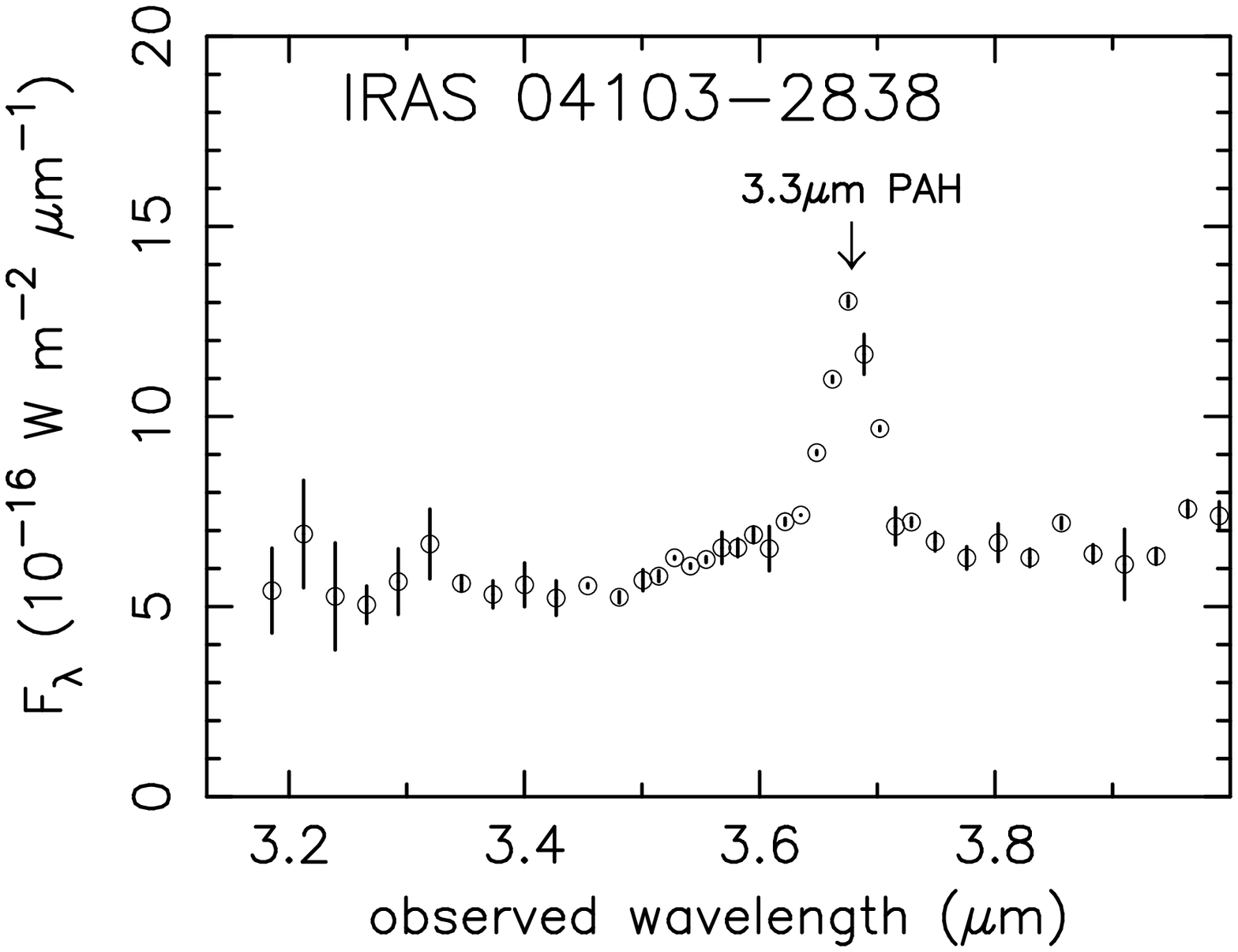}
\FigureFile(80mm,80mm){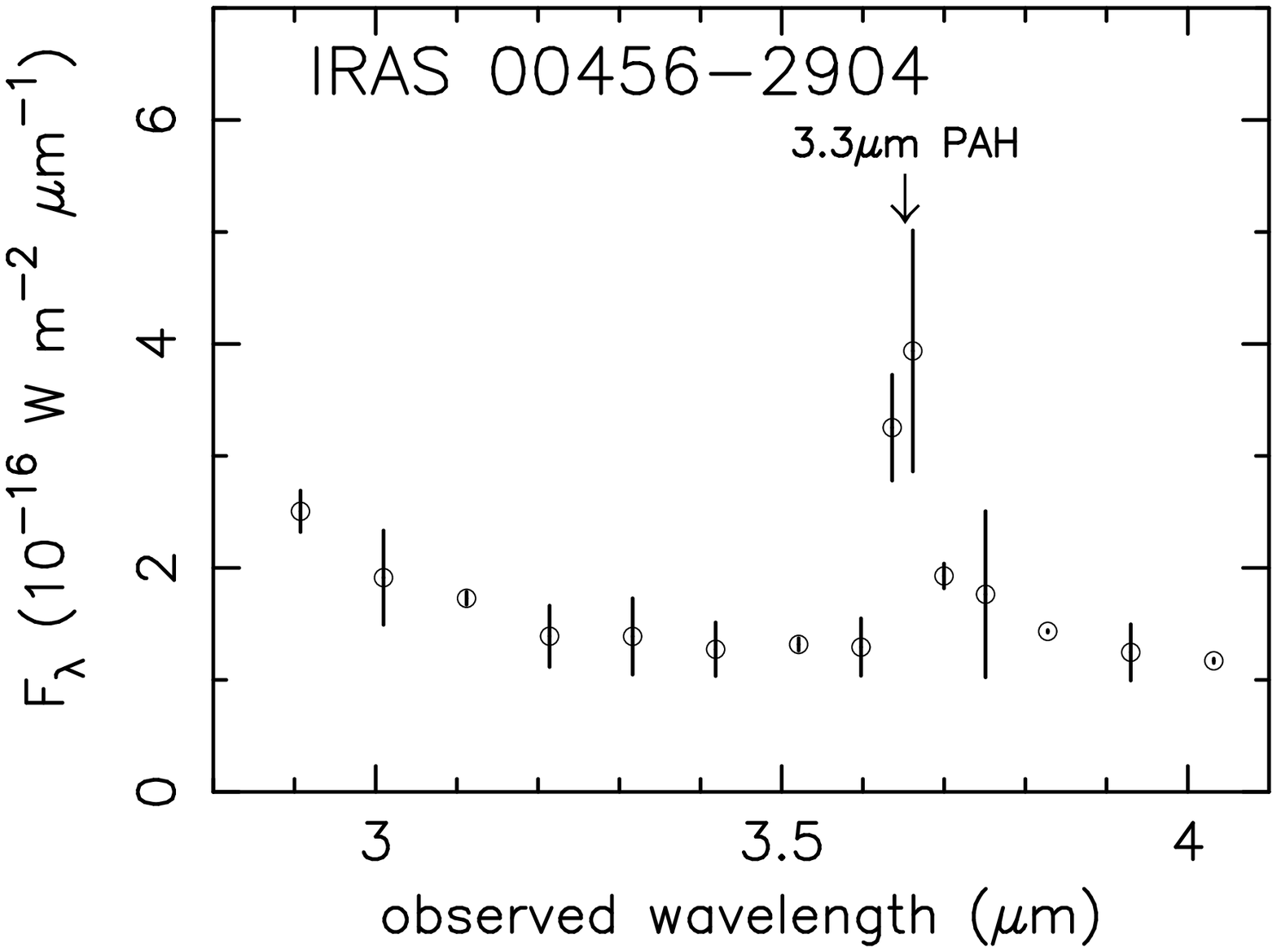}
\FigureFile(80mm,80mm){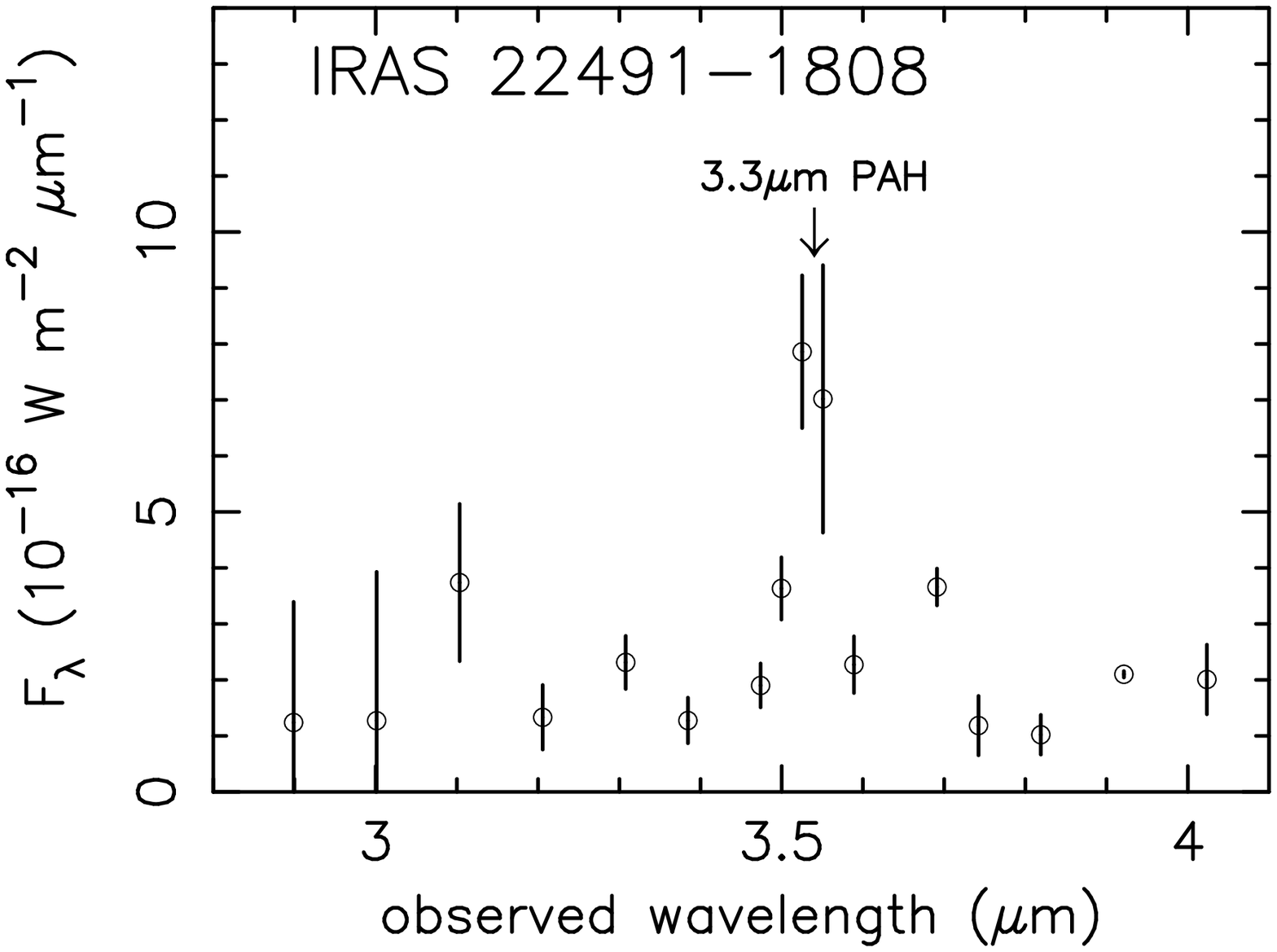}
\FigureFile(80mm,80mm){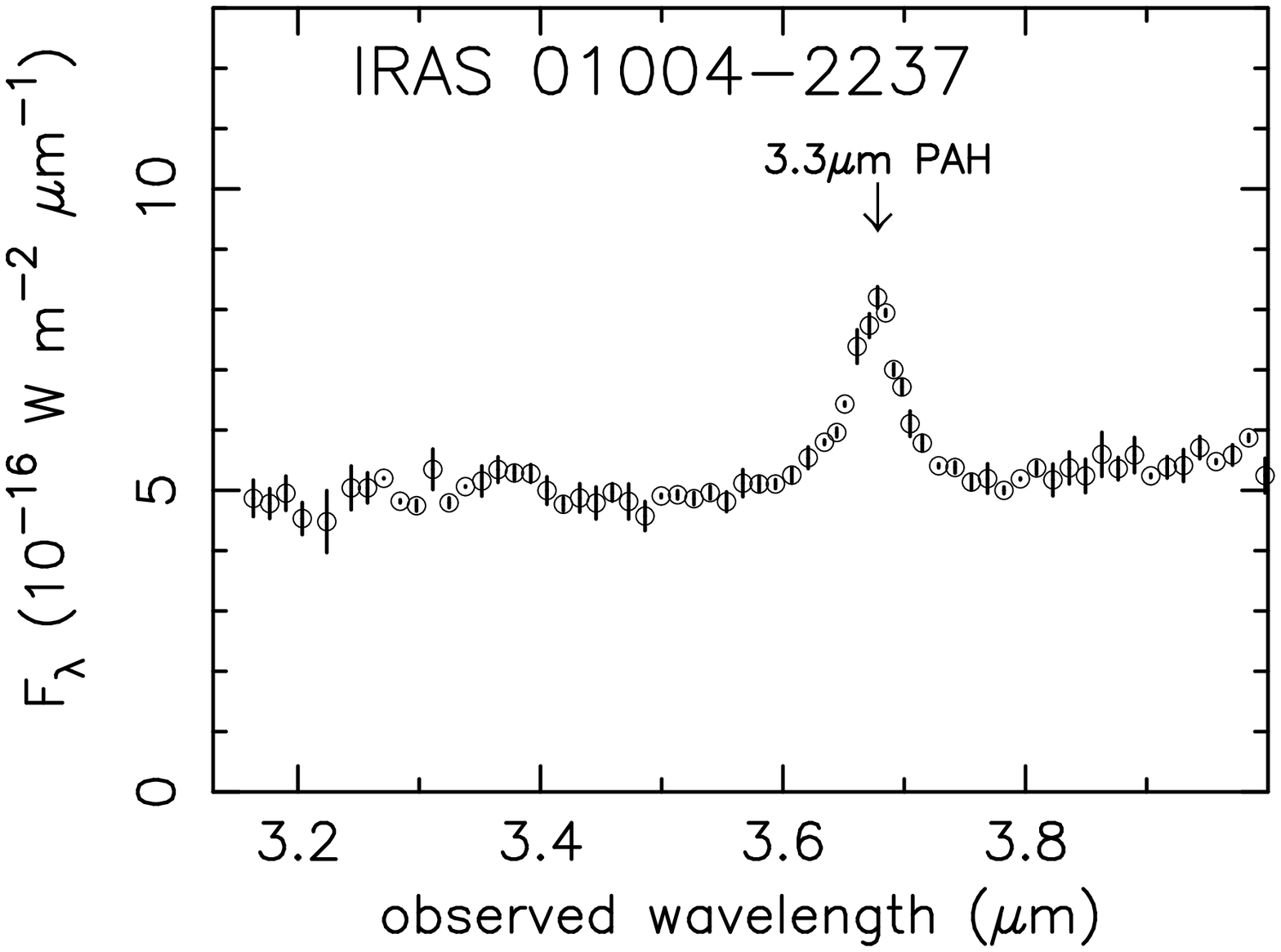}
\FigureFile(80mm,80mm){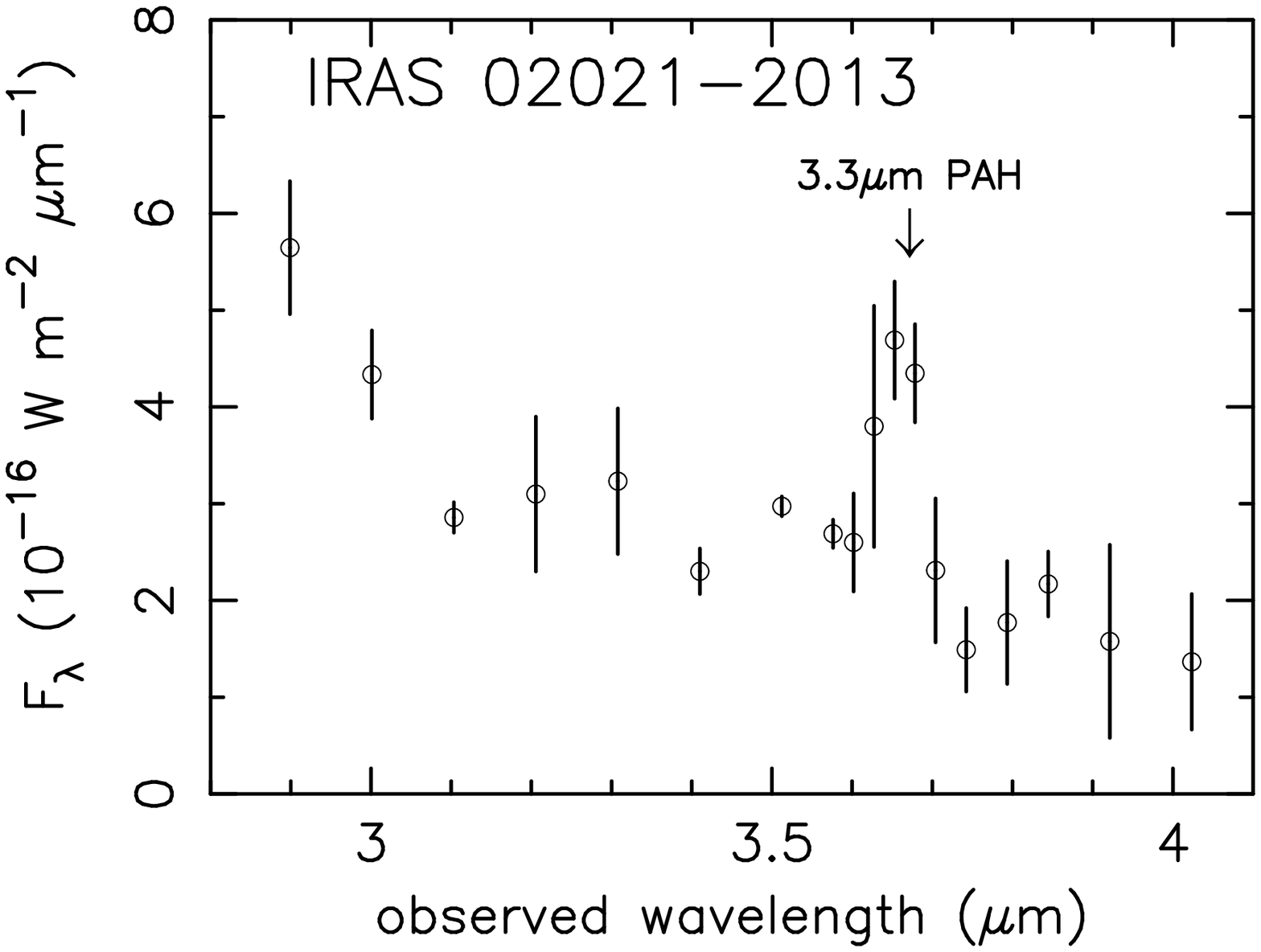}
\end{figure}

\clearpage

\begin{figure}
\FigureFile(80mm,80mm){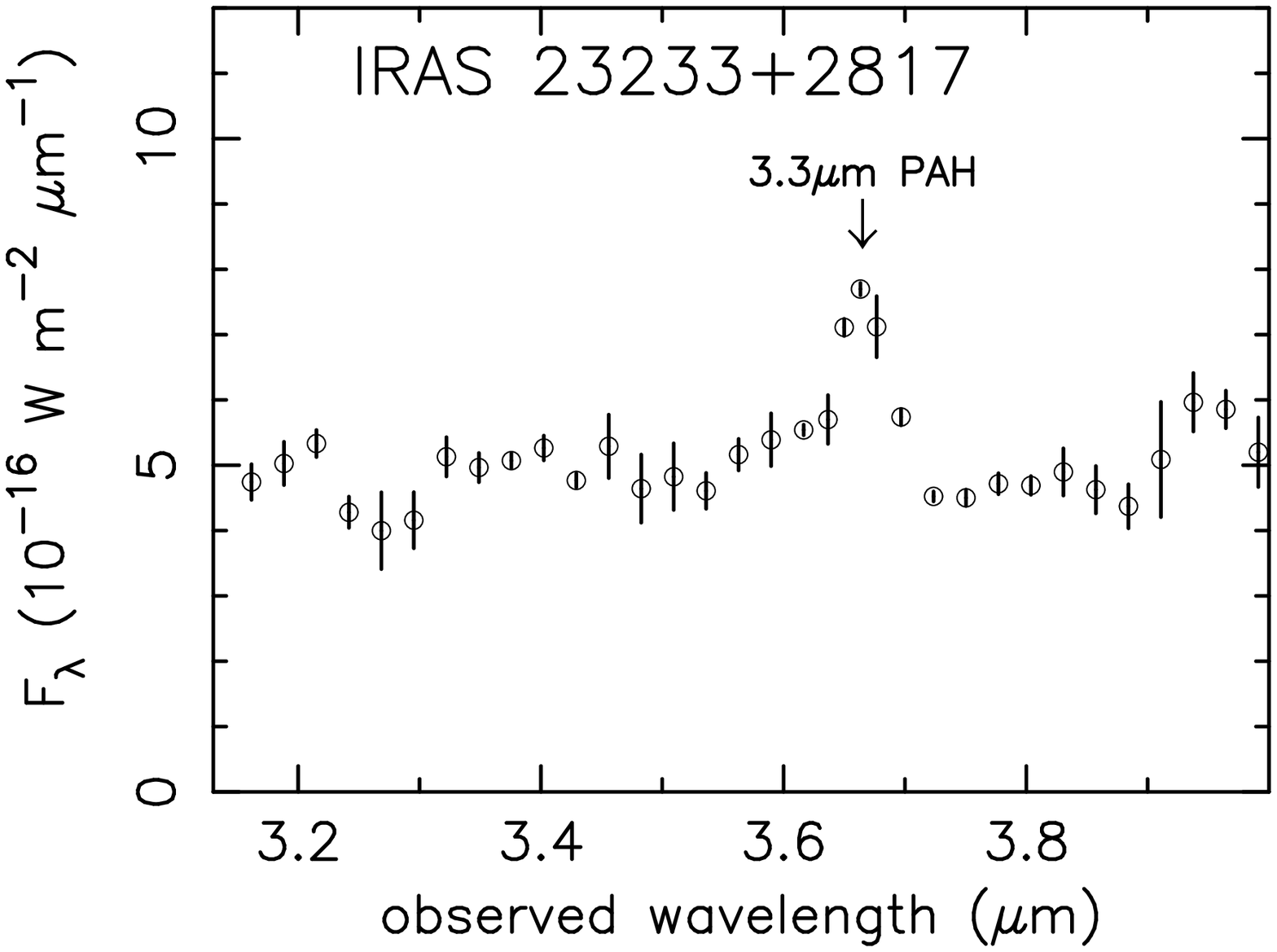}
\FigureFile(80mm,80mm){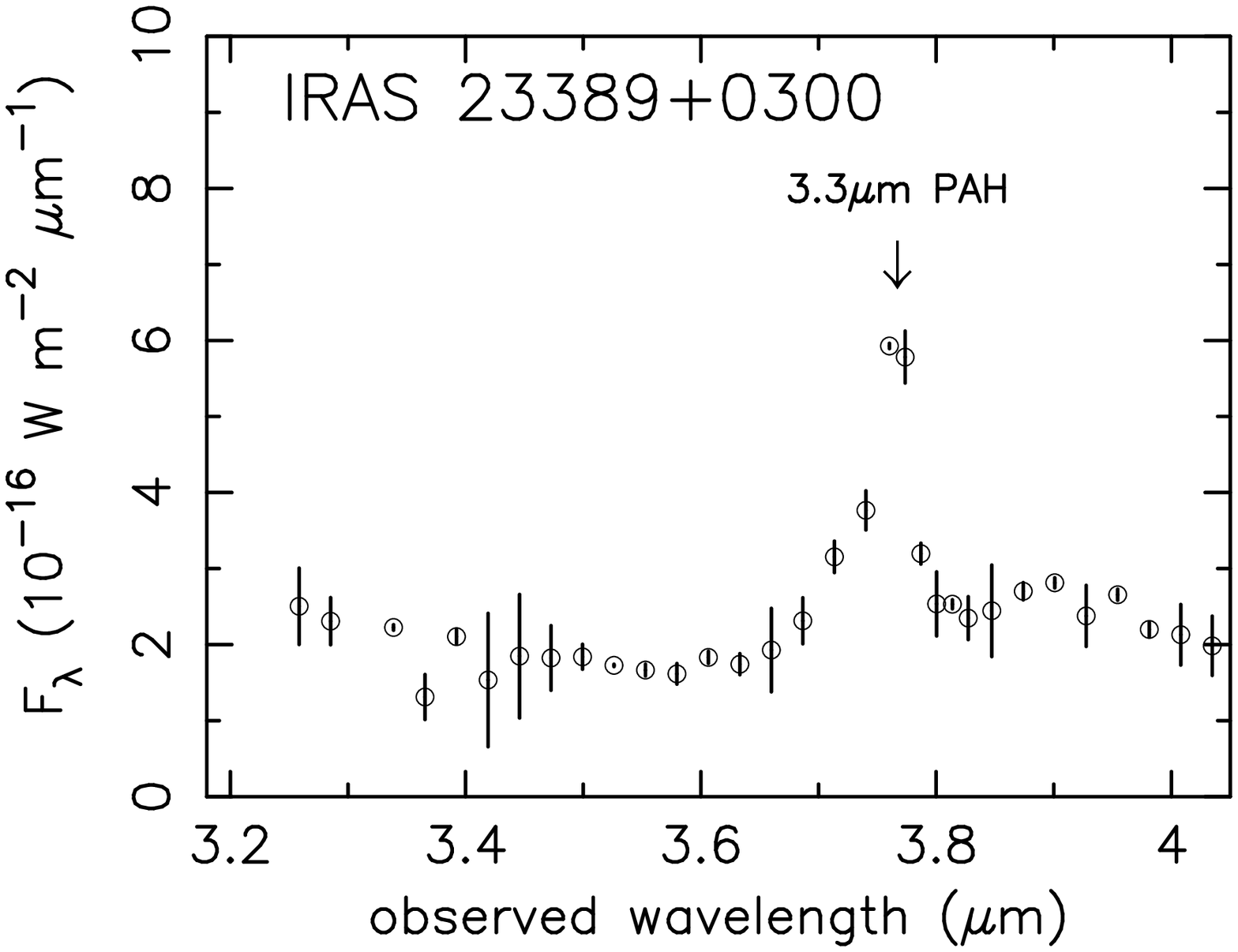}
\caption{
Ground-based infrared $L$-band (2.8--4.1 $\mu$m) spectra of the observed
ULIRGs.  
The abscissa is the observed wavelength in $\mu$m, and the ordinate is
F$_{\lambda}$ in 10$^{-16}$ W m$^{-2}$ $\mu$m$^{-1}$, following
previously shown ground-based $L$-band spectra
\citep{idm06,ima06}. The lower arrows with ``3.3 $\mu$m PAH'' indicate
the expected wavelength of the 3.3 $\mu$m PAH emission ($\lambda_{\rm
rest}$ = 3.29 $\mu$m). 
}
\end{figure}

\end{document}